\documentclass[10pt,journal,compsoc]{IEEEtran}
\usepackage{booktabs} 
\usepackage{amsmath}
\usepackage{pdfsync}
\ifCLASSOPTIONcompsoc
  \usepackage[nocompress]{cite}
\else
  \usepackage{cite}
\fi
\ifCLASSINFOpdf
   \usepackage[pdftex]{graphicx}
\else
\fi
\usepackage{xcolor}
\usepackage{stfloats}
\usepackage{booktabs} 
\usepackage{graphicx}
\usepackage{tikz}
\usetikzlibrary{tikzmark}
\usepackage{caption}
\usepackage{subcaption}
\usepackage{subfloat}
\usepackage{float}
\usepackage{transparent}
\usepackage{xcolor}
\usepackage{titlecaps}
\usepackage{booktabs}
\usepackage{multirow}
\usepackage{ulem} 
\usepackage{array}
\usepackage[ruled]{algorithm2e} 
\usepackage{hyperref}
\usepackage{tabularx} 
\usepackage{soul}

\newcommand{\figref}[1]{Fig.~\ref{#1}}

\newcommand{\shortcite}[1]{\cite{#1}}
\newcommand{\warning}[1]{}
\newcommand{\note}[1]{#1}

\def\sec#1{Sec.~\ref{#1}}
\newcommand{\xzs}[1]{{#1}}

\begin{document}
\title{Neural Enhancement of Analytical Appearance Models}

\author{Xuanzhe~Shen, Xiaohe~Ma, Kun~Zhou~\IEEEmembership{Fellow,~IEEE} and Hongzhi~Wu~\IEEEmembership{Member,~IEEE}
\IEEEcompsocitemizethanks{
\IEEEcompsocthanksitem K. Zhou and H. Wu are the corresponding authors. X. Shen and X. Ma contributed equally. All authors are with State Key Lab of CAD~\&~CG, Zhejiang University, Hangzhou, China, 310058. K. Zhou is also affiliated with Hangzhou Research Institute of AI and Holographic Technology.\protect\\
E-mail: \{kunzhou,hwu\}@acm.org}
}
\markboth{Journal of \LaTeX\ Class Files,~Vol.~14, No.~8, August~2015}%
{Ma \MakeLowercase{\textit{et al.}}: Neural Enhancement of Analytical Appearance Models}
\IEEEtitleabstractindextext{
\begin{abstract}
Traditional analytical reflectance models, while compact and interpretable, lack the capacity to accurately represent physical measurements. Recent neural models, which closely fit input data, are less generalizable and often more expensive to store and evaluate. To combine the strengths and overcome the limitations of these two classes of models, we present neural enhancement, a novel framework to boost an input analytical appearance model, by identifying and replacing its key computational nodes/operators with small-scale multi-layer perceptrons. This allows us to leverage the computational graph structure of the original model, while improving its expressiveness at a modest cost. To make the enhancement computationally tractable, we propose a hypercube-based search to automatically and efficiently identify the node(s) and/or operator(s) to be replaced towards maximal gain in a differentiable fashion. We enhance a number of common analytical BRDF models. The results are, at once accurate, compact and efficient, and compare favorably with state-of-the-art work on fitting measured reflectance\warning{ and bidirectional texture functions}. Finally, our models are fully compatible with\warning{ any} standard rasterization or ray-tracing pipeline.
\end{abstract}
}

\maketitle

\IEEEdisplaynontitleabstractindextext

%
\IEEEpeerreviewmaketitle

\IEEEraisesectionheading{\section{Introduction}\label{sec:introduction}}

%
%
%
%
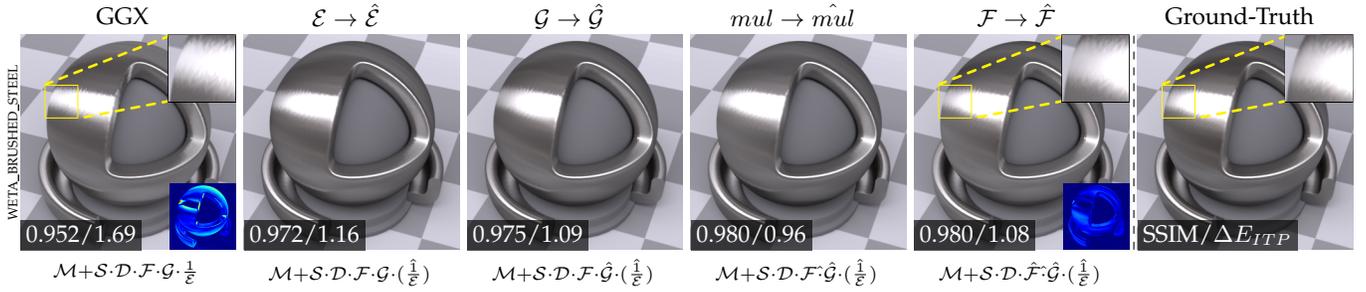
\begin{figure*}
    \centering
    \begin{minipage}{\textwidth}
        \centering
        \begin{minipage}{\textwidth}
            \centering
            \begin{minipage}[c]{0.158\linewidth}
                \centering
                \subcaption{\small GGX}
            \end{minipage}
            \begin{minipage}[c]{0.158\linewidth}
                \centering
                \subcaption{\small $\mathcal{E}\rightarrow \hat{\mathcal{E}}$}
            \end{minipage}
            \begin{minipage}[c]{0.158\linewidth}
                \centering
                \subcaption{\small $\mathcal{G}\rightarrow \hat{\mathcal{G}}$}
            \end{minipage}
            \begin{minipage}[c]{0.158\linewidth}
                \centering
                \subcaption{\small $mul\rightarrow \hat{mul}$}
            \end{minipage}
            \begin{minipage}[c]{0.158\linewidth}
               \centering
                \subcaption{\small $\mathcal{F}\rightarrow \hat{\mathcal{F}}$}
            \end{minipage}
            \begin{minipage}[c]{0.158\linewidth}
                \centering
                \subcaption{\small Ground-Truth}
            \end{minipage}
        \end{minipage}
        \begin{minipage}{0.03in}	
            \rotatebox{90}{\tiny \uppercase{Weta\_brushed\_steel}}
        \end{minipage}	
        \begin{minipage}[c]{0.158\textwidth}
        
            \begin{tikzpicture}
                \node[anchor=south west,inner sep=0] (image) at (0,0) {\includegraphics[width=0.995\linewidth]{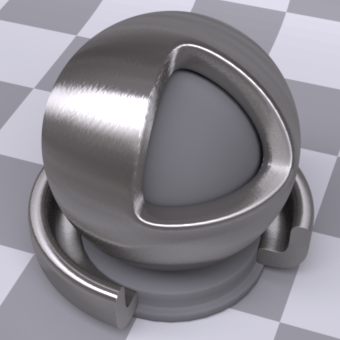}};
                \begin{scope}[shift={(image.north east)}, overlay, remember picture]
                    \draw[yellow, dashed, line width=1pt, line cap=round] (-25pt,-0.6pt) -- ++(-47pt,-18.4pt);
                    \draw[yellow, dashed, line width=1pt, line cap=round] (-25pt,-25.3pt) -- ++(-35pt,-6.7pt);
                \end{scope}
            \end{tikzpicture}
            \put(-81.4,5.5) {\tikz[baseline] \node[fill=black, fill opacity=0.65, text opacity=1, text=white,inner sep=2pt] {\small 0.952/1.69};}
            \put(-72,50){\color{yellow}\framebox(12,12){}}
            \put(-25.4,56){\color{black}\framebox(25,25){}}
            \put(-25.4,56) {\includegraphics[width=25pt]{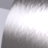}}
            \put(-25,0) {\includegraphics[width=25pt]{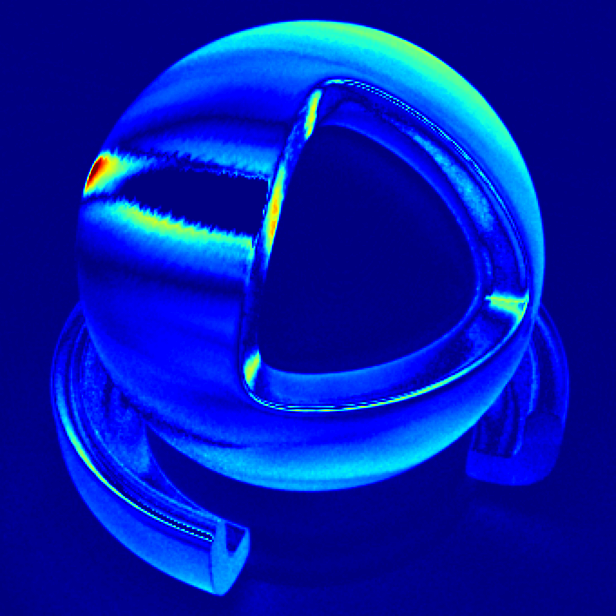}}
        \end{minipage}
        \begin{minipage}[c]{0.158\textwidth}
            \includegraphics[width=0.995\linewidth]{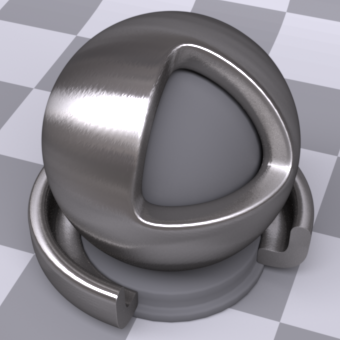}
            \put(-81.4,5.5) {\tikz[baseline] \node[fill=black, fill opacity=0.65, text opacity=1, text=white,inner sep=2pt] {\small 0.972/1.16};}
            \label{fig:sub2}
        \end{minipage}
        \begin{minipage}[c]{0.158\linewidth}
            \includegraphics[width=0.995\linewidth]{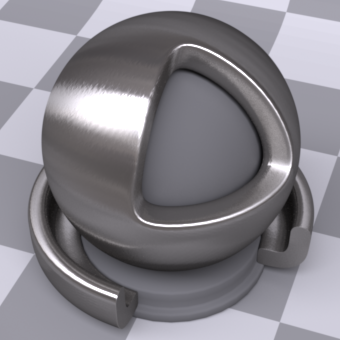}
            \put(-81.4,5.5) {\tikz[baseline] \node[fill=black, fill opacity=0.65, text opacity=1, text=white,inner sep=2pt] {\small 0.975/1.09};}
        \end{minipage}
        \begin{minipage}[c]{0.158\linewidth}
            \includegraphics[width=0.995\linewidth]{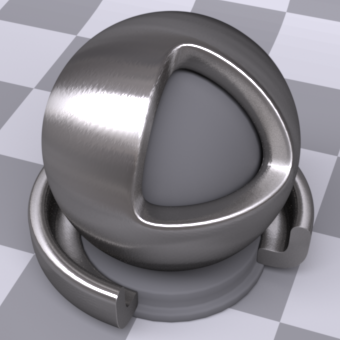}
            \put(-81.4,5.5) {\tikz[baseline] \node[fill=black, fill opacity=0.65, text opacity=1, text=white,inner sep=2pt] {\small 0.980/0.96};}
        \end{minipage}
        \begin{minipage}[c]{0.158\linewidth}
            \begin{tikzpicture}
                \node[anchor=south west,inner sep=0] (image) at (0,0) {\includegraphics[width=0.995\linewidth]{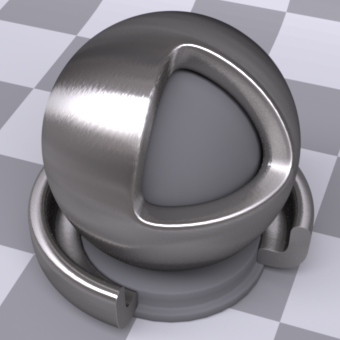}};
                \begin{scope}[shift={(image.north east)}, overlay, remember picture]
                    \draw[yellow, dashed, line width=1pt, line cap=round] (-25pt,-0.6pt) -- ++(-47pt,-18.4pt);
                    \draw[yellow, dashed, line width=1pt, line cap=round] (-25pt,-25.3pt) -- ++(-35pt,-6.7pt);
                \end{scope}
            \end{tikzpicture}
            \put(-81.4,5.5) {\tikz[baseline] \node[fill=black, fill opacity=0.65, text opacity=1, text=white,inner sep=2pt] {\small 0.980/1.08};}
            \put(-72,50){\color{yellow}\framebox(12,12){}}
            \put(-25.4,56){\color{black}\framebox(25,25){}}
            \put(-25.4,56) {\includegraphics[width=25pt]{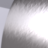}}
            \put(-25,0){\includegraphics[width=25pt]{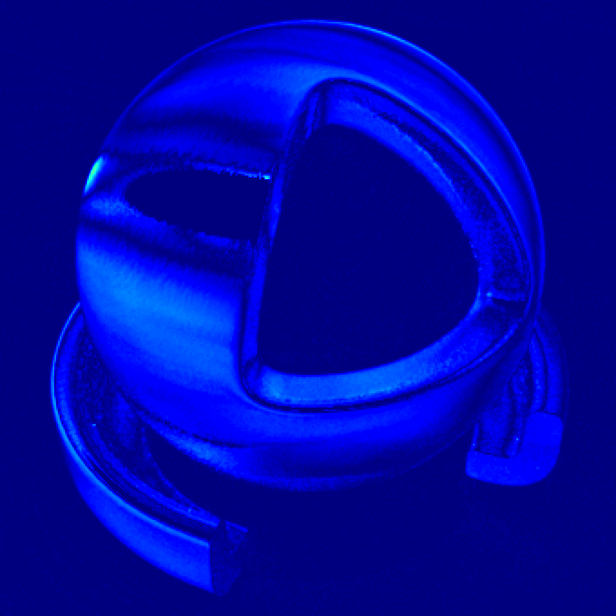}}
        \end{minipage}
        \begin{minipage}[c]{0.158\linewidth}
            \begin{tikzpicture}
                \node[anchor=south west,inner sep=0] (image) at (0,0) {\includegraphics[width=0.995\linewidth]{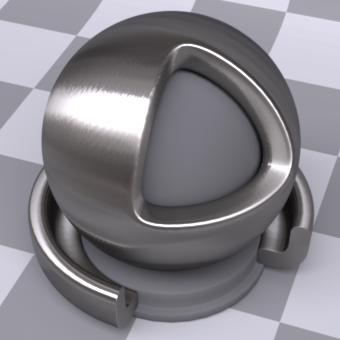}};
                \begin{scope}[shift={(image.north east)}, overlay, remember picture]
                    \draw[yellow, dashed, line width=1pt, line cap=round] (-25pt,-0.6pt) -- ++(-47pt,-18.4pt);
                    \draw[yellow, dashed, line width=1pt, line cap=round] (-25pt,-25.3pt) -- ++(-35pt,-6.7pt);
                \end{scope}
            \end{tikzpicture}
            \put(-81.4,5.5) {\tikz[baseline] \node[fill=black, fill opacity=0.65, text opacity=1, text=white,inner sep=2pt] {\small SSIM/$\Delta E_{ITP}$};}
            \put(-72,50){\color{yellow}\framebox(12,12){}}
            \put(-25.4,56){\color{black}\framebox(25,25){}}
            \put(-25.4,56) {\includegraphics[width=25pt]{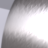}}
        \end{minipage}
    \end{minipage}
    \begin{minipage}{\textwidth}
        \centering
        \begin{minipage}[c]{0.158\linewidth}
            \centering
            \text{ $\scriptstyle \mathcal{M}{+}\mathcal{S} \cdot \mathcal{D {\cdot} {\mathcal{F}}} \cdot {\mathcal{G}} \mathcal{{\cdot}} \frac{1}{{\mathcal{E}}}$}
        \end{minipage}
        \begin{minipage}[c]{0.158\linewidth}
            \centering
            \text{ $\scriptstyle \mathcal{M}{+}\mathcal{S} \cdot \mathcal{D {\cdot} {\mathcal{F}}} {\cdot} {\mathcal{G}} \mathcal{{\cdot}} \hat{(\frac{1}{\mathcal{E}})}$}
        \end{minipage}
        \begin{minipage}[c]{0.158\linewidth}
            \centering
            \text{ $\scriptstyle \mathcal{M}{+}\mathcal{S} \cdot \mathcal{D {\cdot} {\mathcal{F}}} {\cdot} \hat{\mathcal{G}} \mathcal{{\cdot}} \hat{(\frac{1}{\mathcal{E}})}$}
        \end{minipage}
        \begin{minipage}[c]{0.158\linewidth}
            \centering
            \text{ $\scriptstyle \mathcal{M}{+}\mathcal{S} \cdot \mathcal{D {\cdot} {\mathcal{F}}} \hat{\cdot} \hat{\mathcal{G}} \mathcal{{\cdot}} \hat{(\frac{1}{\mathcal{E}})}$}
        \end{minipage}
        \begin{minipage}[c]{0.158\linewidth}
            \centering
            \begin{tikzpicture}
                \node (txt) at (0,0) {\text{ $\scriptstyle \mathcal{M}{+}\mathcal{S} \cdot \mathcal{D {\cdot} \hat{\mathcal{F}}} \hat{\cdot} \hat{\mathcal{G}} \mathcal{{\cdot}} \hat{(\frac{1}{\mathcal{E}})}$}};
                \begin{scope}[shift={(txt.east)}, overlay, remember picture]
                    \draw[dashed] (0.32cm,0.3cm) -- ++(0,2.9cm);
                \end{scope}
            \end{tikzpicture}
        \end{minipage}
        \begin{minipage}[c]{0.158\linewidth}
            \centering
            \text{{}}
        \end{minipage}
    \end{minipage}
    \caption{We enhance the accuracy of an input analytical appearance model, by automatically identifying and replacing original computational nodes/operators with small multi-layer perceptrons (indicated with hats). The 1st to 5th images visualize the boosting process for GGX BRDF model, resulting in a new, hybrid model that more closely matches the measured ground-truth (last image). Quantitative errors in SSIM and $\Delta E_{ITP}$ ($\times10^3$) rae reported in the bottom-left of each related image. We show the error maps of both the original GGX BRDF and the enhanced GGX BRDF, along with zoomed-in results, on the right side of the corresponding images. The vanilla GGX cannot accurately fit grazing-angle anisotropic reflections, unlike our model. Please refer to~\figref{fig:pipeline} for the definition of each term in the analytical model.}
   \label{fig:teaser}
\end{figure*}


\IEEEPARstart{M}{odeling} the visual appearance of objects is a fundamental problem in computer graphics~\cite{Dorsey:2010:Digital}. For reflective appearance, Bidirectional Reflectance Distribution Function (BRDF) is a 4D function that completely describes the reflectance distribution of a material for any lighting and view directions~\cite{nicodemus1965directional}. To represent a BRDF, a straightforward tabulation would impose a heavy burden on storage, editing, and rendering, due to the sheer dimensionality of this function~\cite{Matusik2003:MERL}. Therefore, seeking an optimal appearance representation has been an important on-going quest over the past 50 years.

Starting with the seminal work of~\cite{phong1975illumination}, researchers leverage the knowledge from fields including physics and statistics, to derive a series of highly compact analytical models for representing a BRDF. They range from early phenomenological models~\cite{Ward1992:Ward,Lafortune1997:Lafortune} to first principles ones~\cite{Cook1982:CookTorrance,oren1994generalization,ashikhmin2000anisotropic}, many of which are still actively used in academia and industry today. Although analytical models are compact, efficient, and interpretable/editable, they lack the capacity to fit to measured data with high precision~\cite{Ngan2005:Experimental}.



On the other hand, neural implicit models have gained increasing popularity in the last decade. Since neural networks with sufficient capacities are essentially universal approximators, these models precisely fit measured data with a BRDF-specific network~\cite{Sztrajman2021:NBRDF}, or a large BRDF-independent network in conjunction with optimizeable latent parameters~\cite{FAN2022:NLBRDF}. However, as a data-driven approach, neural models are not easily generalizable. \xzs{Moreover, the lack of semantics in an end-to-end neural approach makes interpretation/editing challenging.}

Our goal is to \xzs{combine the strengths of both analytical and neural models}, by finding a reflectance model that is at once, accurate, compact, and efficient. We are inspired by the historical development of analytical BRDF models: expert researchers often build new models upon existing ones; they first identify under-performing term(s) (e.g., the microfacet normal distribution function, or the Fresnel term), and then replace with a new, carefully hand-crafted version, to improve the overall performance. We would like to harness the power of modern machine learning tools to perform this task in an automatic and systematic manner.

In this paper, we present neural enhancement, a novel framework to enhance an input analytical appearance model, which identifies and replaces bottleneck computational node(s) and/or operator(s) using \note{multilayer perceptrons (MLPs)}, toward maximum gain in fitting accuracy. This allows us to leverage the existing computational graph structure of the original model, while improving its expressiveness at a modest cost. As a straightforward algorithm would result in a combinatorial explosion, we propose a hypercube-based search to test and gradually update only a small subset of all replacement options, making the process computationally tractable.



The effectiveness of neural enhancement is demonstrated on the standard anisotropic Cook-Torrance BRDF model with GGX distribution~\cite{Walter2007:GGX} (parameter\# = 12). Using the training data from multiple \warning{BRDF and bidirectional texture function (BTF)} \note{appearance} datasets, our enhanced model (parameter\# = 39) is a unified representation for accurate fitting of complex measured BRDFs~\cite{Dupuy2018:Adaptive}\warning{as well as captured BTF texels}, whose quality is comparable or better than state-of-the-art BRDF models. \note{Due to the simple structure of MLP,} our model can be directly translated into shader code for standard rasterization or ray-tracing pipeline. The performance of an unoptimized implementation of the enhanced model is comparable to a state-of-the-art neural method with the smallest network size~\cite{Sztrajman2021:NBRDF}. We also apply our idea to improve other analytical models~\cite{Cook1982:CookTorrance,Ward1992:Ward,Brady2014:Genbrdf}, \note{test on bidrectional texture functions (BTFs)} and evaluate the impact of different factors on the final result. We hope that our hybrid approach will inspire interesting future work along this direction.

\section{Related Work}
\label{sec:related_work}

We review related work on bidirectional appearance functions defined at a \textit{single} location. No spatial coherence is exploited, unlike in, e.g., \cite{FAN2023:Biplane,gao2019deep}. While our focus is on BRDF, we also list some BTF-related techniques, due to the similarity in handling a BRDF and a BTF texel: both are parameterized over a pair of lighting and view directions. \note{Another line of work~\cite{shi2020match, guerrero2022matformer} also involves node graphs for appearance, but is orthogonal to ours as their focus is on the spatial domain}. Interested readers are referred to excellent recent surveys for a broader view of the topic~\cite{Weinmann2015COURSE,Weyrich2009SURVEY,Guarnera2016BRDF,Dong2019:Survey,kavoosighafi2024deep}.

\subsection{Analytical BRDF Models}
This class of work can be roughly divided into phenomenological and first principles ones. Phenomenological models rely on careful observations/expertise to develop compact analytical expressions for appearance modeling. For example, the seminal work of~\cite{phong1975illumination} proposes a simple model for computing specular highlights. Ward~\shortcite{Ward1992:Ward} introduces a similar model for anisotropic reflections based on \note{a distribution suggested by Beckmann and Spizzichino~\cite{Beckmann1963}}. Ashikhmin and Shirley~\shortcite{ashikhmin2000anisotropic} further adopt an explicit Fresnel term and a non-constant diffuse one to improve expressiveness.


First principles models are based on manual derivations directly from or inspired by physics. \note{Torrance and Sparrow establish the microfacet theory~\cite{Torrence1967:TorrenceSparrow}, which is later introduced to graphics by Blinn~\cite{Blinn1977:BlinnPhong}, and finally leads to the widely used Cook-Torrance BRDF~\shortcite{Cook1982:CookTorrance}.} Schlick~\shortcite{schlick1994inexpensive} develops efficient approximations for the Fresnel term and the shadowing term of~\cite{smith1967geometrical}. The shadowing function is derived from the numeric integration of arbitrary microfacet distributions in~\cite{michael2000microfacet}. Walter et al.~\shortcite{Walter2007:GGX} rediscover~\cite{trowbridge1975average} as the GGX (Trowbridge-Reitz) distribution, and this model becomes the de-facto standard for physically-based rendering (PBR) nowadays. For brevity, we refer to it as the GGX BRDF model or GGX model in this paper.

Recently, Wu et al.~\shortcite{wu2011sparse} search for a sparse mixture of analytical BRDF models to represent a BTF texel, which can be viewed as an apparent BRDF (ABRDF). Brady et al.~\shortcite{Brady2014:Genbrdf} employ genetic programming in deriving mathematical expressions to characterize complex BRDFs via a large-scale optimization.


Despite considerable continuing research efforts~\cite{kurt2010anisotropic, bagher2012accurate, low2012brdf}, it is fundamentally challenging for an analytical model to fit measured data with high accuracy~\cite{Ngan2005:Experimental}, due to the insufficient capacity of its compact expression. By neural enhancement of an analytical model, our approach increases its expressiveness by replacing it with a few small MLPs, while inheriting many of its strengths, such as compactness, efficiency, and generalizability.


\begin{figure*}[htbp]
    \centering
        \begin{minipage}{7.1in}	
            \centering
            \begin{minipage}{1.0\linewidth}
            \includegraphics[width=\linewidth]{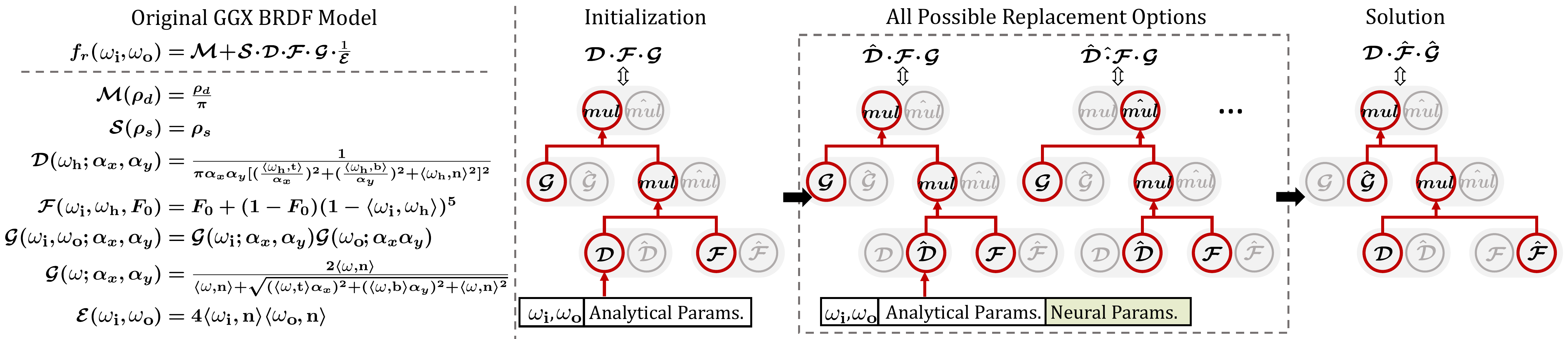}
            \end{minipage}	
        \end{minipage}	
    \caption{Our enhanced model. Here the GGX BRDF model is an input example, defined in the left part of the figure. We start with converting the analytical model into a computational graph of computational nodes and arithmetic operators. Note that only a subgraph is displayed here, due to limited space. Each node/operator can be replaced with a neural module (indicated with a hat), with analytical parameters replaced by the concatenation of analytical and neural ones. Ideally, we would like to pick the model with the lowest loss among all possible replacement options, as the final solution. Params. = parameters.}
\label{fig:pipeline}
\end{figure*}

\subsection{Neural Appearance Models}
These models can be categorized into \textbf{material-specific} and \textbf{material-independent} ones. The first line of work trains a specific network for each material, essentially trading generalization for accuracy. Rainer et al.~\shortcite{Rainer2019:NeuralBTF} introduce a neural encoding of BTF texels with a material-specific autoencoder. Sztrajman et al.~\shortcite{Sztrajman2021:NBRDF} improve highlight reconstruction with lightweight networks using adaptive angular sampling. Trunz et al.~\cite{trunz2022efficient} propose a Shapley-value-based method for latent space pruning. This method does not require retraining and can preserve angular appearance fidelity while achieving compact reflectance representations. In a follow-up paper~\cite{gokbudak2024hypernetworks}, Gokbudak et al. generate the network weights through a hyper-network, enabling reconstruction even under sparse sampling conditions. Zeltner et al.~\cite{zeltner2024real} introduce a real-time neural appearance model that encodes high-resolution Spatially Varying Bidirectional Reflectance Distribution Function (SVBRDF) parameters into a hierarchical latent texture, decoded by two lightweight MLPs for BRDF evaluation and importance sampling. 


Another line of work proposes a general network whose weights are fixed after training. It can represent different materials with optimizable per-material neural parameters. Hu et al.~\shortcite{Hu2020:Deepbrdf} design an autoencoder to discover a low-dimensional latent space from a sequence of BRDF slices. Rainer et al.~\shortcite{Rainer2020:UnifiedBTF} build upon~\cite{Rainer2019:NeuralBTF} to train a unified network architecture, which projects reflectance measurements to a shared latent parameter space. \note{Kuznetsov et al.~\cite{Kuznetsov2021:Neumip} propose a neural framework for multi-scale appearance, which is improved in~\cite{xue2024hierarchical} to better handle complex shadows and specular highlights.} Fischer and Ritschel~\cite{fischer2022metappearance} propose a meta-learning approach for BRDF and SVBRDF estimation that achieves near-overfit quality in a few gradient steps using a learned initialization and step size. Zheng et al.~\shortcite{Zheng2023:CompactBRDF} work in the BRDF function spaces, and exploit neural processes to obtain a compact latent representation. Fan et al.~\shortcite{FAN2022:NLBRDF} perform a learned layering operation on BRDF latent codes via a large network to recover the original function. A biplane representation for BTF is proposed in~\cite{FAN2023:Biplane}, which learns to encode high-frequency details in both spatial and angular domains. \note{
A generalized normal distribution function is introduced to learn high-frequency directional patterns~\cite{Kuznetsov2019GAN}. Shi et al.~\cite{shi2022biscale} train a universal network to predict meso-scale geometry and micro-scale BRDFs based on a large-scale BRDF. Conceptually, our approach can be viewed as a generalized version of the above two, as all nodes/operators can be neural enhanced.}

As a data-driven approach, neural models with sufficient capacities can precisely (over)fit measurements. However, the network size is often large in order to implement a complex mapping, resulting in expensive training and rendering. The generalization ability is also limited, due to the lack of first principles often used in the design of analytical models. Although our model belongs to the class of \textbf{material-independent}, general neural models, it inherits the majority of the computational graph structure of an input analytical model, and only employs a few small MLPs. These two factors considerably alleviate the aforementioned issues.


\subsection{Symbolic Regression}
Symbolic regression aims to discover a symbolic expression $f$ from a number of $(x, f(x))$ pairs~\cite{schmidt2009distilling}. Recently, Petersen et al.~\shortcite{petersen2019deep} use a recurrent neural network to generate a distribution over tractable mathematical expressions, and train the network with a novel risk-seeking policy gradient to improve expression fitting. A method is proposed in~\cite{udrescu2020ai} to identify generalized symmetries in computational graphs using neural network gradients, extending symbolic regression to probability distributions with normalizing flows. Biggio et al.~\shortcite{biggio2021neural} use large-scale pre-training algorithms to procedurally generate an infinite set of equations, and train a transformer to predict symbolic equations from input-output pairs. In comparison, we do not derive pure symbolic expressions from scratch, due to its limited capacity.

\section{Our Model}
\label{sec:our_model}
Our enhanced model can be viewed as a modified version of an analytical one. Given an input analytical model, we first convert it into a computational graph with terminal computational nodes and internal arithmetic operators, as illustrated in~\figref{fig:pipeline}. Next, some of the nodes and/or operators will be identified and replaced (detailed in~\sec{sec:method}) with small MLPs, which we call neural modules, to improve the overall accuracy. 

The nodes represent semantic terms defined in the original model. For example, the nodes of the GGX BRDF model are Lambertian reflectance (denoted as $\mathcal{M}$), specular albedo ($\mathcal{S}$), microfacet distribution function ($\mathcal{D}$), Fresnel approximation ($\mathcal{F}$), geometric factor ($\mathcal{G}$), and the reciprocal normalization term ($\mathbf{\frac{1}{\mathcal{E}}}$), as shown in~\figref{fig:pipeline}. We use the reciprocal term to avoid division as an operator, which tends to be less stable in optimization. The original input are the lighting/view direction ($\mathbf{\omega_i}$/$\mathbf{\omega_o}$), as well as BRDF parameters (diffuse/specular albedo:$\mathbf{\rho_d}$/$\mathbf{\rho_s}$, roughnesses:$\mathbf{\alpha_x}$/$\mathbf{\alpha_y}$, the specular reflectance at normal incidence:$\mathrm{F}_0$, normal:$\mathbf{n}$, and tangent expressed as a rotation angle $\mathbf{t_\theta}$).

For an operator, it takes the results from two children nodes/neural modules as input, and outputs the result of the arithmetic computation it represents. In this paper, addition/multiplication are used as operators.

The neural module is simply a small MLP. Its output dimensionality is designed to match that of the node or operator it replaces. The input configuration of the MLP is determined by the specific node being replaced. If the replaced node takes any input BRDF parameters, these are substituted with a concatenation of all analytical parameters and a newly introduced set of neural parameters. The inclusion of $\mathbf{\omega_i}$,$\mathbf{\omega_o}$ in this concatenation depends on whether they serve as inputs to the original node. For example, we are enhancing the GGX BRDF model, and setting the number of neural parameters to 27; a node takes $\mathbf{\alpha_x}$,$\mathbf{\alpha_y}$ and $\mathbf{\omega_i}$,$\mathbf{\omega_o}$ as input. In this case, the input dimension to the MLP would be $12$(original analytical parameters)+$27$(neural parameters)+$6$(for $\mathbf{\omega_i}$,$\mathbf{\omega_o}$)$=45$. If an operator is replaced, the input remains as two operands from two children. In this paper, the MLP is a 4-layer fully connected network with 16, 32, 16 neurons per hidden layer respectively (\figref{fig:module}). Each FC layer before the last one is followed by a leaky ReLU activation layer. 

Note that our approach is not limited to the above configurations. We experiment with various settings and find they perform well in practice. Please refer to~\sec{sec:evaluations} for the impact of various factors over the final result, including MLP architectures, sizes, and the number of neural parameters. It is also worth mentioning that there are multiple ways to build the computational graph. A node could group together multiple semantic terms as well as their mutual operators. However, doing so draws closer to an end-to-end neural network. On the other hand, a node can be as small as a single input parameter/constant. This, however, significantly increases the computational cost, and makes it difficult to find a suitable neural module for such a simple node. The current granularity of nodes mimics the manual development of analytical models. Additionally, our method is not limited to assigning each node to either a neural module or a single analytical expression. In principle, candidate components from multiple analytical models and the neural module could jointly compete at each node. However, to avoid a rapid increase in complexity, we currently enhance one analytical model at a time. We leave it to future work to automatically determine the optimal granularity for each node and to enable more comprehensive enhancement that incorporates multiple analytical models.

\section{Neural Enhancement}
\label{sec:method}

In this section, we describe how to identify and replace nodes/operators with neural modules (\sec{sec:our_model}) in the computational graph of an input analytical model, towards maximum gain in fitting quality. A na\"ive straightforward approach would enumerate all $2^N$ different combinations of neural replacement options, train an enhanced model for each configuration, and pick the one that produces the lowest error. Here $N$ is the total number of nodes and operators in the graph. To avoid this combinatorial explosion, we propose a hypercube-based search algorithm at the cost of landing on a local minimum.


\subsection{Enhancement State}
Before describing our algorithm, we first introduce a related concept. For a node/operator, we use a bit to indicate whether it is the original one (=0) or replaced with a neural module (=1). Therefore, the enhancement state of all nodes and operators in a computational graph can be represented as an N-bit vector. The collection of all possible states forms an N-dimensional hypercube~\cite{koshy2004discrete}. Generally speaking, if there are $m-1$ different neural modules to choose from for each node/operator, we can extend the enhancement state to an N-dimensional vector of $m$-ary digits.

\subsection{Algorithm}
\label{sec:HOS}

\begin{figure}[t]
    \centering
    \includegraphics[width = 1.0\linewidth]{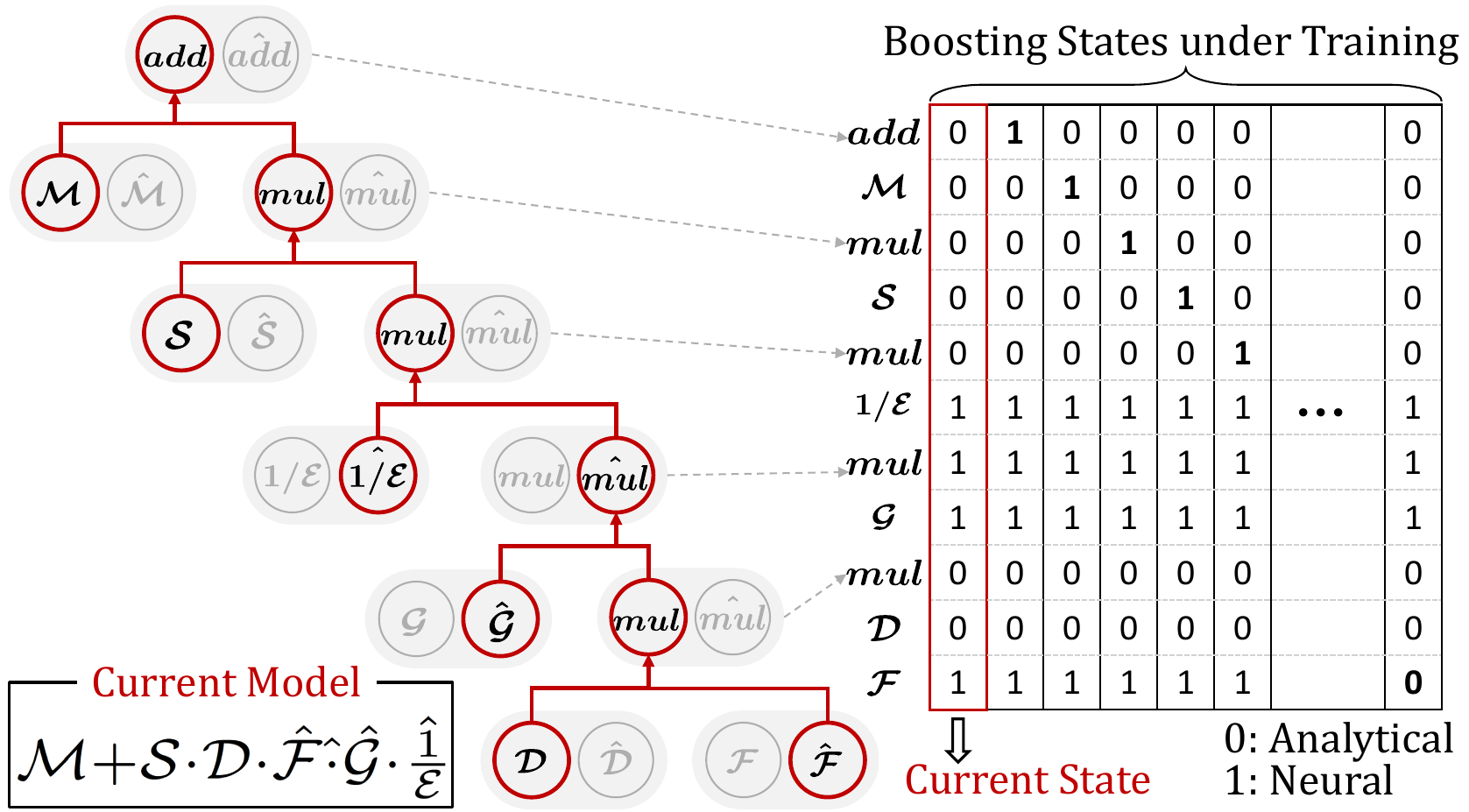}
    \caption{Correspondence between enhancement state and computational graph. Here we take GGX BRDF as an example. Its computational graph has 11 nodes/operators in total, resulting in an 11-bit vector that represents the enhancement state: for each bit, 0 indicates that the original node/operator is used, and 1 for a neural module. The table on the right part shows all states being evaluated at the moment, whose Hamming distance is less than or equal to 1 with the current state (the first column marked in red).}
    
\label{fig:training_states}
\end{figure}

We denote the current enhancement state as $B_c$, which is initialized to be an all-zero vector corresponding to the input analytical model. To efficiently reduce computational complexity, instead of training all possible models, each time we only consider a model (whose state is denoted as $B$) that differs from the current model in at most one node/operator. This corresponds to a Hamming distance of less or equal to 1 between $B$ and $B_c$ (i.e., a node/operator may be replaced by a neural module; conversely, a module may be switched back to the original node/operator). There are N+1 such models. Each B is independently built using the computational graph and optimized BRDF/neural parameters of $B_c$, then a neural module is added or reverted, or the computational graph remains unchanged, based on the Hamming distance between $B_c$ and $B$. We train each of these models by jointly optimizing the weights in all neural modules, neural parameters, and analytical BRDF parameters of $B_c$ and all $B$. After a user-specified number of epochs (30 in main experiments), we pick the model with the lowest error as the current one, and repeat the above process until the enhancement state does not change.~\figref{fig:training_states} illustrates the correspondence between the state and the computational graph, along with the relationship between $B_c$ and $B$.

Note that we are not limited to using a Hamming distance threshold of 1. With more computational resources, one can certainly increase the number of models being evaluated simultaneously with a larger threshold (e.g., a threshold of 2 corresponds to $1+N+\frac{N(N-1)}{2}$ models). Moreover, our hypercube-based search is flexible to include additional constraints. For example, the user might specify if certain nodes/operators remain analytical or neural, which correspond to fixed bits in the enhancement state vector. One can also impose an upper limit on the number of neural modules, which can be computed by counting the number of 1's in the state.

\begin{figure}[t]
    \centering
        \includegraphics[width = 1.0\linewidth]{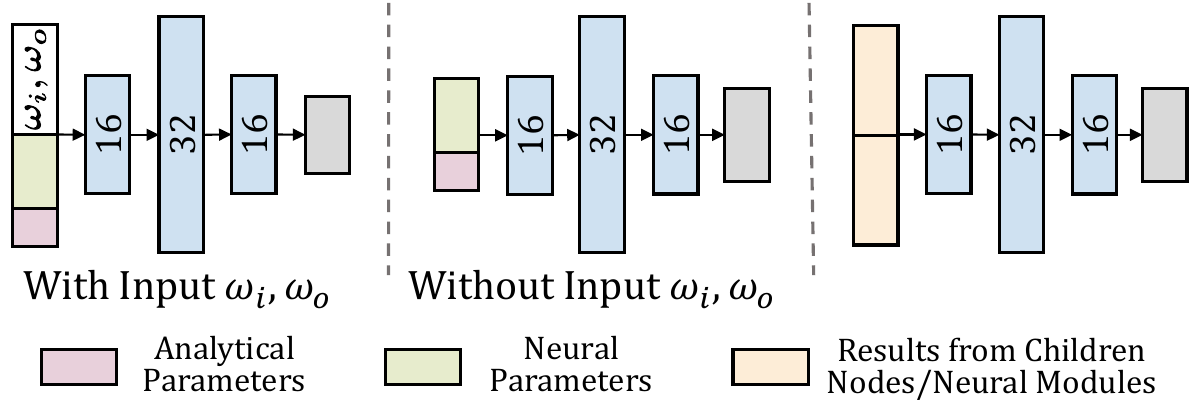}
    \caption{Visualization of neural module architectures. Each module is a 4-layer fully connected network, and each FC layer before the last one is followed by a leaky ReLU activation. For replacement to a node, if $\omega_i, \omega_o$ are part of the original input, we concatenate them with analytical BRDF paremeters and neural parameters as the input to the module (left); otherwise, the concatenation of analytical BRDF parameters and neural parameters are fed to the network directly (center). For the replacement of an operator, the neural module takes the results from two children nodes/modules as input (right). The output of a module is the same as the node/operator being replaced. }
\label{fig:module}
\end{figure}

\subsection{Training}
\label{sec:training}
To train a model, our loss function measures the L1 difference between the \note{log-transformed} ground-truth and predicted BRDF value as follows, similar to previous work~\cite{Sztrajman2021:NBRDF,Zheng2023:CompactBRDF}:

\begin{equation}
\begin{split}
L = & \sum_{\mathbf{k}} \left| \log \left( 1 + \tilde{f}_r(\mathbf{\omega_i^k}, \mathbf{\omega_o^k}) (\tilde{\mathbf{n}}, \mathbf{\omega_i^k}) \right) \right. \\
    & - \left. \log \left( 1 + f_r(\mathbf{\omega_i^k}, \mathbf{\omega_o^k}) (\tilde{\mathbf{n}}, \mathbf{\omega_i^k}) \right) \right|,
\end{split}
\label{eq:LOSS}
\end{equation}
where $\tilde{f}_r$/$f_r$ is the ground-truth/predicted BRDF value with the current model, respectively. $\mathbf{\tilde{n}}$ is the ground-truth normal, and $\mathbf{\omega_i^k}/\mathbf{\omega_o^k}$ is a sampled lighting/view direction, respectively.

We use the RMSProp optimizer with an alpha of 0.9 and the learning rate of $10^{-3}$, as it provides more stable convergence and slightly better reconstruction quality than Adam in our preliminary tests. The batch size is 100K. Xavier initialization is applied to all weights in neural modules. Similar to~\cite{FAN2022:NLBRDF}, all neural parameters are initialized to 0.5. For an analytical BRDF parameter, it is initialized with the mean value of the valid range in OpenSVBRDF statistics~\cite{ma2023opensvbrdf}.



\section{Runtime}
\label{sec:runtime}
\subsection{Fitting}
\label{sec:fitting}

To use the neural enhanced model to fit a BRDF or a BTF texel, we fix the weights in all neural modules, and optimize the neural parameters as well as analytical BRDF parameters, with respect to measurements at different view and lighting directions. The initialization of these parameters is the same as in~\sec{sec:training}. The fitting process is performed with a learning rate of $10^{-3}$ over 1,000 epochs.


\subsection{Rendering}
For rendering with our model, we start with the existing shader code of the original analytical model. Next, according to the final enhancement state, we modify the parts of the code with simple implementations of small MLPs, which mainly involve vector addition and multiplication operations. The modified shader takes neural parameters and analytical parameters as input, and outputs a BRDF value for a given pair of $(\omega_i, \omega_o)$. \note{For importance sampling, we directly re-fit the data samples to the original analytical model based on the code of~\cite{Dupuy2015PhotorealisticSR}}, and use the corresponding importance sampling method, similar to existing work~\cite{Sztrajman2021:NBRDF,Zheng2023:CompactBRDF}.  It will be interesting future work to explore neural importance sampling (i.e., NeuSample~\cite{xu2023neusample}) without refitting: one possible approach is to learn to predict analytical lobe mixtures for our enhanced model. We render a standard, public-domain material ball of~\cite{mazzone2023shaderball}, and evaluate the reconstruction quality using SSIM and $\Delta E_{ITP}$~\cite{Kavoosighafi2025metric}.





\section{Data Preparation}
\label{sec:data}
We train on a variety of synthetic and captured data sources. The total number of training samples is over 28 million, consisting of the following three parts.

\textbf{Isotropic BRDFs.} We use 100 measured BRDFs in the MERL database~\cite{Matusik2003:MERL}. We draw $10^5$ random directional pairs according to~\cite{Rusinkiewicz1998:New}, and use them to query the measurement values from MERL. The total number of training samples is $100 \times 100,000 = 10.0$M.


\textbf{Anisotropic BRDFs.} We sample two types of anisotropic BRDF from OpenSVBRDF~\cite{ma2023opensvbrdf}. First, we randomly sample 100 parametric GGX BRDFs based on statistics across the dataset. Then we render $10^5$ values per BRDF as virtual measurements using the same sampling method mentioned above. Similar to~\cite{Ma:2023:MoE}, each measurement is perturbed with a multiplicative Gaussian noise ($\mu$ = 1, $\sigma$ = 0.1) to simulate the acquisition noise.

Second, we randomly sample 300 neural representations from the dataset. We recover the lumitexel parameterized by their device from the neural representation, which contains 12,288 $\omega_i$ and one fixed $\omega_o$. We further remove the device-specific form-factor from the lumitexel to obtain the BRDF values as training samples. The total number of training samples from OpenSVBRDF is $100 \times 10^5 + 300 \times 12,288 = 13.7$M.

\textbf{BTF.} We randomly sample 200 ABRDFs from the first 11 samples of each category in the UBO BTF dataset~\cite{weinmann2014material}, with $151 \times 151$ pairs of $(\omega_i,\omega_o)$ for each ABRDF. The total number of training samples is $200 \times 151 \times 151 = 4.5$M.


\section{Results \& Discussions}
\label{sec:comparison}
All experiments are conducted on a workstation with dual Intel Xeon 4210 CPUs, 256GB DDR4 memory, and a single NVIDIA GeForce RTX 3090 GPU. It takes about 20 hours to learn our enhanced model with 210 epochs. We render the images with Mitsuba~\cite{jakob2022mitsuba3}, using BRDF importance sampling with 4,096 samples per pixel.

Our main result is the enhanced GGX model defined as $\mathcal{M}{+}\mathcal{S\cdot} \mathcal{D \cdot} \hat{\mathcal{F}} \hat{\cdot} \mathcal{\hat{G} \cdot} \hat{(\frac{1}{\mathcal{E}})}$. Here a term with a hat indicates a replacement with a neural module. Altogether 3 computational nodes and 1 operator are replaced with neural modules, resulting in approximately 7K trainable weights. The parameters of the new model are original 12 analytical BRDF parameters ($\mathbf{n}$, $\mathbf{t}_{\theta}$, $\rho_d$, $\rho_s$, $\alpha_x$, $\alpha_y$, $F_0$) and 27 neural parameters, the total number of which is 39. In addition, the network size of the enhanced GGX model is 26.45KB.

Performance-wise, it takes 27.3s to fit our enhanced GGX model with respect to $10^5$ measurements at different view and lighting directions. It takes 34.2s for fitting the same data with the original GGX model. The rendering speed in Mitsuba is 13.68$\times10^6$/21.83$\times10^6$ rays/s with our enhanced model and the original one, respectively. \note{While optimizing rendering performance is not the goal of this paper, we can add an extra term (which approximates runtime efficiency) to the loss function to account for it in future work.}



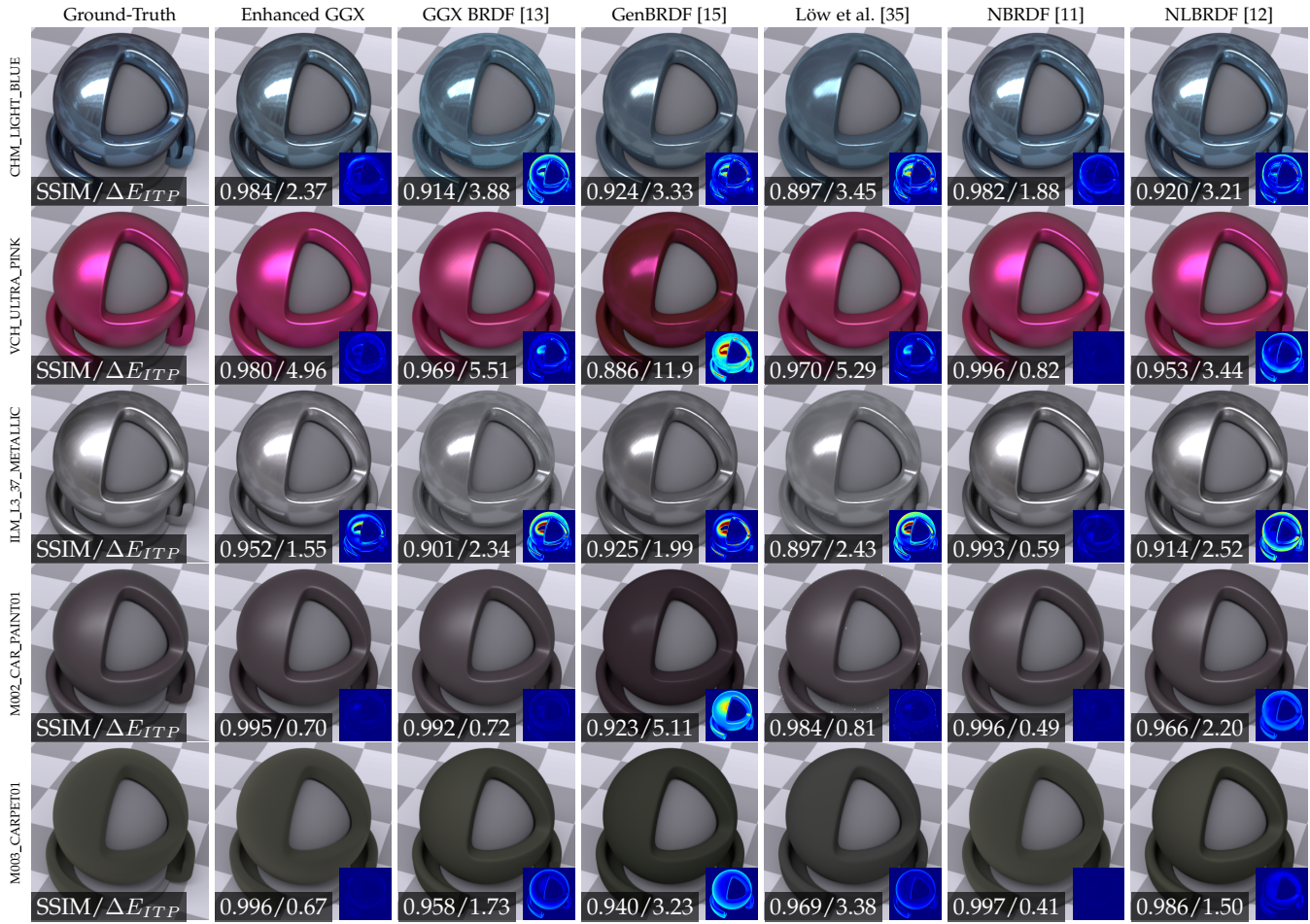
\begin{figure*}[htb]
    \centering
    \begin{minipage}{7.1in}
    \begin{minipage}{0.03in}	
            \centering
                \vspace{0.2in}
                \rotatebox{90}{\scriptsize \textsc{ }}
            \end{minipage}
        \hspace{-0.1in}
        \begin{minipage}{7.1in}
            \centering
            \begin{minipage}{0.95in}
                \centering
                \subcaption{\scriptsize Ground-Truth}
            \end{minipage}			
            \begin{minipage}{0.95in}
                \centering
                \subcaption{\scriptsize Enhanced GGX}
            \end{minipage}		
            \begin{minipage}{0.95in}
                \centering
                \subcaption{\scriptsize GGX BRDF~\cite{Walter2007:GGX}}
            \end{minipage}		
            \begin{minipage}{0.95in}
                \centering
                \subcaption{\scriptsize GenBRDF~\cite{Brady2014:Genbrdf}}
            \end{minipage}	
            \begin{minipage}{0.95in}
                \centering
                \subcaption{\scriptsize L{\"o}w et al.~\cite{low2012brdf}}
            \end{minipage}	
            \begin{minipage}{0.95in}
                \centering
                \subcaption{\scriptsize NBRDF~\cite{Sztrajman2021:NBRDF}}
            \end{minipage}
            \begin{minipage}{0.95in}
                \centering
                \subcaption{\scriptsize NLBRDF~\cite{FAN2022:NLBRDF}}
            \end{minipage}
        \end{minipage}
    \end{minipage}
    
    \begin{minipage}{7.1in}
        \begin{minipage}{0.03in}	
            \rotatebox{90}{\tiny \uppercase{Chm\_light\_blue}}
        \end{minipage}	
        \hspace{-0.1in}
        \begin{minipage}{7.1in}
            \centering
            \begin{minipage}{0.95in}
            \centering
            \includegraphics[width=0.95in]{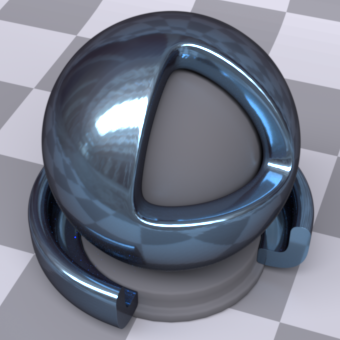}
            \put(-68.6,4.8) {\tikz[baseline] \node[fill=black, fill opacity=0.65, text opacity=1, text=white,inner sep=2pt] {\small SSIM/$\Delta E_{ITP}$};}
            \end{minipage}
            \begin{minipage}{0.95in}
            \centering
            \includegraphics[width=0.95in]{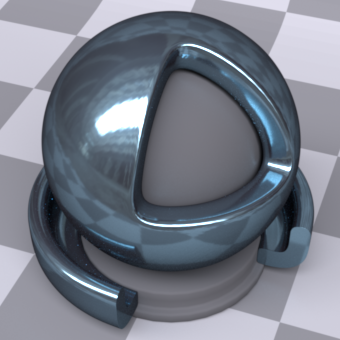}
             \put(-68.6,4.8) {\tikz[baseline] \node[fill=black, fill opacity=0.65, text opacity=1, text=white,inner sep=2pt] {\small 0.984/2.37};}
             \put(-20,0){\includegraphics[width=20pt]{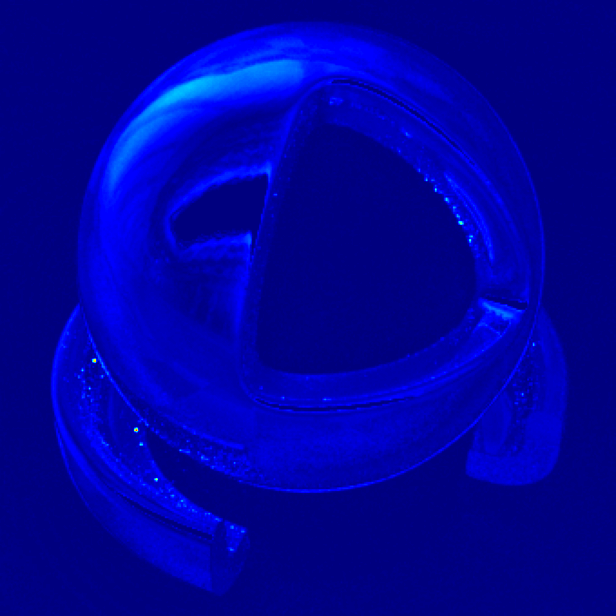}}
            \end{minipage}
            \begin{minipage}{0.95in}
            \centering
            \includegraphics[width=0.95in]{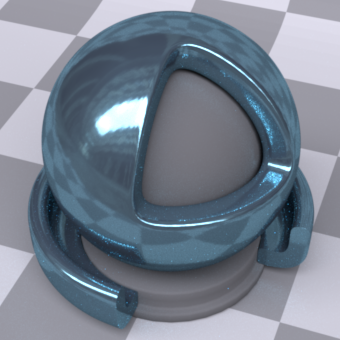}
             \put(-68.6,4.8) {\tikz[baseline] \node[fill=black, fill opacity=0.65, text opacity=1, text=white,inner sep=2pt] {\small 0.914/3.88};}
             \put(-20,0){\includegraphics[width=20pt]{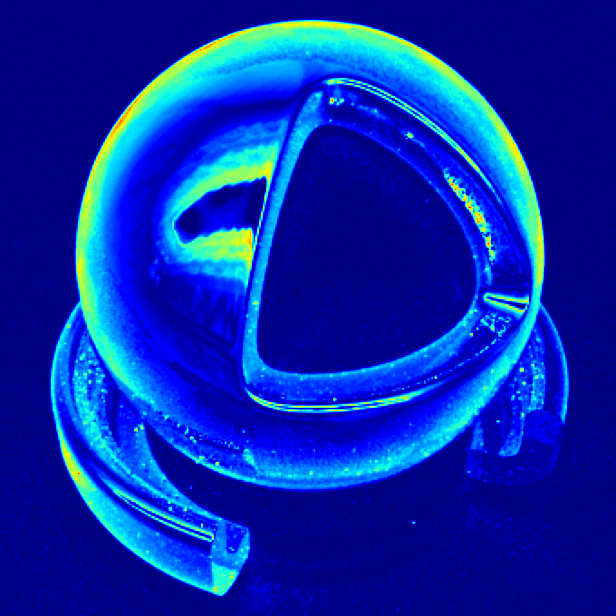}}
            \end{minipage}
            \begin{minipage}{0.95in}
            \centering
            \includegraphics[width=0.95in]{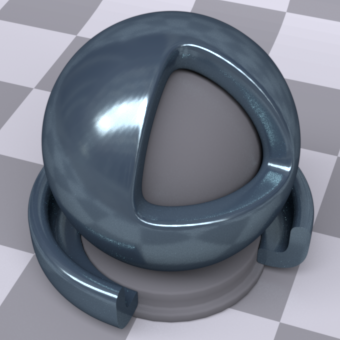}
             \put(-68.6,4.8) {\tikz[baseline] \node[fill=black, fill opacity=0.65, text opacity=1, text=white,inner sep=2pt] {\small 0.924/3.33};}
             \put(-20,0){\includegraphics[width=20pt]{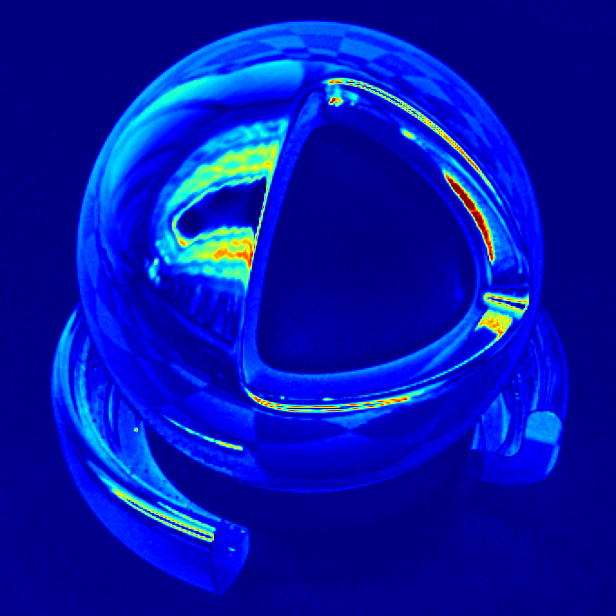}}
            \end{minipage}
            \begin{minipage}{0.95in}
            \centering
            \includegraphics[width=0.95in]{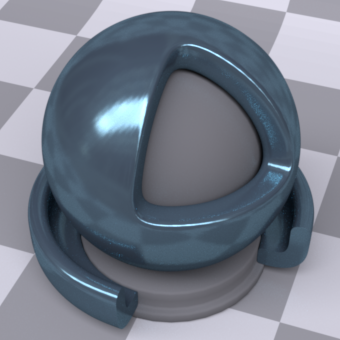}
             \put(-68.6,4.8) {\tikz[baseline] \node[fill=black, fill opacity=0.65, text opacity=1, text=white,inner sep=2pt] {\small 0.897/3.45};}
             \put(-20,0){\includegraphics[width=20pt]{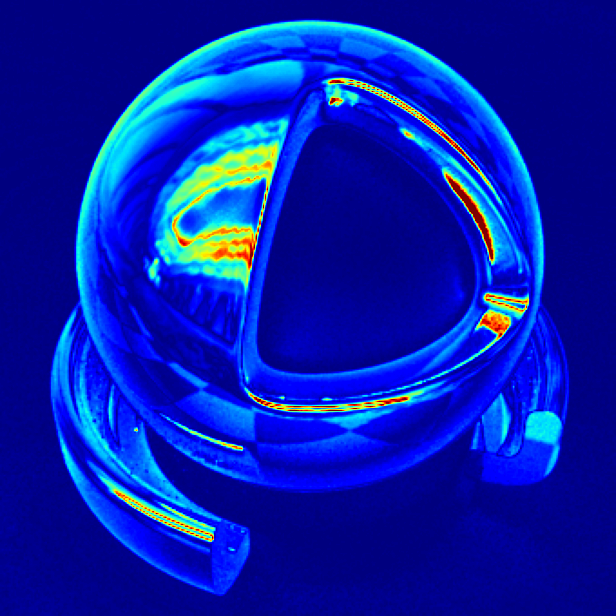}}
            \end{minipage}
             \begin{minipage}{0.95in}
            \centering
            \includegraphics[width=0.95in]{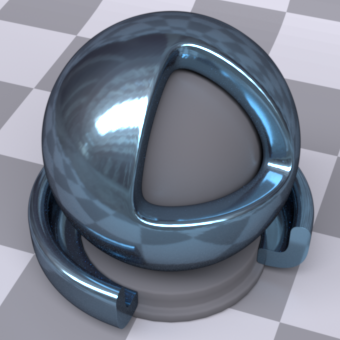}
             \put(-68.6,4.8) {\tikz[baseline] \node[fill=black, fill opacity=0.65, text opacity=1, text=white,inner sep=2pt] {\small 0.982/1.88};}
             \put(-20,0){\includegraphics[width=20pt]{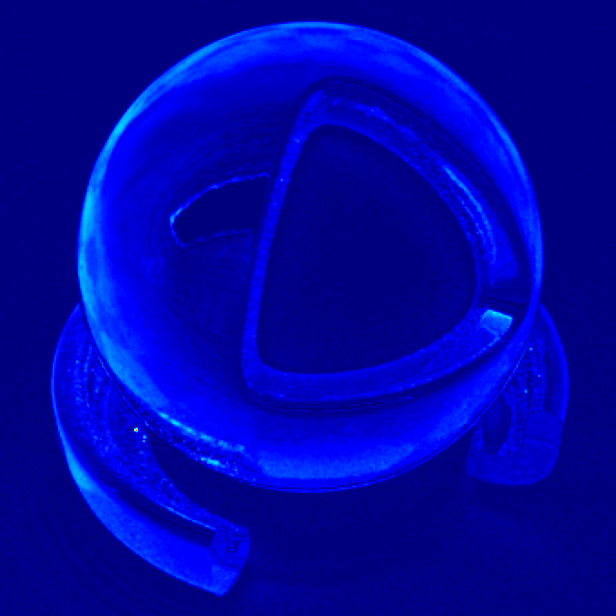}}
            \end{minipage}
             \begin{minipage}{0.95in}
            \centering
            \includegraphics[width=0.95in]{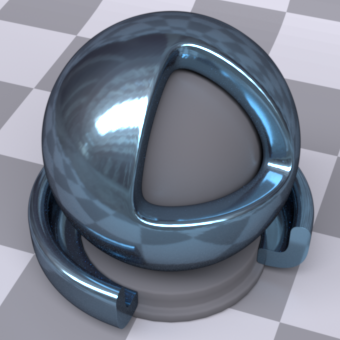}
             \put(-68.6,4.8) {\tikz[baseline] \node[fill=black, fill opacity=0.65, text opacity=1, text=white,inner sep=2pt] {\small 0.920/3.21};}
             \put(-20,0){\includegraphics[width=20pt]{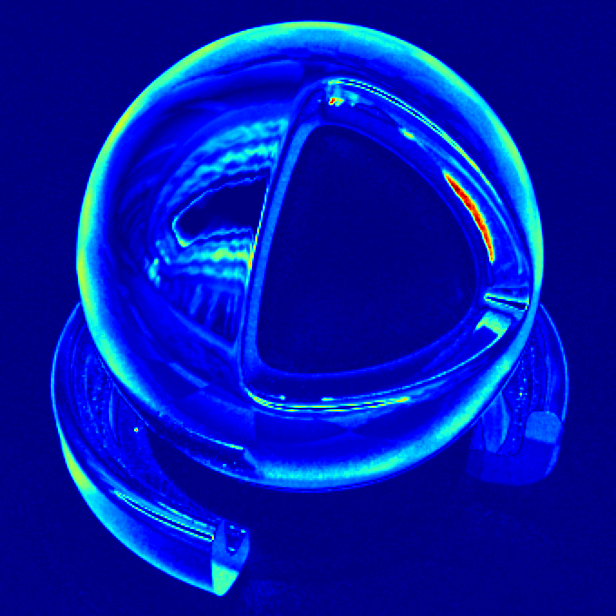}}
            \end{minipage}
        \end{minipage}
    \end{minipage}

    \begin{minipage}{7.1in}
        \begin{minipage}{0.03in}	
            \centering
            \rotatebox{90}{\tiny \uppercase{Vch\_ultra\_pink}}
        \end{minipage}	
        \hspace{-0.1in}
        \begin{minipage}{7.1in}
            \centering
            \begin{minipage}{0.95in}  
            \includegraphics[width=0.95in]{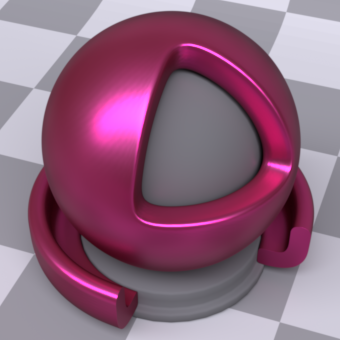} 
            \put(-68.6,4.8) {\tikz[baseline] \node[fill=black, fill opacity=0.65, text opacity=1, text=white,inner sep=2pt] {\small SSIM/$\Delta E_{ITP}$};} 
             \end{minipage}
             \begin{minipage}{0.95in}              
            \includegraphics[width=0.95in]{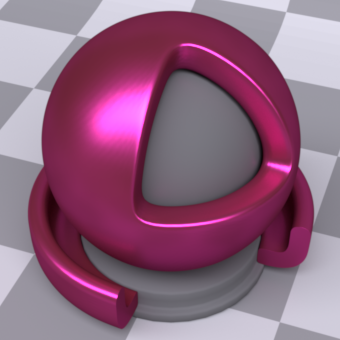}
             \put(-68.6,4.8) {\tikz[baseline] \node[fill=black, fill opacity=0.65, text opacity=1, text=white,inner sep=2pt] {\small 0.980/4.96};}    
             \put(-20,0){\includegraphics[width=20pt]{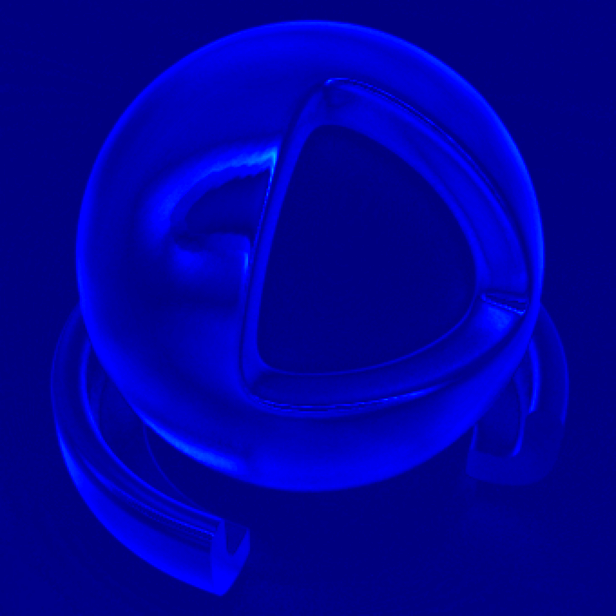}}
             \end{minipage}
             \begin{minipage}{0.95in}               
            \includegraphics[width=0.95in]{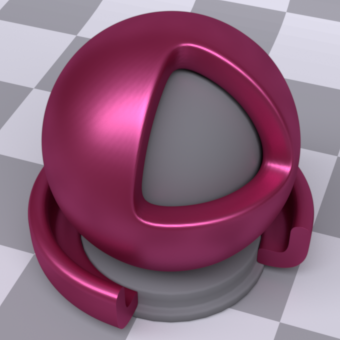}
             \put(-68.6,4.8) {\tikz[baseline] \node[fill=black, fill opacity=0.65, text opacity=1, text=white,inner sep=2pt] {\small 0.969/5.51};}    
             \put(-20,0){\includegraphics[width=20pt]{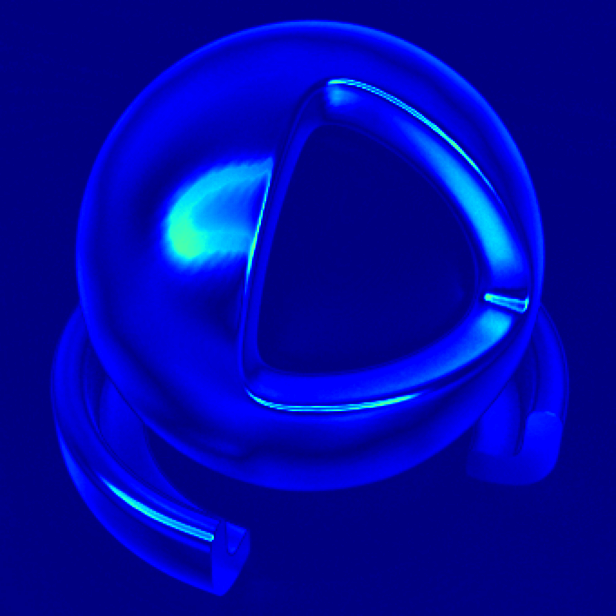}}
             \end{minipage}
             \begin{minipage}{0.95in}               
            \includegraphics[width=0.95in]{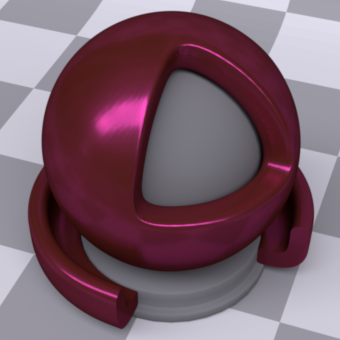}
             \put(-68.6,4.8) {\tikz[baseline] \node[fill=black, fill opacity=0.65, text opacity=1, text=white,inner sep=2pt] {\small 0.886/11.9};}  
             \put(-20,0){\includegraphics[width=20pt]{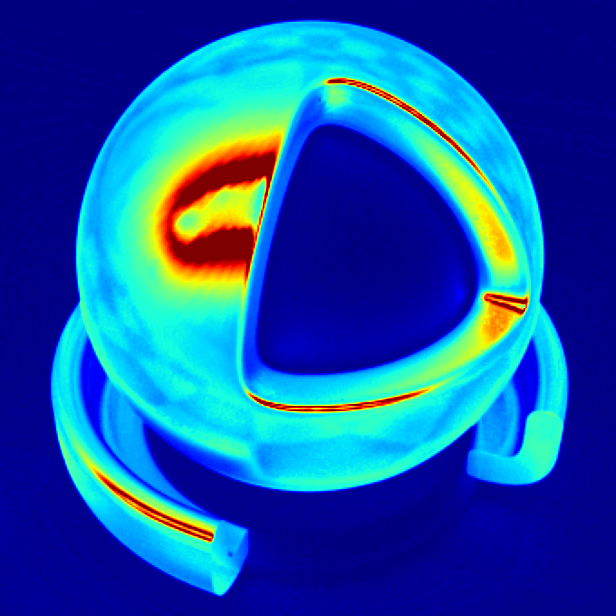}}
             \end{minipage}
             \begin{minipage}{0.95in}                 
            \includegraphics[width=0.95in]{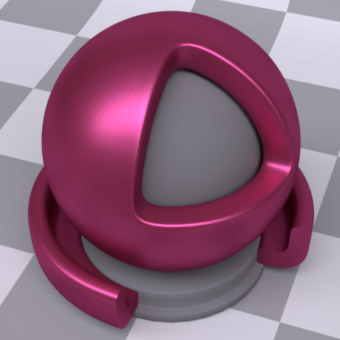}
             \put(-68.6,4.8) {\tikz[baseline] \node[fill=black, fill opacity=0.65, text opacity=1, text=white,inner sep=2pt] {\small 0.970/5.29};}   
             \put(-20,0){\includegraphics[width=20pt]{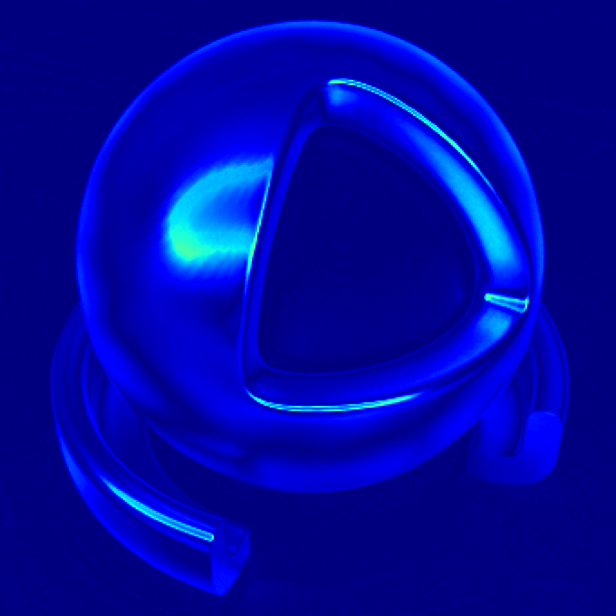}}
             \end{minipage}
             \begin{minipage}{0.95in}                
            \includegraphics[width=0.95in]{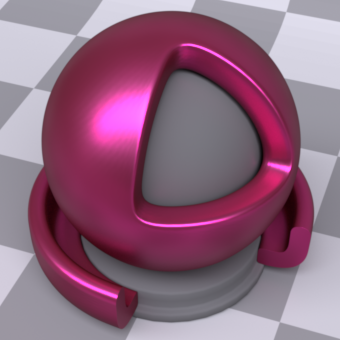}
             \put(-68.6,4.8) {\tikz[baseline] \node[fill=black, fill opacity=0.65, text opacity=1, text=white,inner sep=2pt] {\small 0.996/0.82};}    
             \put(-20,0){\includegraphics[width=20pt]{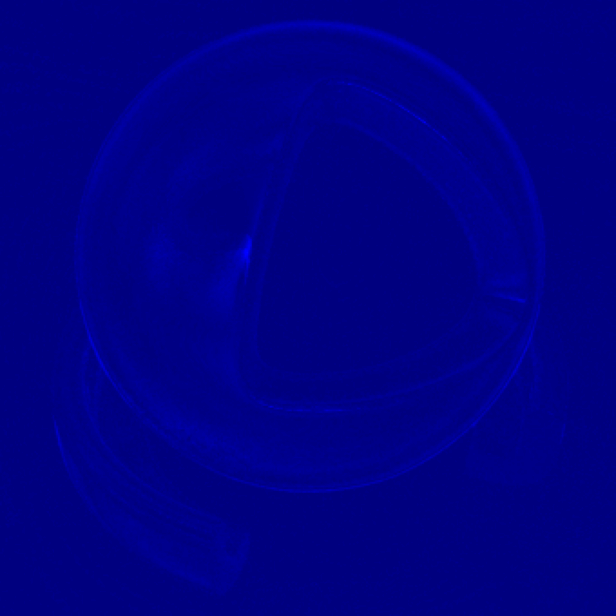}}
             \end{minipage}
             \begin{minipage}{0.95in}               
            \includegraphics[width=0.95in]{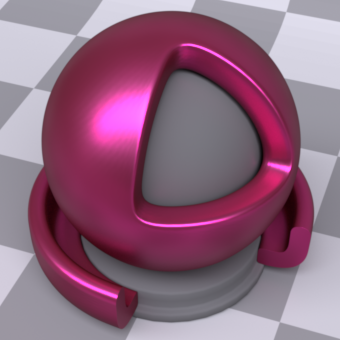}
             \put(-68.6,4.8) {\tikz[baseline] \node[fill=black, fill opacity=0.65, text opacity=1, text=white,inner sep=2pt] {\small 0.953/3.44};}
             \put(-20,0){\includegraphics[width=20pt]{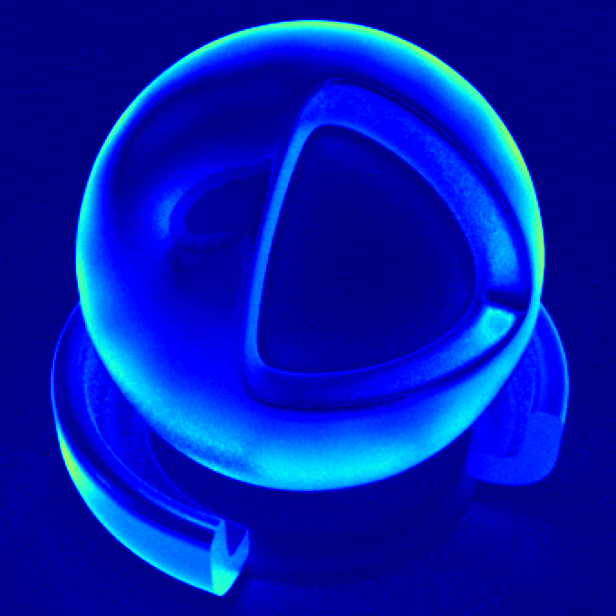}}
             \end{minipage}
        \end{minipage}
    \end{minipage}

    \begin{minipage}{7.1in}
        \begin{minipage}{0.03in}	
            \centering
            \rotatebox{90}{\tiny \uppercase{Ilm\_l3\_37\_metallic}}
        \end{minipage}	
        \hspace{-0.1in}
        \begin{minipage}{7.1in}
            \centering
            \begin{minipage}{0.95in}
            \includegraphics[width=0.95in]{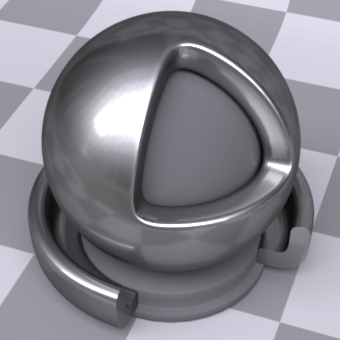}
            \put(-68.6,4.8) {\tikz[baseline] \node[fill=black, fill opacity=0.65, text opacity=1, text=white,inner sep=2pt] {\small SSIM/$\Delta E_{ITP}$};}
             \end{minipage}
             \begin{minipage}{0.95in}
            \includegraphics[width=0.95in]{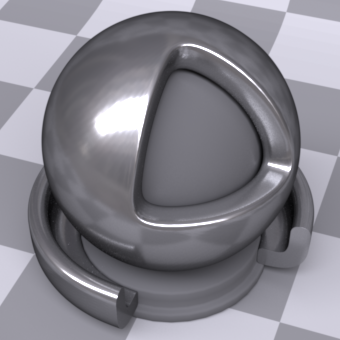}
             \put(-68.6,4.8) {\tikz[baseline] \node[fill=black, fill opacity=0.65, text opacity=1, text=white,inner sep=2pt] {\small 0.952/1.55};}
             \put(-20,0){\includegraphics[width=20pt]{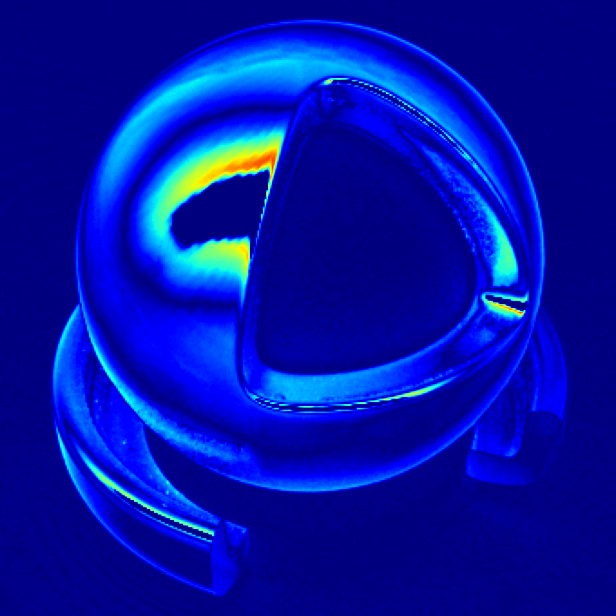}}
             \end{minipage}
             \begin{minipage}{0.95in}
            \includegraphics[width=0.95in]{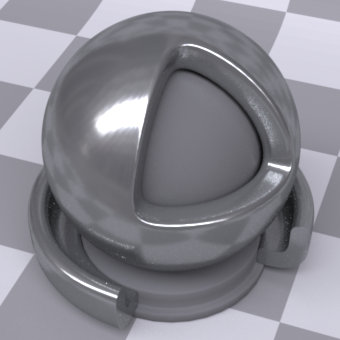}
             \put(-68.6,4.8) {\tikz[baseline] \node[fill=black, fill opacity=0.65, text opacity=1, text=white,inner sep=2pt] {\small 0.901/2.34};}
             \put(-20,0){\includegraphics[width=20pt]{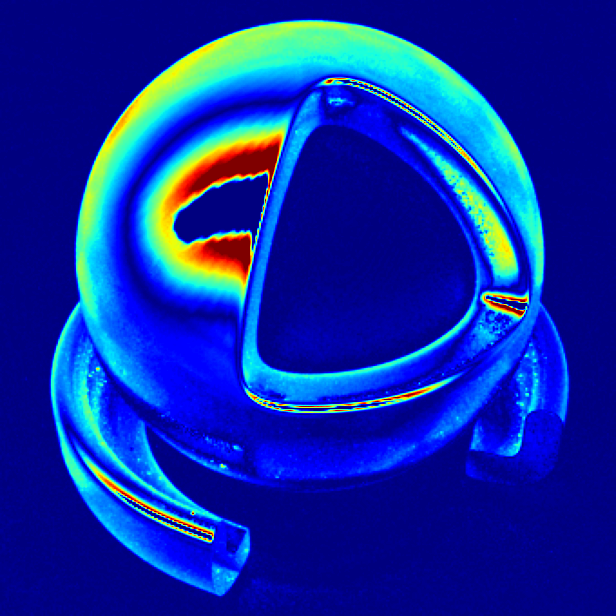}}
             \end{minipage}
             \begin{minipage}{0.95in}
            \includegraphics[width=0.95in]{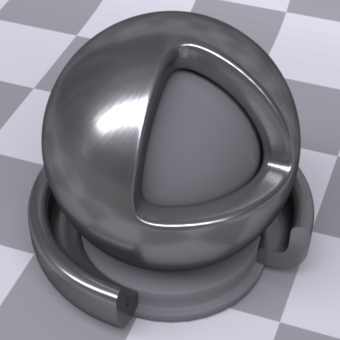}
             \put(-68.6,4.8) {\tikz[baseline] \node[fill=black, fill opacity=0.65, text opacity=1, text=white,inner sep=2pt] {\small 0.925/1.99};}
             \put(-20,0){\includegraphics[width=20pt]{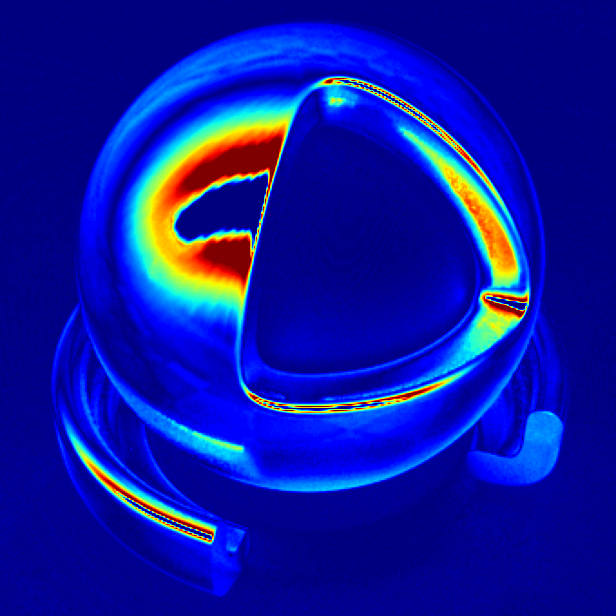}}
             \end{minipage}
             \begin{minipage}{0.95in}
            \includegraphics[width=0.95in]{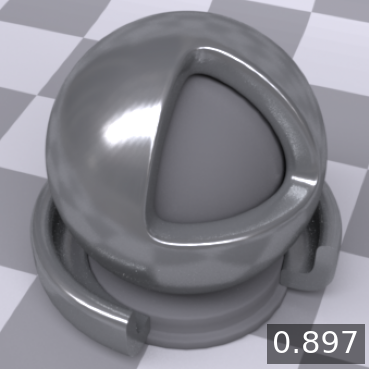}
             \put(-68.6,4.8) {\tikz[baseline] \node[fill=black, fill opacity=0.65, text opacity=1, text=white,inner sep=2pt] {\small 0.897/2.43};}
             \put(-20,0){\includegraphics[width=20pt]{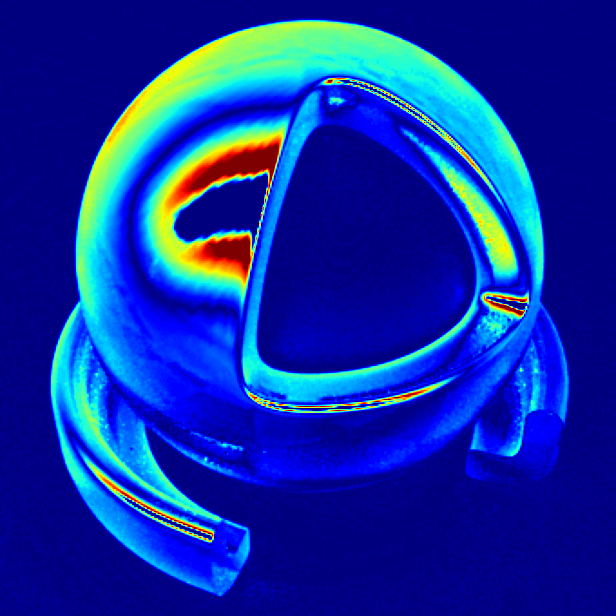}}
             \end{minipage}
             \begin{minipage}{0.95in}
            \includegraphics[width=0.95in]{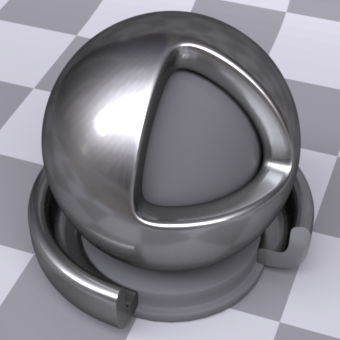}
             \put(-68.6,4.8) {\tikz[baseline] \node[fill=black, fill opacity=0.65, text opacity=1, text=white,inner sep=2pt] {\small 0.993/0.59};}
             \put(-20,0){\includegraphics[width=20pt]{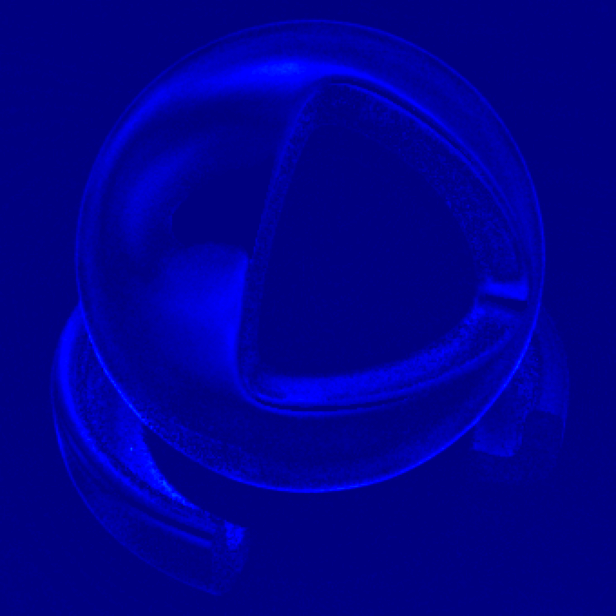}}
             \end{minipage}
             \begin{minipage}{0.95in}
            \includegraphics[width=0.95in]{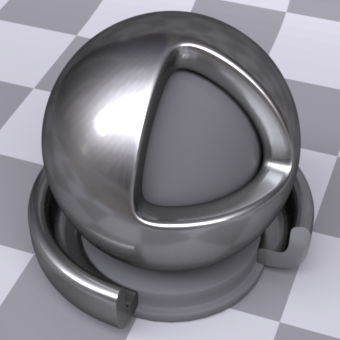}
             \put(-68.6,4.8) {\tikz[baseline] \node[fill=black, fill opacity=0.65, text opacity=1, text=white,inner sep=2pt] {\small 0.914/2.52};}
             \put(-20,0){\includegraphics[width=20pt]{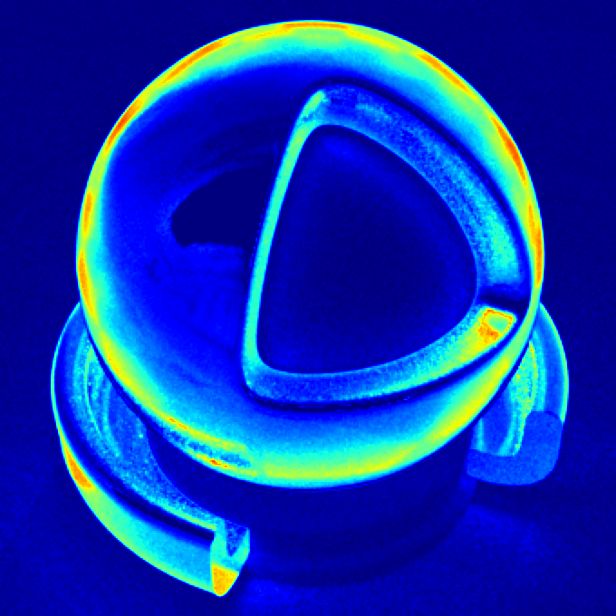}}
             \end{minipage}
        \end{minipage}
    \end{minipage}

    \begin{minipage}{7.1in}
        \begin{minipage}{0.03in}	
            \centering
            \rotatebox{90}{\tiny \uppercase{\note{M002\_car\_paint01}}}
        \end{minipage}	
        \hspace{-0.1in}
        \begin{minipage}{7.1in}
            \centering
            \begin{minipage}{0.95in}
            \includegraphics[width=0.95in]{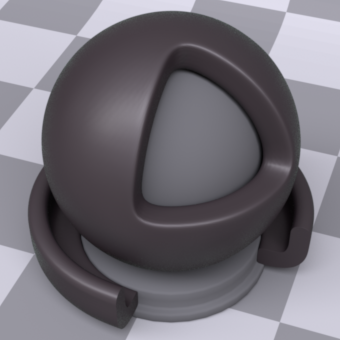}
            \put(-68.6,4.8) {\tikz[baseline] \node[fill=black, fill opacity=0.65, text opacity=1, text=white,inner sep=2pt] {\small SSIM/$\Delta E_{ITP}$};}
             \end{minipage}
             \begin{minipage}{0.95in}
            \includegraphics[width=0.95in]{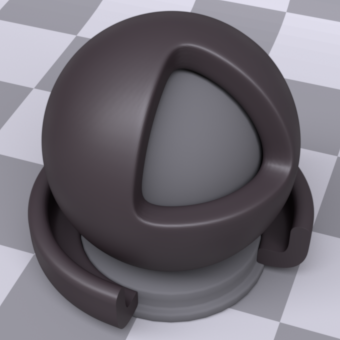}
             \put(-68.6,4.8) {\tikz[baseline] \node[fill=black, fill opacity=0.65, text opacity=1, text=white,inner sep=2pt] {\small 0.995/0.70};}    
             \put(-20,0){\includegraphics[width=20pt]{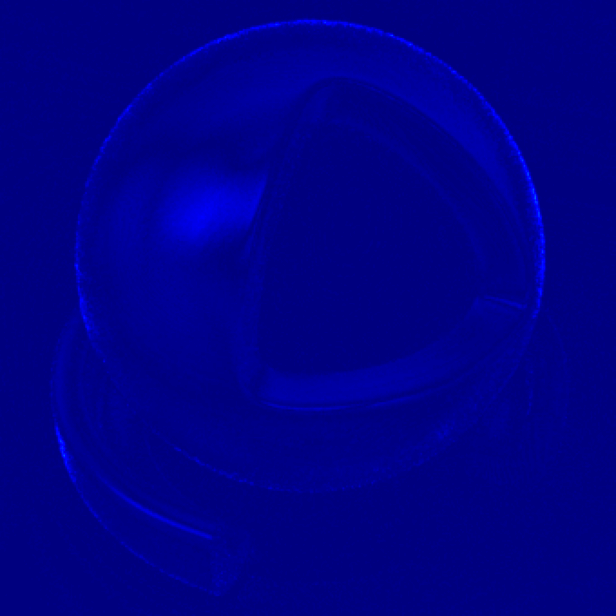}}
             \end{minipage}
             \begin{minipage}{0.95in}        
            \includegraphics[width=0.95in]{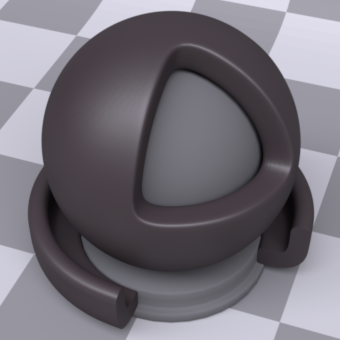}
             \put(-68.6,4.8) {\tikz[baseline] \node[fill=black, fill opacity=0.65, text opacity=1, text=white,inner sep=2pt] {\small 0.992/0.72};}    
             \put(-20,0){\includegraphics[width=20pt]{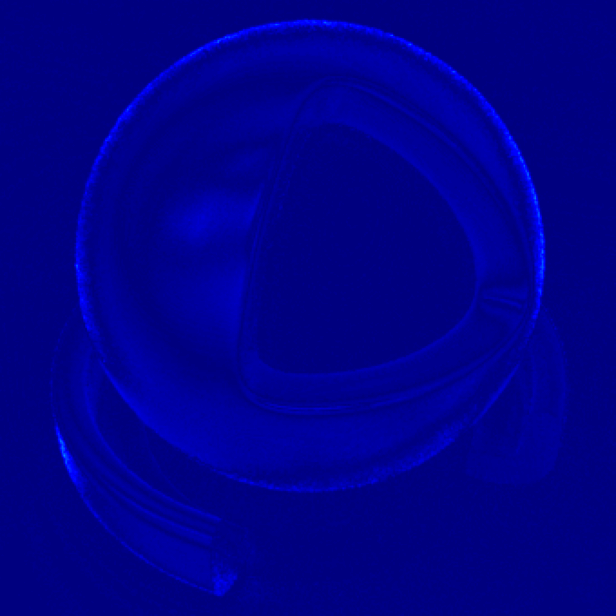}}
             \end{minipage}
             \begin{minipage}{0.95in}        
            \includegraphics[width=0.95in]{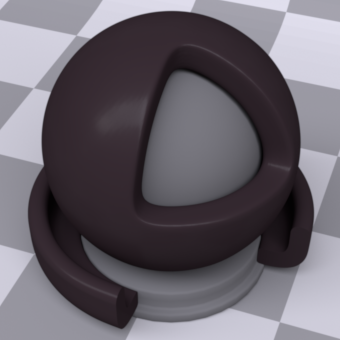}
             \put(-68.6,4.8) {\tikz[baseline] \node[fill=black, fill opacity=0.65, text opacity=1, text=white,inner sep=2pt] {\small 0.923/5.11};}  
             \put(-20,0){\includegraphics[width=20pt]{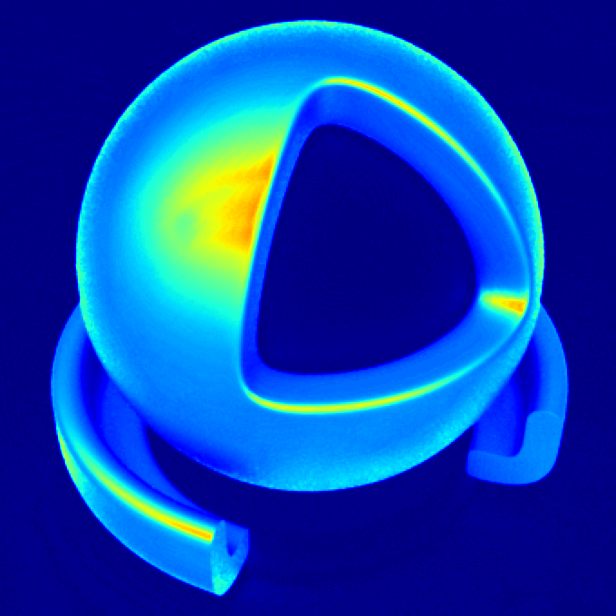}}
             \end{minipage}
             \begin{minipage}{0.95in}          
            \includegraphics[width=0.95in]{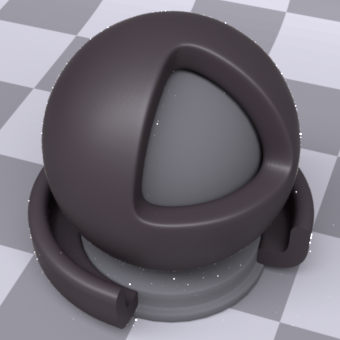}
             \put(-68.6,4.8) {\tikz[baseline] \node[fill=black, fill opacity=0.65, text opacity=1, text=white,inner sep=2pt] {\small 0.984/0.81};}  
             \put(-20,0){\includegraphics[width=20pt]{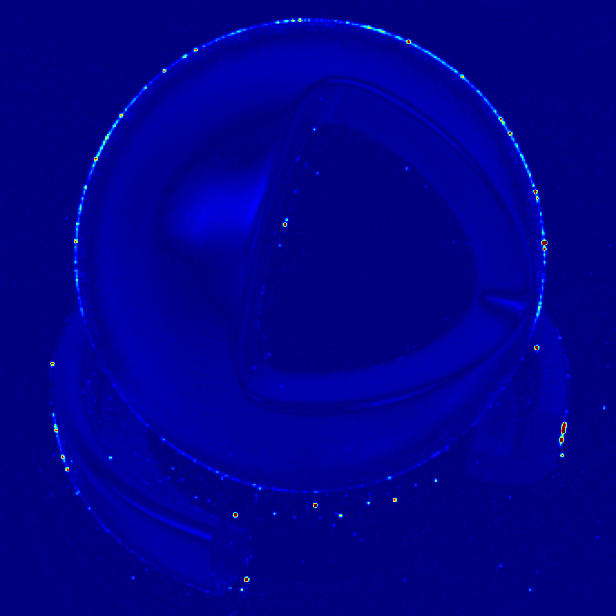}}
             \end{minipage}
             \begin{minipage}{0.95in}          
            \includegraphics[width=0.95in]{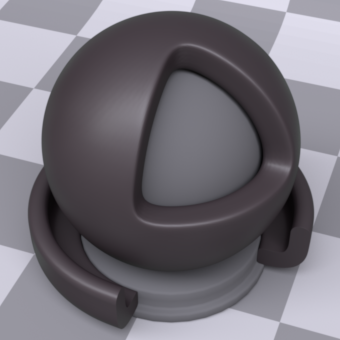}
             \put(-68.6,4.8) {\tikz[baseline] \node[fill=black, fill opacity=0.65, text opacity=1, text=white,inner sep=2pt] {\small 0.996/0.49};}  
             \put(-20,0){\includegraphics[width=20pt]{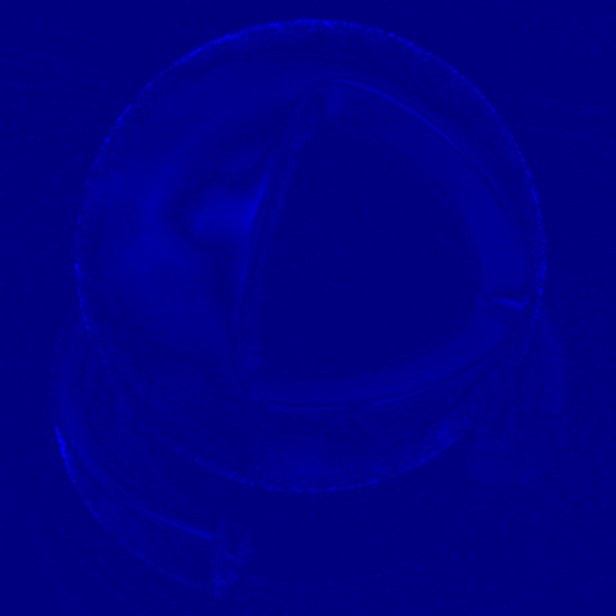}}
             \end{minipage}
             \begin{minipage}{0.95in}         
            \includegraphics[width=0.95in]{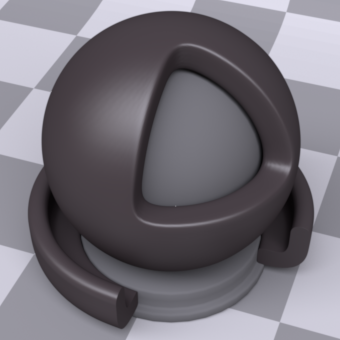}
             \put(-68.6,4.8) {\tikz[baseline] \node[fill=black, fill opacity=0.65, text opacity=1, text=white,inner sep=2pt] {\small 0.966/2.20};}
             \put(-20,0){\includegraphics[width=20pt]{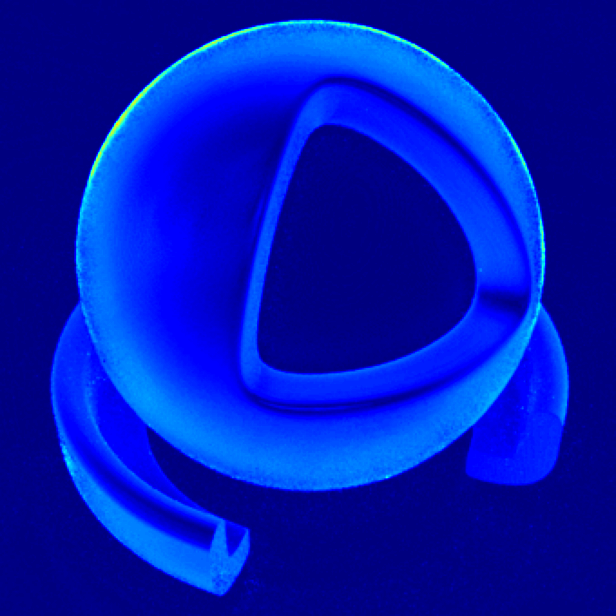}}
             \end{minipage}
        \end{minipage}
    \end{minipage}

    \begin{minipage}{7.1in}
        \begin{minipage}{0.03in}	
            \centering
            \rotatebox{90}{\tiny \uppercase{\note{M003\_carpet01}}}
        \end{minipage}	
        \hspace{-0.1in}
        \begin{minipage}{7.1in}
            \centering
            \begin{minipage}{0.95in} 
            \includegraphics[width=0.95in]{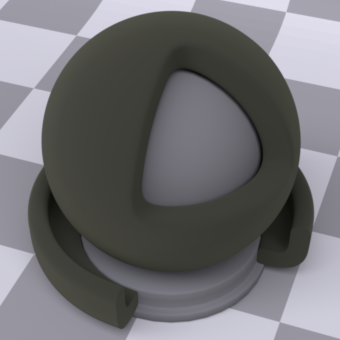}
            \put(-68.6,4.8) {\tikz[baseline] \node[fill=black, fill opacity=0.65, text opacity=1, text=white,inner sep=2pt] {\small SSIM/$\Delta E_{ITP}$};}
             \end{minipage}
             \begin{minipage}{0.95in}                 
            \includegraphics[width=0.95in]{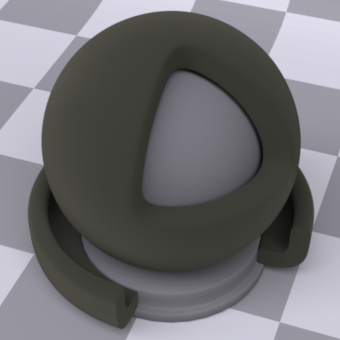}
             \put(-68.6,4.8) {\tikz[baseline] \node[fill=black, fill opacity=0.65, text opacity=1, text=white,inner sep=2pt] {\small 0.996/0.67};}   
             \put(-20,0){\includegraphics[width=20pt]{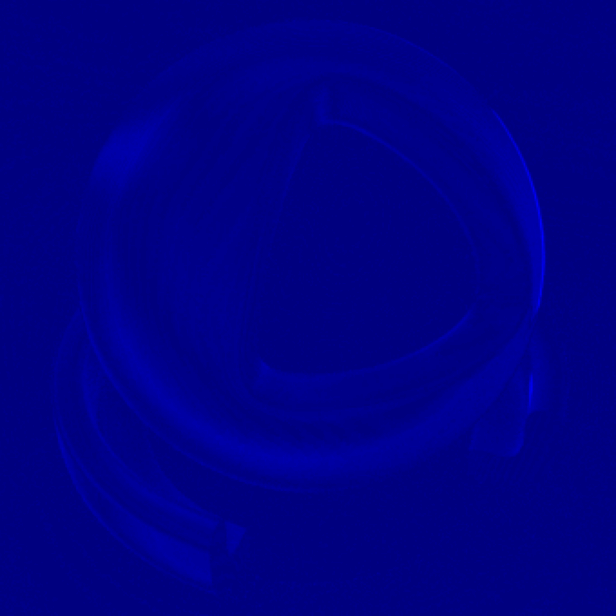}}
             \end{minipage}
             \begin{minipage}{0.95in}                
            \includegraphics[width=0.95in]{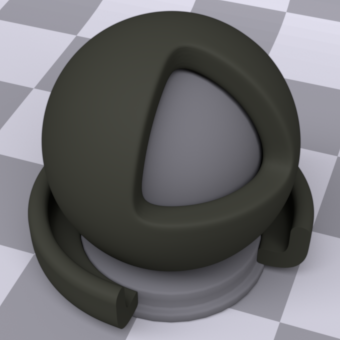}
             \put(-68.6,4.8) {\tikz[baseline] \node[fill=black, fill opacity=0.65, text opacity=1, text=white,inner sep=2pt] {\small 0.958/1.73};} 
             \put(-20,0){\includegraphics[width=20pt]{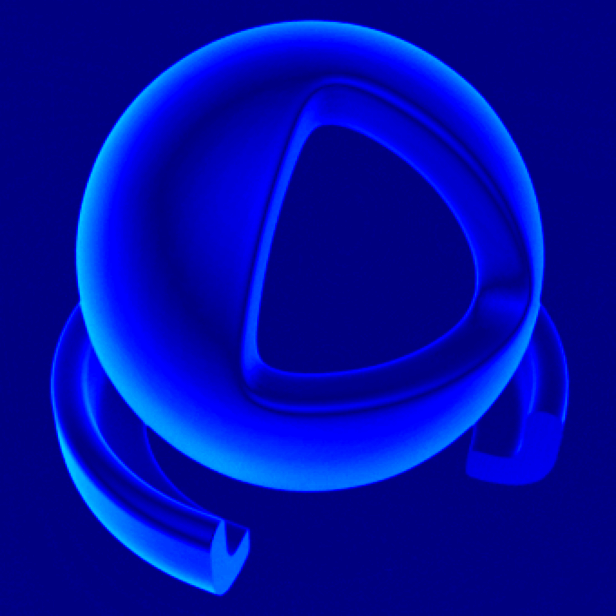}}
             \end{minipage}
             \begin{minipage}{0.95in}                  
            \includegraphics[width=0.95in]{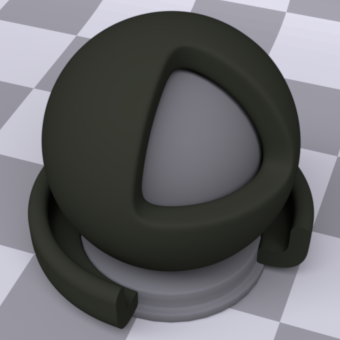}
             \put(-68.6,4.8) {\tikz[baseline] \node[fill=black, fill opacity=0.65, text opacity=1, text=white,inner sep=2pt] {\small 0.940/3.23};}   
             \put(-20,0){\includegraphics[width=20pt]{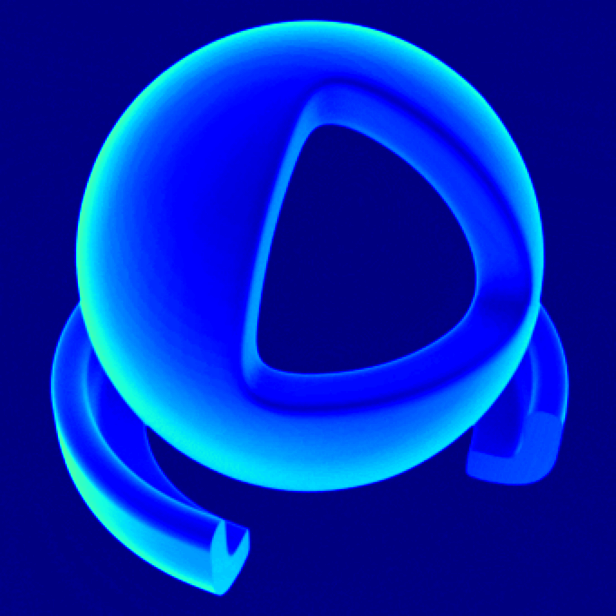}}
             \end{minipage}
             \begin{minipage}{0.95in}                
            \includegraphics[width=0.95in]{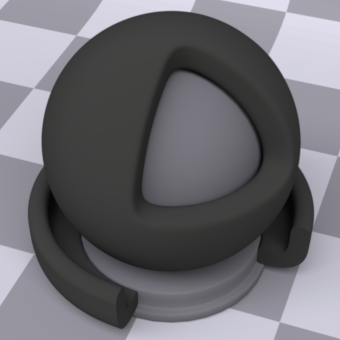}
             \put(-68.6,4.8) {\tikz[baseline] \node[fill=black, fill opacity=0.65, text opacity=1, text=white,inner sep=2pt] {\small 0.969/3.38};}   
             \put(-20,0){\includegraphics[width=20pt]{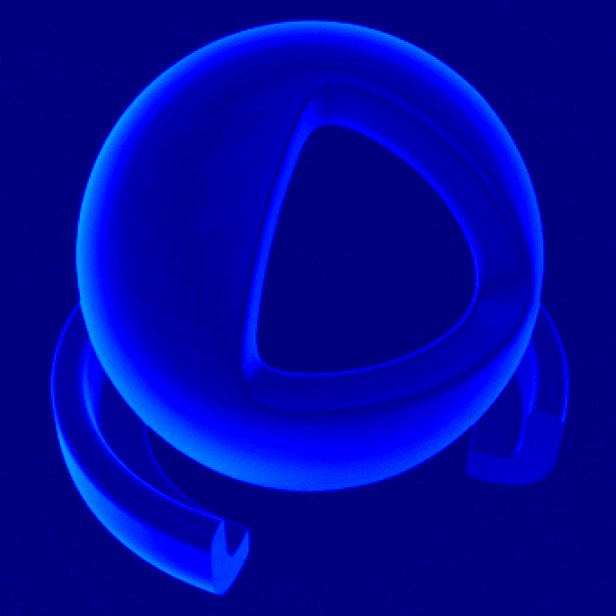}}
             \end{minipage}
             \begin{minipage}{0.95in}                
            \includegraphics[width=0.95in]{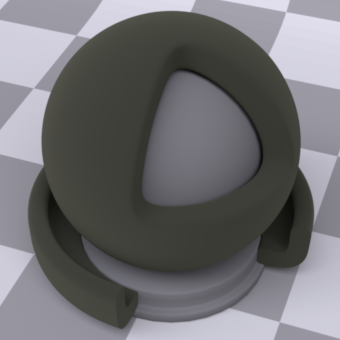}
             \put(-68.6,4.8) {\tikz[baseline] \node[fill=black, fill opacity=0.65, text opacity=1, text=white,inner sep=2pt] {\small 0.997/0.41};} 
             \put(-20,0){\includegraphics[width=20pt]{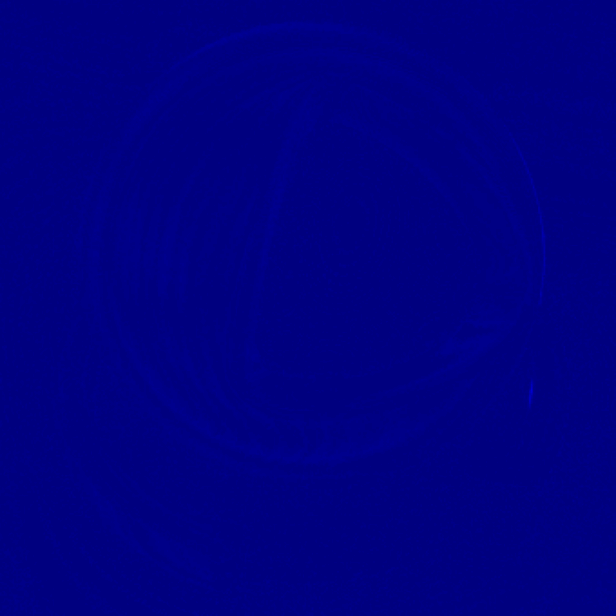}}
             \end{minipage}
             \begin{minipage}{0.95in}                  
            \includegraphics[width=0.95in]{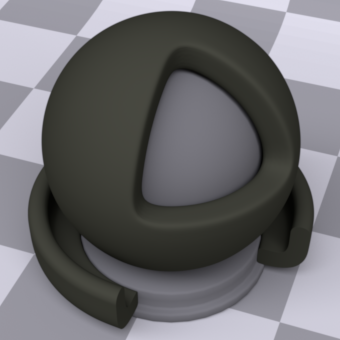}
             \put(-68.6,4.8) {\tikz[baseline] \node[fill=black, fill opacity=0.65, text opacity=1, text=white,inner sep=2pt] {\small 0.986/1.50};}
             \put(-20,0){\includegraphics[width=20pt]{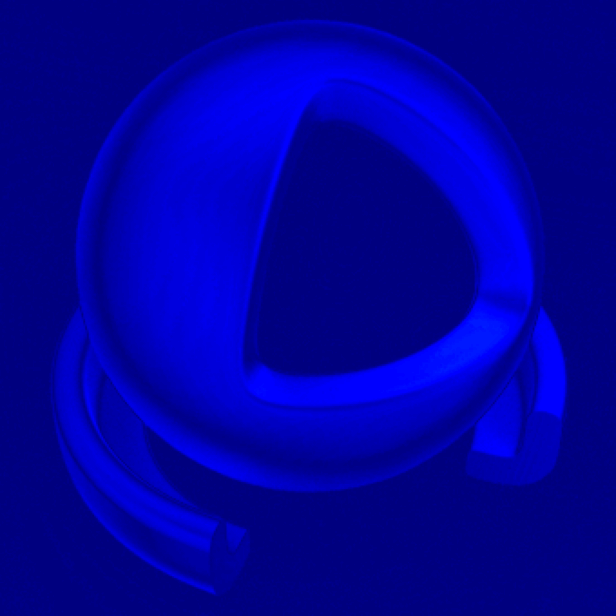}}
             \end{minipage}
        \end{minipage}
    \end{minipage}

   \caption{\note{Comparisons with state-of-the-art BRDF representations on measured BRDFs from EPFL~\cite{Dupuy2018:Adaptive} (top 3 rows) and UTIA~\cite{Filip2014:Template} (bottom 2 rows) datasets. From the first column to last, the ground-truths, the results of our enhanced GGX, the original GGX BRDF~\cite{Walter2007:GGX}, the FULL model in~\cite{Brady2014:Genbrdf}, L{\"o}w et al.~\cite{low2012brdf}, NBRDF~\cite{Sztrajman2021:NBRDF}, and NLBRDF~\cite{FAN2022:NLBRDF}. Quantitative errors in SSIM and $\Delta E_{ITP}$ ($\times 10^3$) are reported at the bottom-left of each related image}, while the error map is visualized at the bottom-right.}
   \label{fig:isocompare}
\end{figure*}


\subsection{Comparisons}
\begin{figure*}[htbp]
    \centering
        \begin{minipage}{7.1in}	
            \centering
            \begin{minipage}{1.0\linewidth}
            \includegraphics[width=\linewidth]{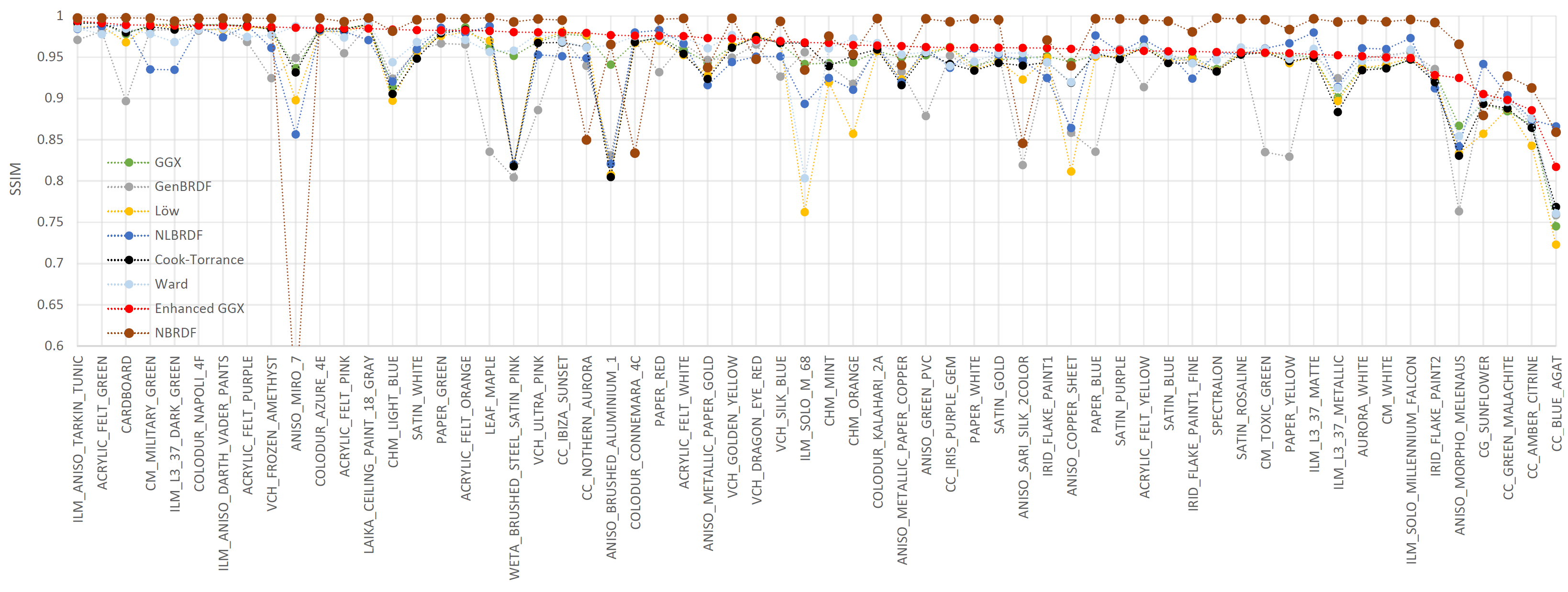}
            \end{minipage}	
        \end{minipage}	
    \caption{We computed the SSIM of our enhanced GGX, the original GGX, GenBRDF~\cite{Brady2014:Genbrdf}, the original Cook-Torrance~\cite{Cook1982:CookTorrance}, Löw~\cite{low2012brdf}, Ward BRDF~\cite{Ward1992:Ward} models, NLBRDF~\cite{FAN2022:NLBRDF} and NBRDF~\cite{Sztrajman2021:NBRDF}, across validation materials in~\cite{Dupuy2018:Adaptive}. Our enhanced GGX model outperforms analytical BRDFs in most cases. It also achieves better results compared to the material-independent NLBRDF, which has over $10^9$ network parameters. Please refer to the supplemental material for more details including rendered results.}
    \label{fig:all_ssim}
    
\end{figure*}

\note{We compare our enhanced GGX model with state-of-the-art techniques using high-quality measured materials, including 51 isotropic and 11 anisotropic BRDFs from the EPFL dataset~\cite{Dupuy2018:Adaptive} (a total of 62 materials) and the materials from the UTIA dataset~\cite{Kuznetsov2019GAN}}, as our validation data. Such data are never seen by any of the models for fairness. Moreover, while many of the existing methods are specialized in handling a certain type of BRDF or BTF, our model is \textbf{trained once for all}. We do not fine-tune the model for different tasks. In~\figref{fig:all_ssim} and the supplemental material, we quantitatively and qualitatively show the fitting quality of various models across all isotropic and anisotropic validation materials from~\cite{Dupuy2018:Adaptive}. Please also refer to the accompanying video for animated results.

\textbf{Isotropic BRDFs.}
In ~\figref{fig:isocompare}, we compare the reconstruction quality of isotropic BRDFs against state-of-the-art analytical models~\cite{low2012brdf, Brady2014:Genbrdf, Walter2007:GGX} and neural ones~\cite{Sztrajman2021:NBRDF, FAN2022:NLBRDF}. We collect $10^5$ BRDF values at different pairs of $\omega_i,\omega_o$, according to the sampling method described in~\sec{sec:data}, and use the same data as input to each model. As shown in the figure, even the latest analytical models cannot always produce accurate fitting results. The network of NLBRDF~\cite{FAN2022:NLBRDF} with over $10^9$ weights deviates considerably from the ground-truth in certain cases. \note{This is due to the inherent spectral bias and lack of positional encoding in NLBRDF, making it challenging to reconstruct high-frequency highlights~\cite{rahaman2019spectral,mildenhall2021nerf}. Consequently, mere network scaling yields smooth local minima, whereas our analytical backbone serves as a robust inductive bias for capturing sharp details in the angular domain.} The top performers are our model and a state-of-the-art dedicated \warning{BRDF-specific} network, NBRDF~\cite{Sztrajman2021:NBRDF}. \note{While NBRDF trains a different network for each material, our approach optimizes the parameters for a fixed network}. Our enhanced GGX substantially improves the original GGX, producing a more accurate fit to the ground-truth appearance. 



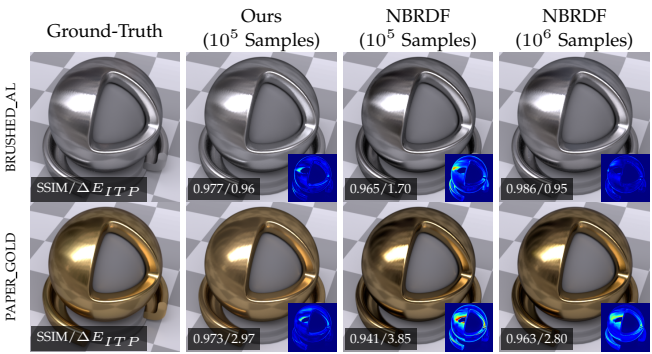
\begin{figure}[htbp]
    \centering
    \begin{minipage}{3.4in}
    \begin{minipage}{0.02in}	
            \centering
                \vspace{0.2in}
                \rotatebox{90}{\scriptsize \textsc{ }}
            \end{minipage}
        \begin{minipage}{3.4in}
            \centering
            
            \begin{minipage}{0.23\linewidth}
                \centering
                \subcaption{\scriptsize Ground-Truth}
            \end{minipage}			
            \begin{minipage}{0.23\linewidth}
                \centering
                \subcaption{\scriptsize \shortstack{Ours\\($10^5$ Samples)}}
            \end{minipage}		
            \begin{minipage}{0.23\linewidth}
                \centering
                \subcaption{\scriptsize \shortstack{NBRDF\\($10^5$ Samples)}}
            \end{minipage}		
            \begin{minipage}{0.23\linewidth}
                \centering
                \subcaption{\scriptsize \shortstack{NBRDF\\($10^6$ Samples)}}
            \end{minipage}	
        \end{minipage}
    \end{minipage}

    \centering		
    \begin{minipage}{3.4in}
        \begin{minipage}{0.02in}	
            \centering
                \rotatebox{90}{\tiny \uppercase{Brushed\_al}}
            \end{minipage}	
             \begin{minipage}{3.4in}	
                \centering
            \begin{minipage}{0.23\linewidth}
                \includegraphics[width=\linewidth]{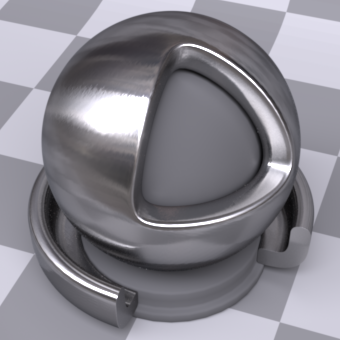}
                \put(-56,4.8) {\tikz[baseline] \node[fill=black, fill opacity=0.65, text opacity=1, text=white,inner sep=2pt] {\tiny SSIM/$\Delta E_{ITP}$};}
            \end{minipage}
            \begin{minipage}{0.23\linewidth}
                \centering
                \includegraphics[width=\linewidth]{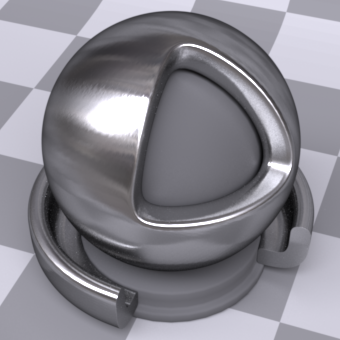}
                \put(-56,4.8) {\tikz[baseline] \node[fill=black, fill opacity=0.65, text opacity=1, text=white,inner sep=2pt] {\tiny 0.977/0.96};} 
                \put(-18,0){\includegraphics[width=18pt]{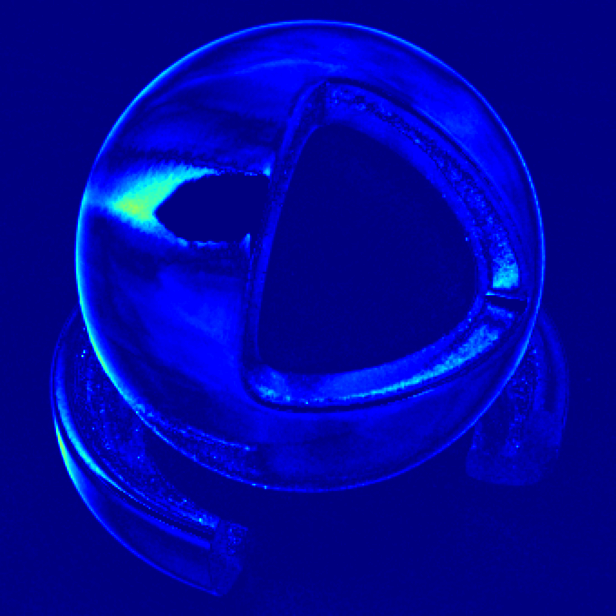}}
            \end{minipage}
            \begin{minipage}{0.23\linewidth}
                \centering
                \includegraphics[width=\linewidth]{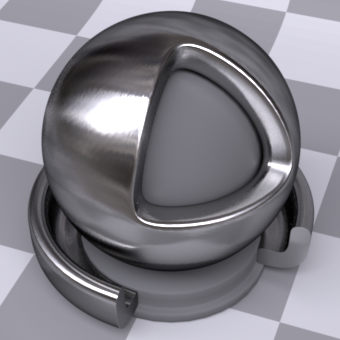}
                \put(-56,4.8) {\tikz[baseline] \node[fill=black, fill opacity=0.65, text opacity=1, text=white,inner sep=2pt] {\tiny  0.965/1.70};} 
                \put(-18,0){\includegraphics[width=18pt]{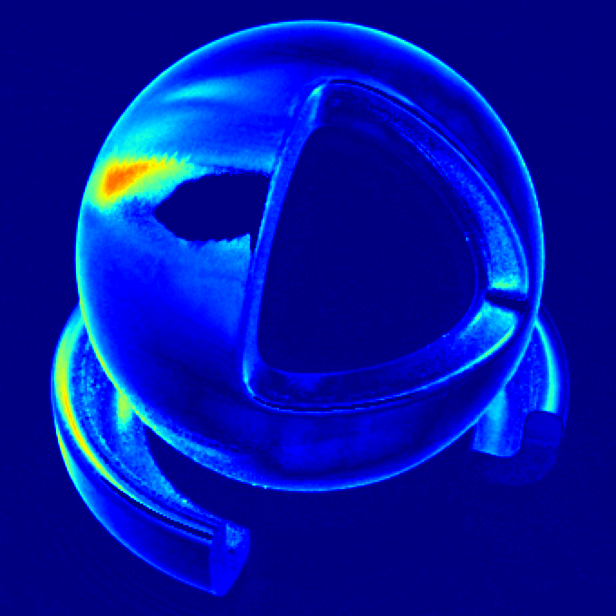}}
            \end{minipage}
            \begin{minipage}{0.23\linewidth}
                \centering
                \includegraphics[width=\linewidth]{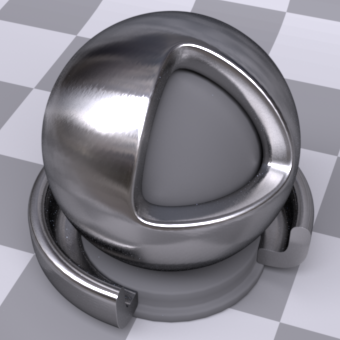}
                \put(-56,4.8) {\tikz[baseline] \node[fill=black, fill opacity=0.65, text opacity=1, text=white,inner sep=2pt] {\tiny 0.986/0.95};} 
                \put(-18,0){\includegraphics[width=18pt]{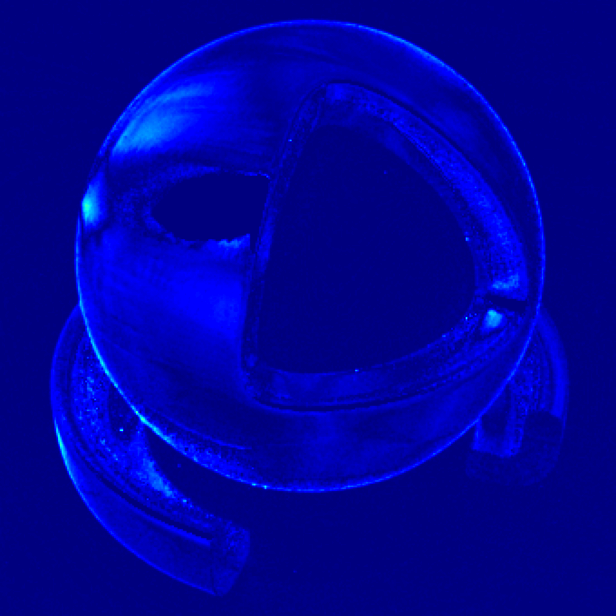}}
            \end{minipage}
        \end{minipage}	
    \end{minipage}	

    \begin{minipage}{3.4in}
        \begin{minipage}{0.02in}	
            \centering
                \rotatebox{90}{\tiny \uppercase{paper\_gold}}
        \end{minipage}	
         \begin{minipage}{3.4in}	
            \centering
            \begin{minipage}{0.23\linewidth}
            \includegraphics[width=\linewidth]{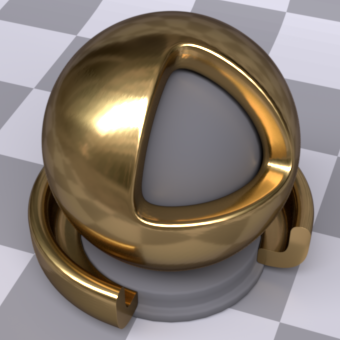}
            \put(-56,4.8) {\tikz[baseline] \node[fill=black, fill opacity=0.65, text opacity=1, text=white,inner sep=2pt] {\tiny SSIM/$\Delta E_{ITP}$};}
            \end{minipage}
            \begin{minipage}{0.23\linewidth}
            \includegraphics[width=\linewidth]{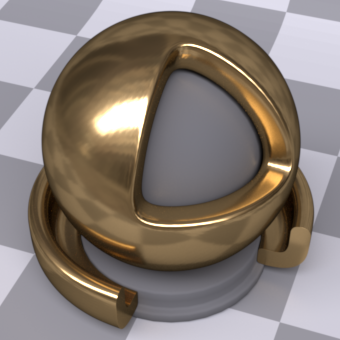}
            \put(-56,4.8) {\tikz[baseline] \node[fill=black, fill opacity=0.65, text opacity=1, text=white,inner sep=2pt] {\tiny 0.973/2.97};} 
            \put(-18,0){\includegraphics[width=18pt]{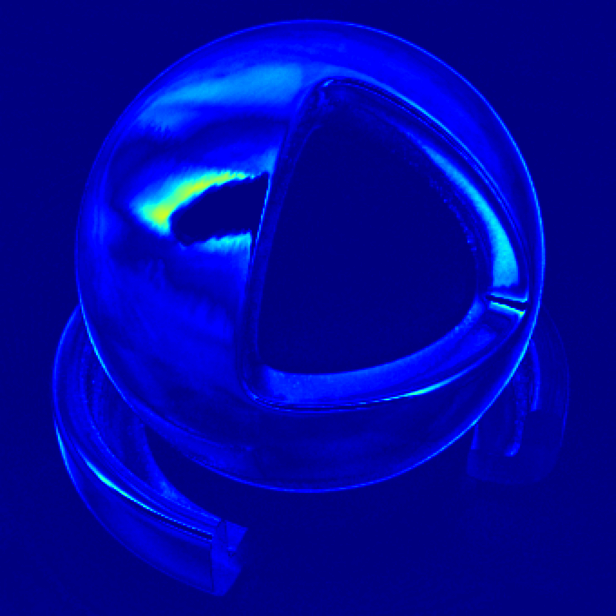}}
            \end{minipage}	
            \begin{minipage}{0.23\linewidth}
            \includegraphics[width=\linewidth]{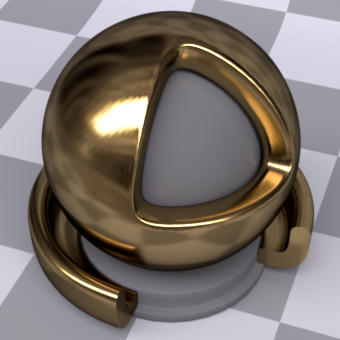}
            \put(-56,4.8) {\tikz[baseline] \node[fill=black, fill opacity=0.65, text opacity=1, text=white,inner sep=2pt] {\tiny 0.941/3.85};} 
            \put(-18,0){\includegraphics[width=18pt]{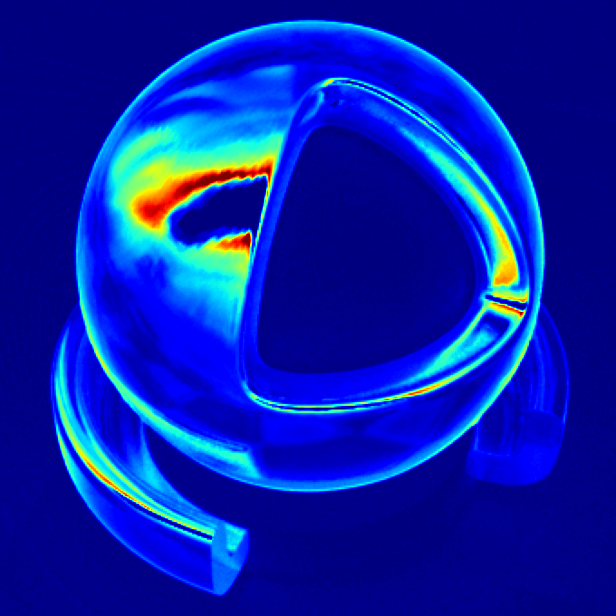}}
            \end{minipage}	
            \begin{minipage}{0.23\linewidth}
            \includegraphics[width=\linewidth]{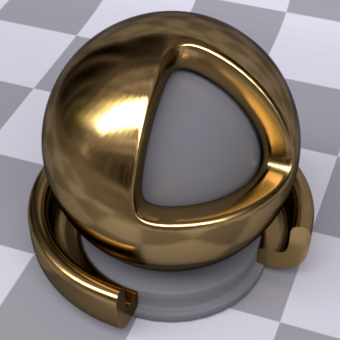}
            \put(-56,4.8) {\tikz[baseline] \node[fill=black, fill opacity=0.65, text opacity=1, text=white,inner sep=2pt] {\tiny 0.963/2.80};} 
            \put(-18,0){\includegraphics[width=18pt]{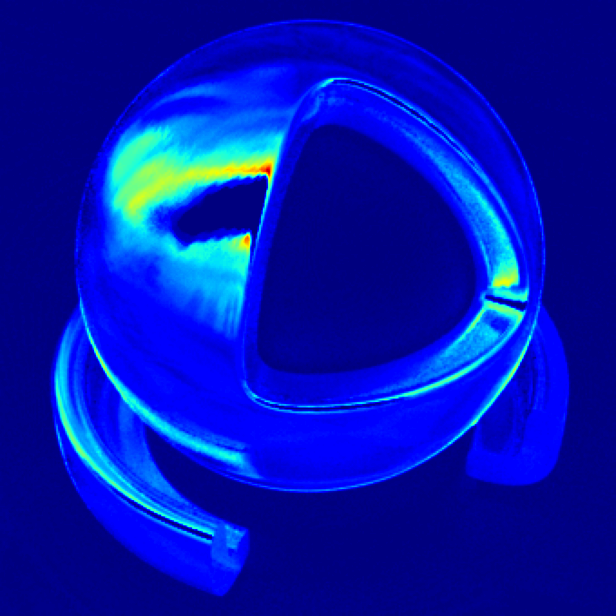}}
            \end{minipage}	
        \end{minipage}	
    \end{minipage}	
   \caption{Comparisons with NBRDF~\cite{Sztrajman2021:NBRDF} on anisotropic BRDFs from EPFL dataset~\cite{Dupuy2018:Adaptive}. From the first column to last, ground-truths, reconstruction results of our enhanced GGX using $10^5$ input measurements, results of NBRDF using $10^5$ measurements, and using $10^6$ measurements, respectively. \note{Quantitative errors in SSIM and $\Delta E_{ITP}$ ($\times 10^3$) are reported at the bottom-left of each related image}, and the error map is shown at the bottom-right.} 
   
    \label{fig:compare_anisotropic}
\end{figure}

\textbf{Anisotropic BRDFs.}
In~\figref{fig:compare_anisotropic}, we compare the reconstruction quality of anisotropic BRDFs against NBRDF~\cite{Sztrajman2021:NBRDF}. Note that we no longer compare with most of the methods in~\figref{fig:isocompare}, because they are limited to isotropic BRDFs only. Since our model preserves some of the prior knowledge from its input analytical model, our generalization ability is stronger than a purely data-driven approach: we achieve higher-quality fitting results than theirs, when using the same number of input measurements. 

\begin{figure}
    \centering	
    \begin{minipage}{3.4in}
        \begin{minipage}{0.02in}	
            \centering
                \vspace{0.2in}
                \rotatebox{90}{\scriptsize \textsc{ }}
        \end{minipage}
        \begin{minipage}{3.4in}
            \centering
            \begin{minipage}{0.23\linewidth}
                \centering
            \subcaption{\tiny \shortstack{\uppercase{Chm\_orange}}}
            \end{minipage}			
            \begin{minipage}{0.23\linewidth}
                \centering
                \subcaption{\tiny \shortstack{\uppercase{Silk\_blue}}}
            \end{minipage}		
            \begin{minipage}{0.23\linewidth}
                \centering
                \subcaption{\tiny \shortstack{\uppercase{Chm\_light\_blue}}}
            \end{minipage}		
            \begin{minipage}{0.23\linewidth}
                \centering
                \subcaption{\tiny \shortstack{\uppercase{Frozen\_amethyst}}}
            \end{minipage}	
        \end{minipage}
    \end{minipage}
    
    \begin{minipage}{3.4in}
        \begin{minipage}{0.02in}	
            \centering
            \rotatebox{90}{\text{\scriptsize Ground-Truth}}
        \end{minipage}	
        \begin{minipage}{3.4in}	
            \centering
            \begin{minipage}{0.23\linewidth}
                \centering
                \includegraphics[width=\linewidth]{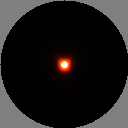}
            \end{minipage}
            \begin{minipage}{0.23\linewidth}
                \centering
                \includegraphics[width=\linewidth]{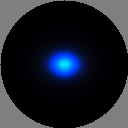}
            \end{minipage}
            \begin{minipage}{0.23\linewidth}
                \centering
                \includegraphics[width=\linewidth]{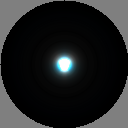}
            \end{minipage}
            \begin{minipage}{0.23\linewidth}
                \centering
                \includegraphics[width=\linewidth]{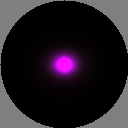}
            \end{minipage}
        \end{minipage}
    \end{minipage}
    
    \begin{minipage}{3.4in}
        \begin{minipage}{0.02in}	
            \centering
            \rotatebox{90}{\scriptsize Ours}
        \end{minipage}	
        \begin{minipage}{3.4in}	
            \centering
            \begin{minipage}{0.23\linewidth}
                \includegraphics[width=\linewidth]{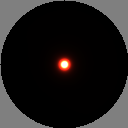}
                \put(-40.5,2.5) {\small \color{white} 0.0015}
            \end{minipage}
            \begin{minipage}{0.23\linewidth}
                \includegraphics[width=\linewidth]{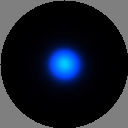}
                \put(-40.5,2.5) {\small \color{white} 0.0056}
            \end{minipage}
            \begin{minipage}{0.23\linewidth}
                \includegraphics[width=\linewidth]{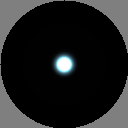}
                \put(-40.5,2.5) {\small \color{white} 0.0101}
            \end{minipage}
            \begin{minipage}{0.23\linewidth}
                \includegraphics[width=\linewidth]{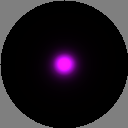}
                \put(-40.5,2.5) {\small \color{white} 0.0038}
            \end{minipage}
        \end{minipage}
    \end{minipage}

    \begin{minipage}{3.4in}
        \begin{minipage}{0.02in}	
            \centering
             \rotatebox{90}{\scriptsize NBRDF}
        \end{minipage}	
        \begin{minipage}{3.4in}	
            \centering
            \begin{minipage}{0.23\linewidth}
                \includegraphics[width=\linewidth]{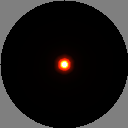}
                \put(-40.5,2.5) {\small \color{white} 0.0075}
            \end{minipage}
            \begin{minipage}{0.23\linewidth}
                \includegraphics[width=\linewidth]{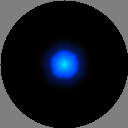}
                \put(-40.5,2.5) {\small \color{white} 0.0095}
            \end{minipage}
            \begin{minipage}{0.23\linewidth}
                \includegraphics[width=\linewidth]{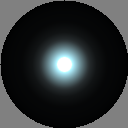}
                \put(-40.5,2.5) {\small \color{white} 0.0478}
            \end{minipage}
            \begin{minipage}{0.23\linewidth}
                \includegraphics[width=\linewidth]{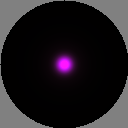}
                \put(-40.5,2.5) {\small \color{white} 0.0094}
            \end{minipage}
        \end{minipage}
    \end{minipage}
    
   \caption{Comparisons with NBRDF~\cite{Sztrajman2021:NBRDF} on different sampling strategies. Each image is a BRDF slice with a fixed $w_o$. The first row are ground-truths, followed by the results of our enhanced GGX and NBRDF. The first two columns use anisotropic sampling angles $(\theta_h, \phi_h, \theta_d, \phi_d)$ as input for isotropic BRDF. The last two columns show results using the retro-reflection directional samples as input. Numerical error is reported at the bottom of each related image, computed as the mean absolute value of the logarithmic loss.} 
    \label{fig:compare_anisotropic_aspects}
\end{figure}

We perform further comparisons with NBRDF~\cite{Sztrajman2021:NBRDF} in~\figref{fig:compare_anisotropic_aspects}. We find that their approach is sensitive to the sampling strategy. Using anisotropic sampling angles ($\theta_{h},\phi_{h},\theta_{d},\phi_{d}$) for isotropic data leads to poor quality in their results. In addition, fitting the retro-reflection directional samples from EPFL dataset~\cite{Dupuy2018:Adaptive} leads to the results clearly deviating from the ground-truth (e.g., mismatching highlight shape). In both cases, we produce fitting results closer to the ground-truth, demonstrating our robustness to different sampling strategies. 



\subsection{Evaluations}
\label{sec:evaluations}

\textbf{Enhancement Process.}
In~\figref{fig:teaser}, we further visualize the process of neural enhancement of the GGX model. The order of enhanced terms is $\mathcal{E}$, $\mathcal{G}$, $\mathcal{\times} ($of $ \mathcal{F \times G})$, and $\mathcal{F}$. This progression shares similarities with the historical development of BRDF models, such as the improvement of $\mathcal{F}$, $\mathcal{G}$, and $\mathcal{E}$ in papers like~\cite{schlick1994inexpensive} and~\cite{ashikmin2000microfacet}. Our framework automates this previously manual process of identifying and improving various terms in an existing analytical model.

\begin{figure*}[htbp]
    \centering
    \begin{minipage}{7.1in}
    \begin{minipage}{0.03in}	
            \centering
                \vspace{0.2in}
                \rotatebox{90}{\scriptsize \textsc{ }}
            \end{minipage}
        \hspace{-0.1in}
        \begin{minipage}{7.1in}
            \centering
            \begin{minipage}{0.95in}
                \centering
                {\scriptsize Ground-Truth}
            \end{minipage}			
            \begin{minipage}{0.95in}
                \centering
                {\scriptsize Enhanced \\ \vspace{-0.05in} Cook-Torrance}
            \end{minipage}		
            \begin{minipage}{0.95in}
                \centering
                {\scriptsize Cook-Torrance~\cite{Cook1982:CookTorrance}}
            \end{minipage}		
            \begin{minipage}{0.95in}
                \centering
                {\scriptsize Enhanced Ward}
            \end{minipage}	
            \begin{minipage}{0.95in}
                \centering
                {\scriptsize Ward~\cite{Ward1992:Ward}}
            \end{minipage}	
            \begin{minipage}{0.95in}
                \centering
                {\scriptsize Enhanced GenBRDF}
            \end{minipage}
            \begin{minipage}{0.95in}
                \centering
                {\scriptsize GenBRDF~\cite{Brady2014:Genbrdf}}
            \end{minipage}
        \end{minipage}
    \end{minipage}
    
    \begin{minipage}{7.1in}
        \begin{minipage}{0.03in}	
            \centering
            \rotatebox{90}{\tiny \uppercase{Ilm\_l3\_37\_metallic}}
        \end{minipage}	
        \hspace{-0.1in}
        \begin{minipage}{7.1in}
            \centering
            \begin{minipage}{0.95in} 
            \centering
            \includegraphics[width=0.95in]{fig_new/gt_EPFL/ilm_l3_37_metallic_rgb.png} 
            \put(-68.6,4.8) {\tikz[baseline] \node[fill=black, fill opacity=0.65, text opacity=1, text=white,inner sep=2pt] {\tiny SSIM/$\Delta E_{ITP}$};}
             \end{minipage}
             \begin{minipage}{0.95in}              
            \centering
            \includegraphics[width=0.95in]{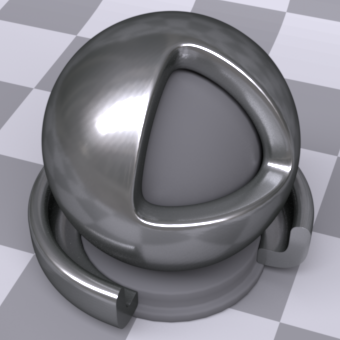}
             \put(-68.6,4.8) {\tikz[baseline] \node[fill=black, fill opacity=0.65, text opacity=1, text=white,inner sep=2pt] {\tiny 0.965/1.32};}  
             \put(-20,0){\includegraphics[width=20pt]{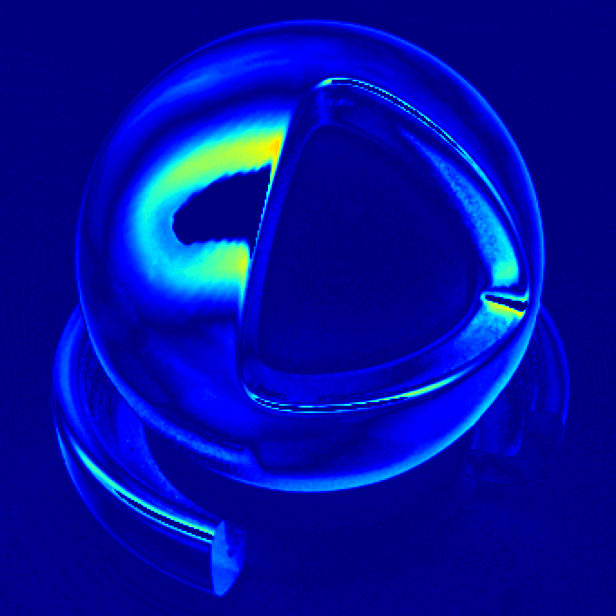}}
             \end{minipage}
             \begin{minipage}{0.95in}
            \centering
            \includegraphics[width=0.95in]{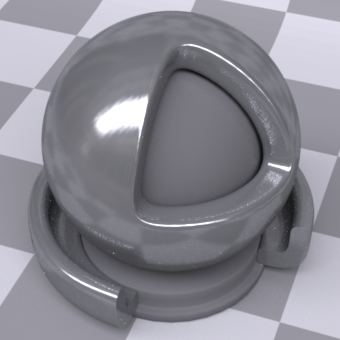}
             \put(-68.6,4.8) {\tikz[baseline] \node[fill=black, fill opacity=0.65, text opacity=1, text=white,inner sep=2pt] {\tiny 0.883/2.67};}  
             \put(-20,0){\includegraphics[width=20pt]{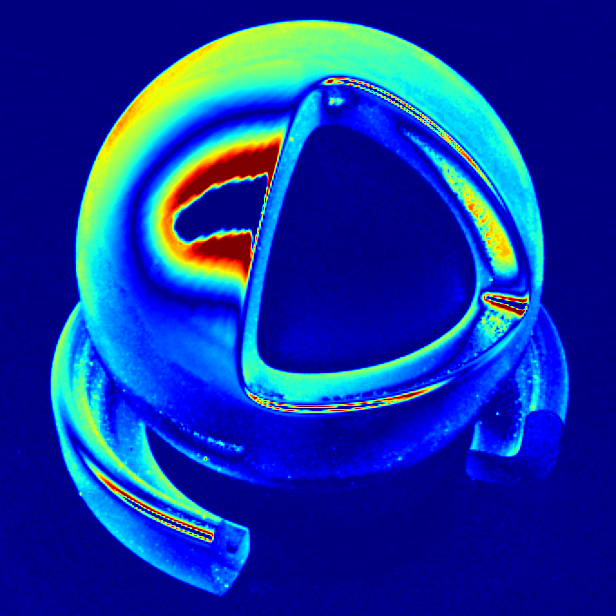}}
             \end{minipage}
             \begin{minipage}{0.95in}
            \centering
            \includegraphics[width=0.95in]{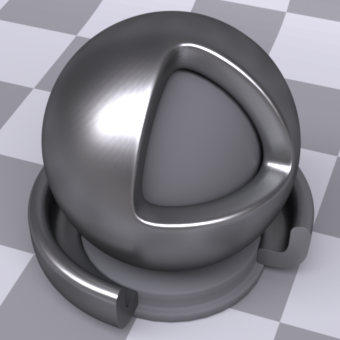}
             \put(-68.6,4.8) {\tikz[baseline] \node[fill=black, fill opacity=0.65, text opacity=1, text=white,inner sep=2pt] {\tiny 0.951/1.35};}  
             \put(-20,0){\includegraphics[width=20pt]{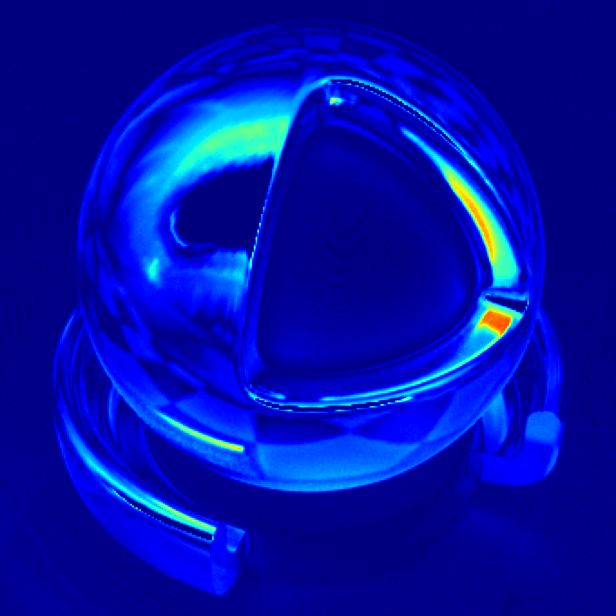}}
             \end{minipage}
             \begin{minipage}{0.95in}            
            \centering
            \includegraphics[width=0.95in]{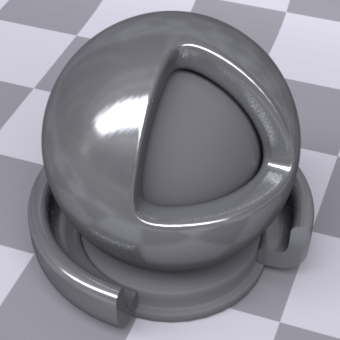}
             \put(-68.6,4.8) {\tikz[baseline] \node[fill=black, fill opacity=0.65, text opacity=1, text=white,inner sep=2pt] {\tiny 0.913/2.16};} 
             \put(-20,0){\includegraphics[width=20pt]{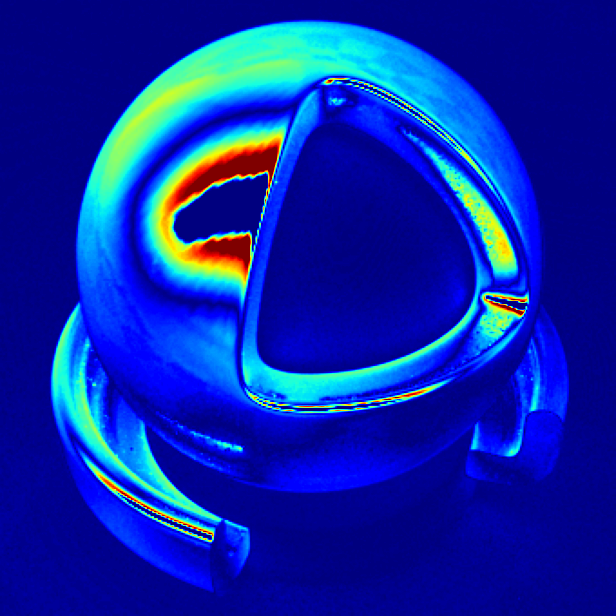}}
             \end{minipage}
             \begin{minipage}{0.95in}  
            \centering
            \includegraphics[width=0.95in]{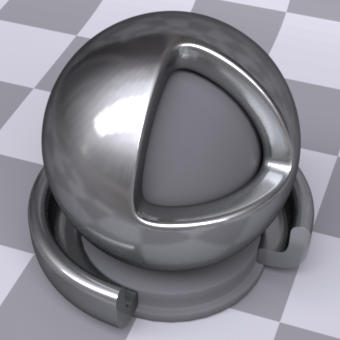}
             \put(-68.6,4.8) {\tikz[baseline] \node[fill=black, fill opacity=0.65, text opacity=1, text=white,inner sep=2pt] {\tiny 0.945/1.71};}   
             \put(-20,0){\includegraphics[width=20pt]{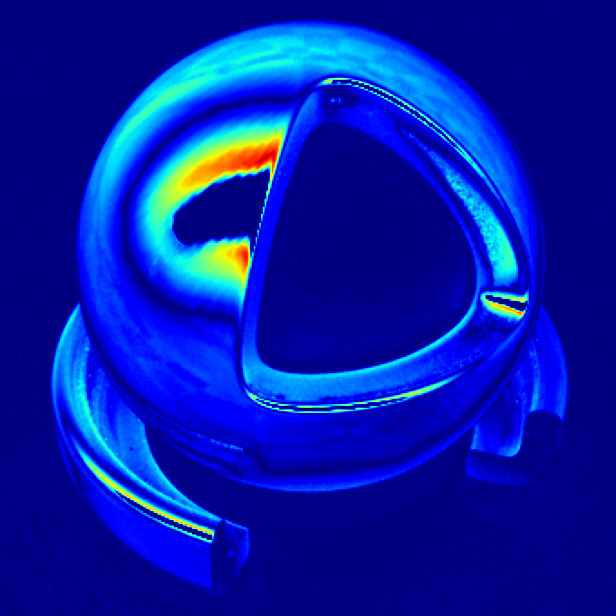}}
             \end{minipage}
             \begin{minipage}{0.95in}     
            \centering
            \includegraphics[width=0.95in]{fig_new/analytical_gen/ilm_l3_37_metallic_rgb.png}
             \put(-68.6,4.8) {\tikz[baseline] \node[fill=black, fill opacity=0.65, text opacity=1, text=white,inner sep=2pt] {\tiny 0.925/1.99};} 
             \put(-20,0){\includegraphics[width=20pt]{fig_new/analytical_gen_err/ilm_l3_37_metallic_rgb.png}}
             \end{minipage}
        \end{minipage}
    \end{minipage}
    
    \begin{minipage}{7.1in}
        \begin{minipage}{0.03in}	
            \centering
            \rotatebox{90}{\tiny \uppercase{Leaf\_maple}}
        \end{minipage}	
        \hspace{-0.1in}
        \begin{minipage}{7.1in}
            \centering
            \begin{minipage}{0.95in}  
            \includegraphics[width=0.95in]{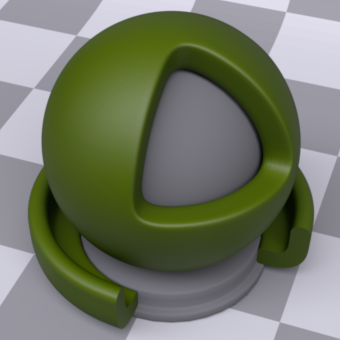}
            \put(-68.6,4.8) {\tikz[baseline] \node[fill=black, fill opacity=0.65, text opacity=1, text=white,inner sep=2pt] {\tiny SSIM/$\Delta E_{ITP}$};}
             \end{minipage}
             \begin{minipage}{0.95in}      
            \includegraphics[width=0.95in]{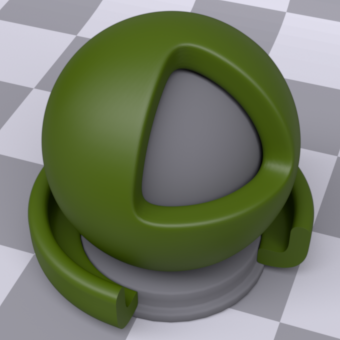}
             \put(-68.6,4.8) {\tikz[baseline] \node[fill=black, fill opacity=0.65, text opacity=1, text=white,inner sep=2pt] {\tiny 0.985/5.03};} 
             \put(-20,0){\includegraphics[width=20pt]{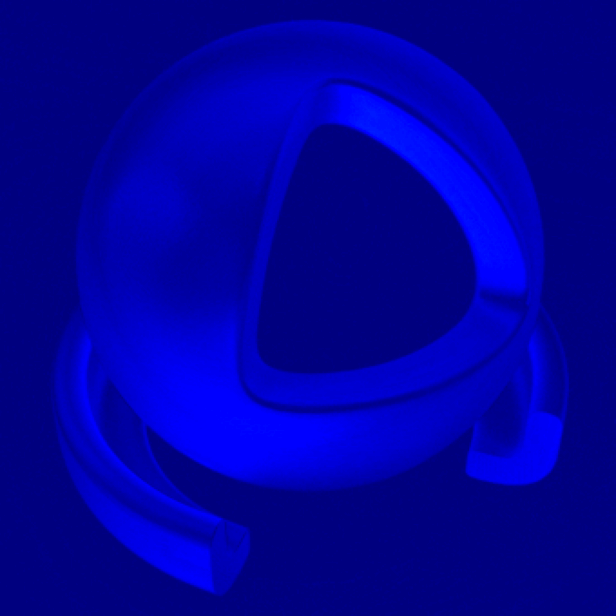}}
             \end{minipage}
             \begin{minipage}{0.95in}      
            \includegraphics[width=0.95in]{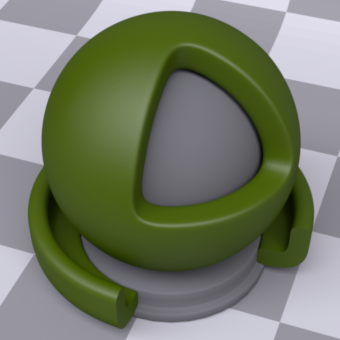}
             \put(-68.6,4.8) {\tikz[baseline] \node[fill=black, fill opacity=0.65, text opacity=1, text=white,inner sep=2pt] {\tiny 0.982/5.55};} 
             \put(-20,0){\includegraphics[width=20pt]{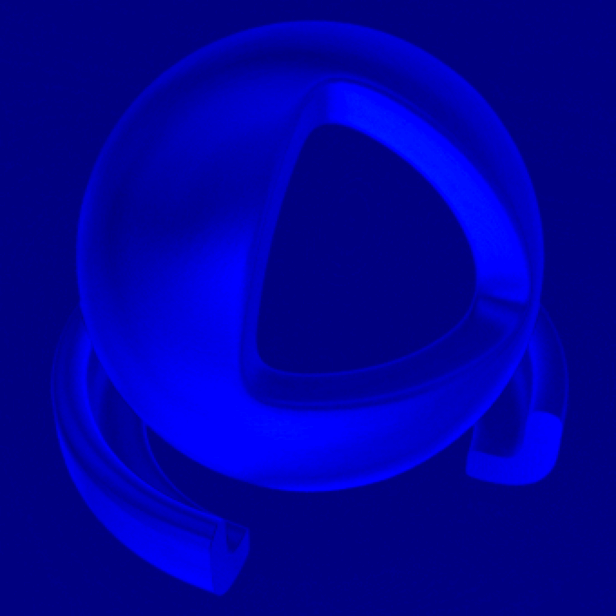}}
             \end{minipage}
             \begin{minipage}{0.95in}      
            \includegraphics[width=0.95in]{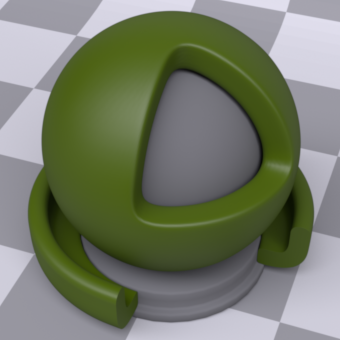}
             \put(-68.6,4.8) {\tikz[baseline] \node[fill=black, fill opacity=0.65, text opacity=1, text=white,inner sep=2pt] {\tiny 0.978/4.38};} 
             \put(-20,0){\includegraphics[width=20pt]{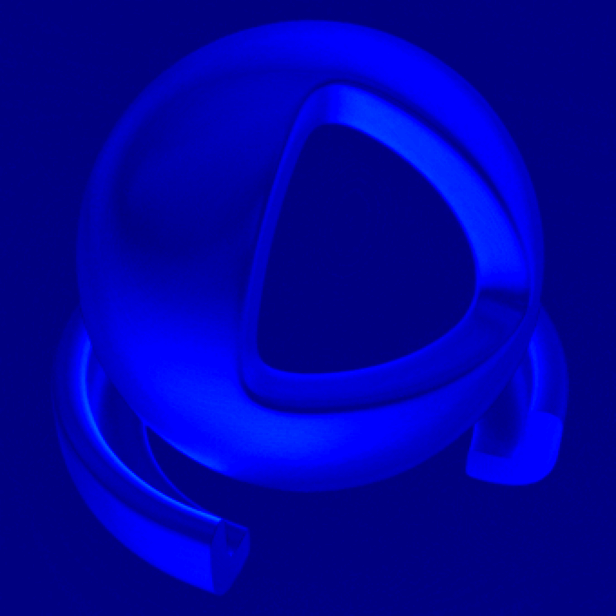}}
             \end{minipage}
             \begin{minipage}{0.95in}      
            \includegraphics[width=0.95in]{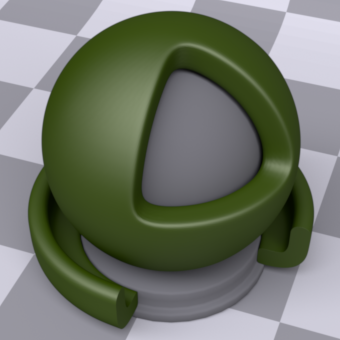}
             \put(-68.6,4.8) {\tikz[baseline] \node[fill=black, fill opacity=0.65, text opacity=1, text=white,inner sep=2pt] {\tiny 0.957/9.94};} 
             \put(-20,0){\includegraphics[width=20pt]{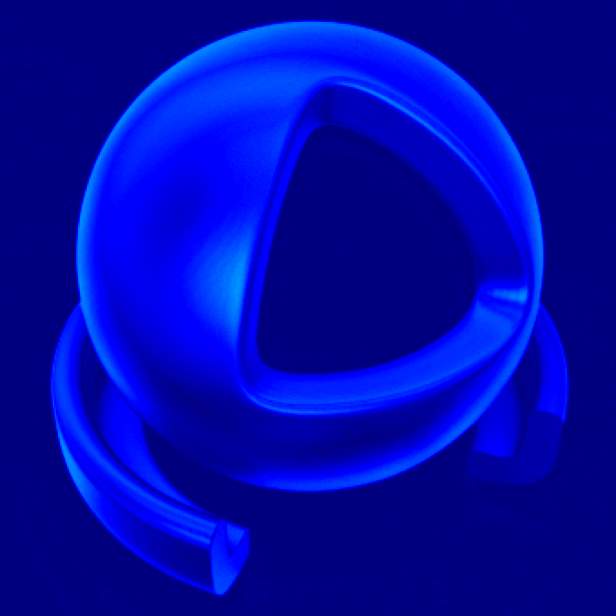}}
             \end{minipage}
             \begin{minipage}{0.95in}      
            \includegraphics[width=0.95in]{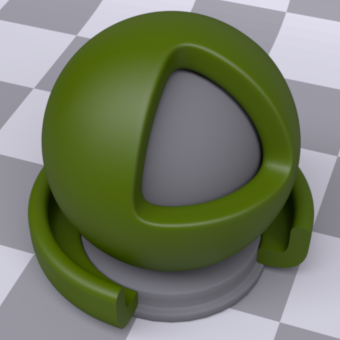}
             \put(-68.6,4.8) {\tikz[baseline] \node[fill=black, fill opacity=0.65, text opacity=1, text=white,inner sep=2pt] {\tiny 0.980/5.17};} 
             \put(-20,0){\includegraphics[width=20pt]{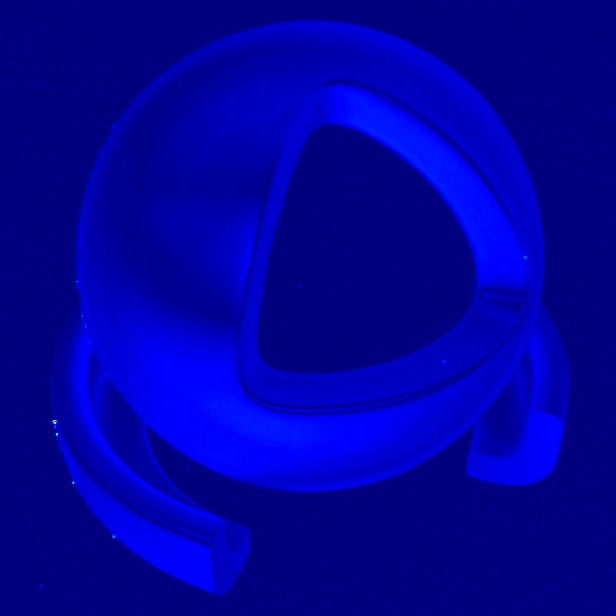}}
             \end{minipage}
             \begin{minipage}{0.95in}      
            \includegraphics[width=0.95in]{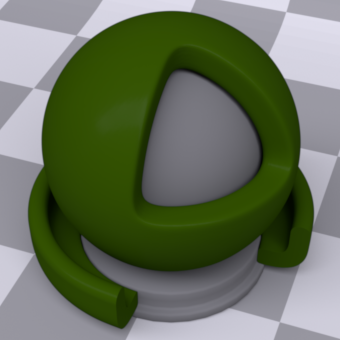}
             \put(-68.6,4.8) {\tikz[baseline] \node[fill=black, fill opacity=0.65, text opacity=1, text=white,inner sep=2pt] {\tiny 0.835/26.1};} 
             \put(-20,0){\includegraphics[width=20pt]{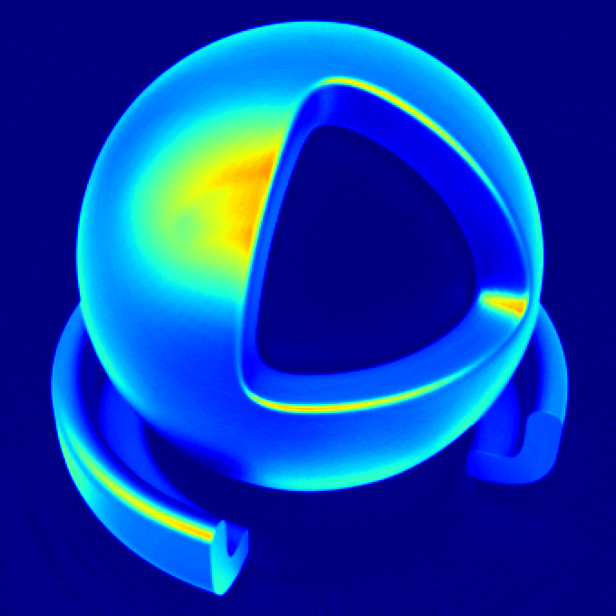}}
             \end{minipage}
        \end{minipage}
    \end{minipage}

    \begin{minipage}{7.1in}
        \begin{minipage}{0.03in}	
            \centering
            \rotatebox{90}{\tiny \uppercase{Felt\_orange}}
        \end{minipage}	
        \hspace{-0.1in}
        \begin{minipage}{7.1in}
            \centering
            \begin{minipage}{0.95in}
            \includegraphics[width=0.95in]{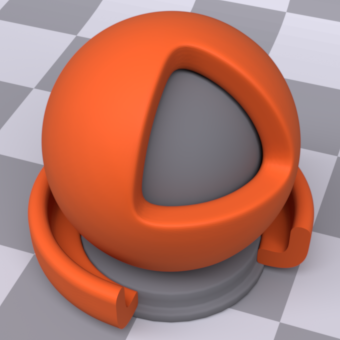}
            \put(-68.6,4.8) {\tikz[baseline] \node[fill=black, fill opacity=0.65, text opacity=1, text=white,inner sep=2pt] {\tiny SSIM/$\Delta E_{ITP}$};}
             \end{minipage}
             \begin{minipage}{0.95in}      
            \includegraphics[width=0.95in]{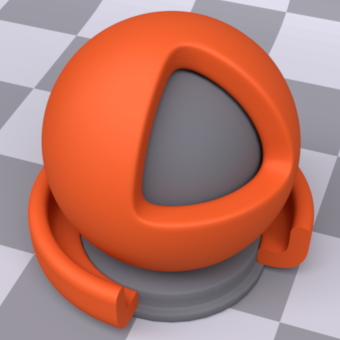}
             \put(-68.6,4.8) {\tikz[baseline] \node[fill=black, fill opacity=0.65, text opacity=1, text=white,inner sep=2pt] {\tiny 0.987/3.22};} 
             \put(-20,0){\includegraphics[width=20pt]{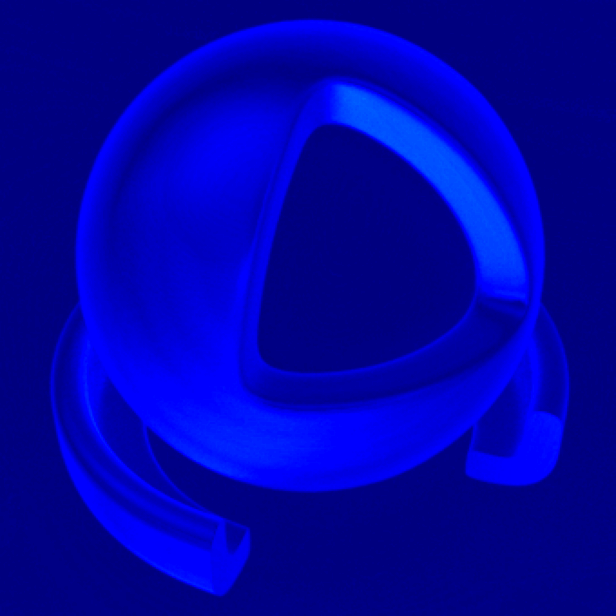}}
             \end{minipage}
             \begin{minipage}{0.95in}      
            \includegraphics[width=0.95in]{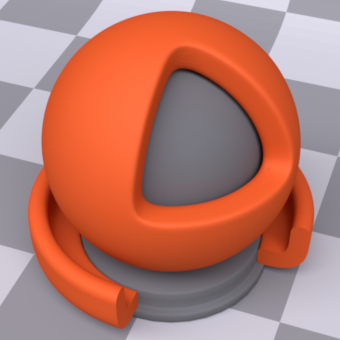}
             \put(-68.6,4.8) {\tikz[baseline] \node[fill=black, fill opacity=0.65, text opacity=1, text=white,inner sep=2pt] {\tiny 0.983/3.26};} 
             \put(-20,0){\includegraphics[width=20pt]{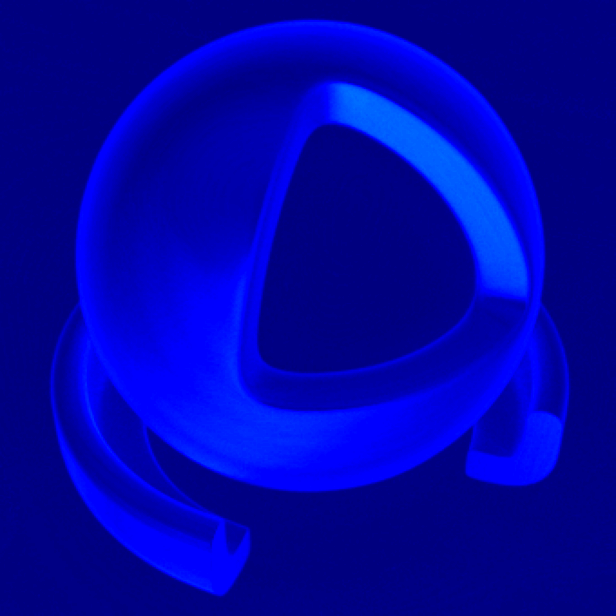}}
             \end{minipage}
             \begin{minipage}{0.95in}      
            \includegraphics[width=0.95in]{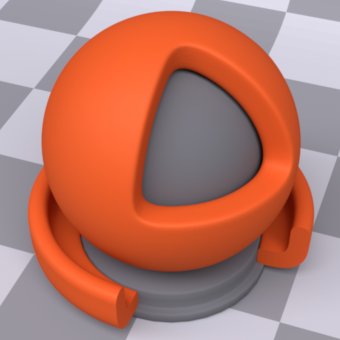}
             \put(-68.6,4.8) {\tikz[baseline] \node[fill=black, fill opacity=0.65, text opacity=1, text=white,inner sep=2pt] {\tiny 0.986/3.08};} 
             \put(-20,0){\includegraphics[width=20pt]{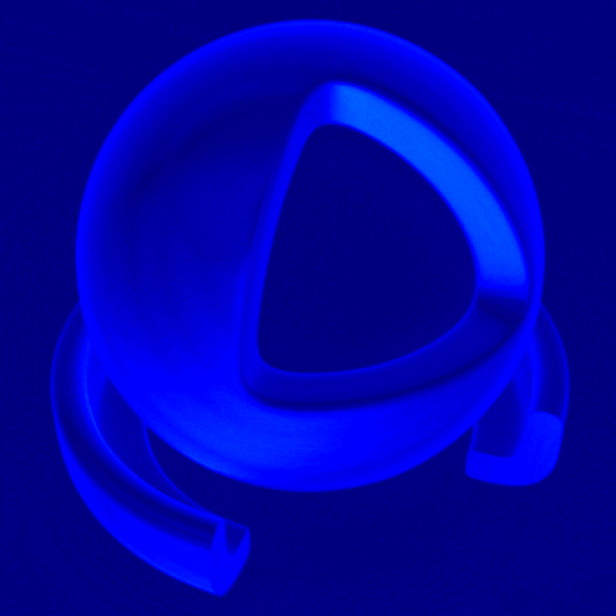}}
             \end{minipage}
             \begin{minipage}{0.95in}      
            \includegraphics[width=0.95in]{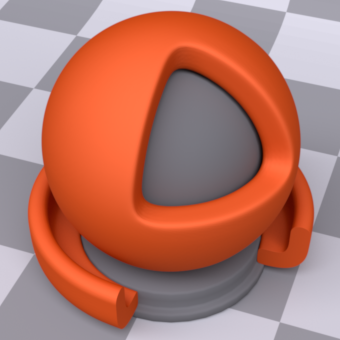}
             \put(-68.6,4.8) {\tikz[baseline] \node[fill=black, fill opacity=0.65, text opacity=1, text=white,inner sep=2pt] {\tiny 0.972/5.21};} 
             \put(-20,0){\includegraphics[width=20pt]{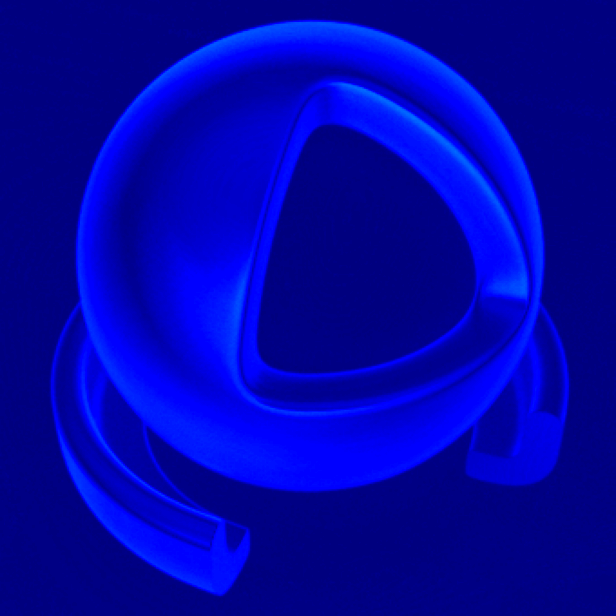}}
             \end{minipage}
             \begin{minipage}{0.95in}      
            \includegraphics[width=0.95in]{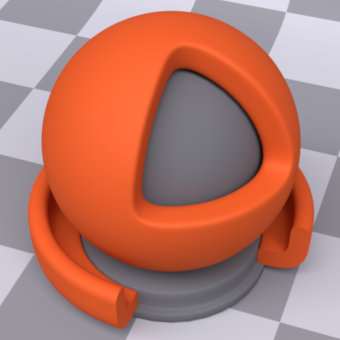}
             \put(-68.6,4.8) {\tikz[baseline] \node[fill=black, fill opacity=0.65, text opacity=1, text=white,inner sep=2pt] {\tiny 0.984/2.67};} 
             \put(-20,0){\includegraphics[width=20pt]{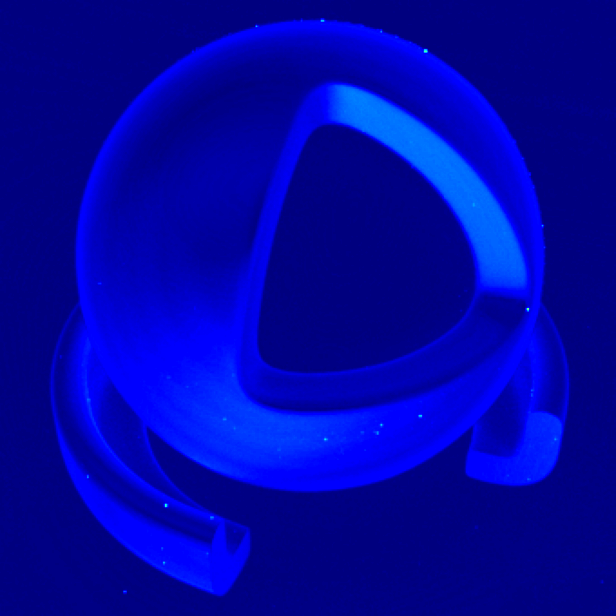}}
             \end{minipage}
             \begin{minipage}{0.95in}      
            \includegraphics[width=0.95in]{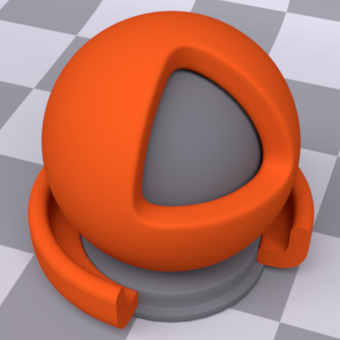}
             \put(-68.6,4.8) {\tikz[baseline] \node[fill=black, fill opacity=0.65, text opacity=1, text=white,inner sep=2pt] {\tiny 0.966/9.89};} 
             \put(-20,0){\includegraphics[width=20pt]{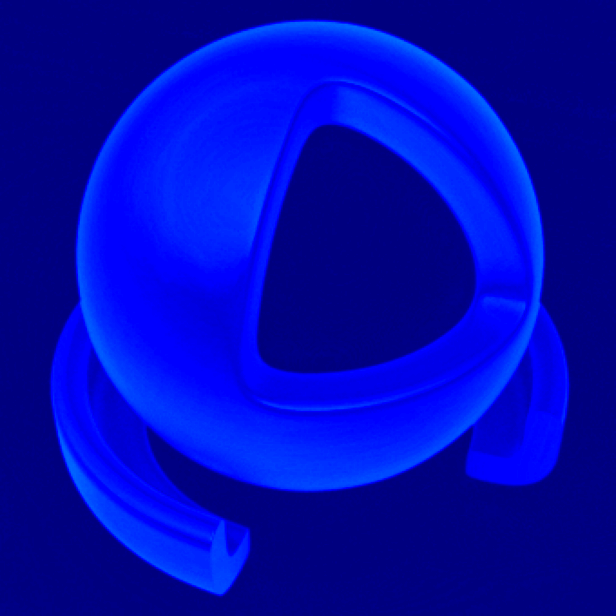}}
             \end{minipage} 
        \end{minipage}
    \end{minipage}
    
    \begin{minipage}{7.1in}
        \begin{minipage}{0.03in}	
            \rotatebox{90}{\tiny \uppercase{Spectralon}}
        \end{minipage}	
        \hspace{-0.1in}
        \begin{minipage}{7.1in}
            \centering
            \begin{minipage}{0.95in}
            \includegraphics[width=0.95in]{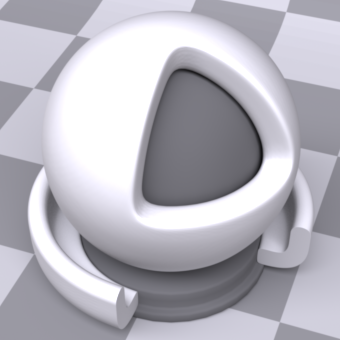}  
            \put(-68.6,4.8) {\tikz[baseline] \node[fill=black, fill opacity=0.65, text opacity=1, text=white,inner sep=2pt] {\tiny SSIM/$\Delta E_{ITP}$};}          
            \end{minipage}
            \begin{minipage}{0.95in}
            \includegraphics[width=0.95in]{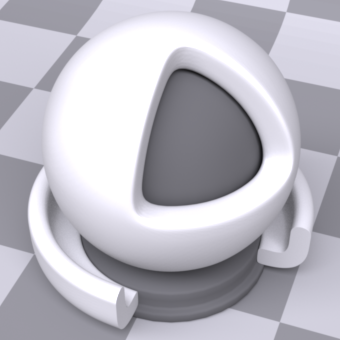}
             \put(-68.6,4.8) {\tikz[baseline] \node[fill=black, fill opacity=0.65, text opacity=1, text=white,inner sep=2pt] {\tiny 0.977/0.92};}            
             \put(-20,0){\includegraphics[width=20pt]{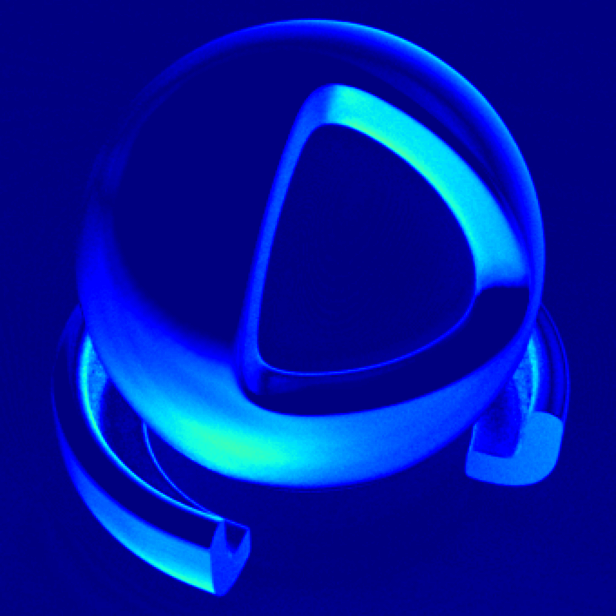}}
            \end{minipage}
            \begin{minipage}{0.95in}
            \includegraphics[width=0.95in]{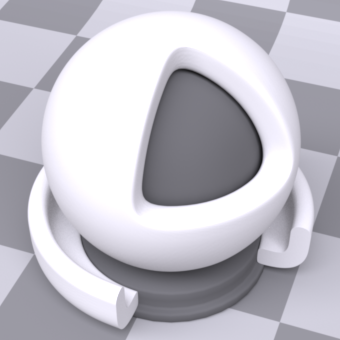}
             \put(-68.6,4.8) {\tikz[baseline] \node[fill=black, fill opacity=0.65, text opacity=1, text=white,inner sep=2pt] {\tiny 0.933/1.73};}            
             \put(-20,0){\includegraphics[width=20pt]{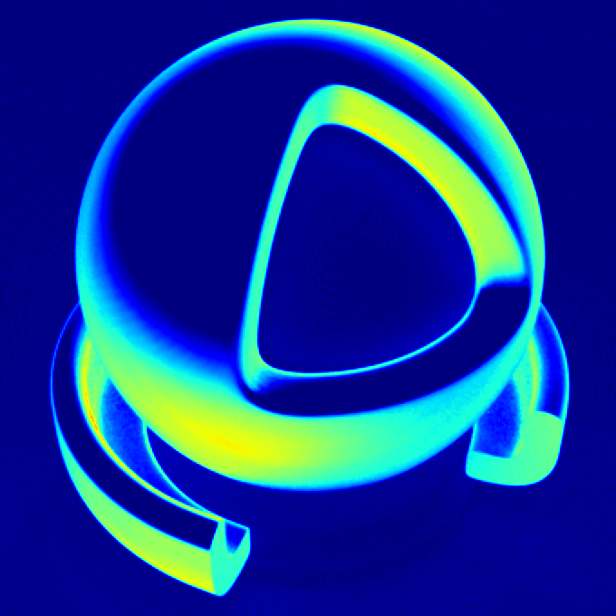}}
            \end{minipage}
            \begin{minipage}{0.95in}
            \includegraphics[width=0.95in]{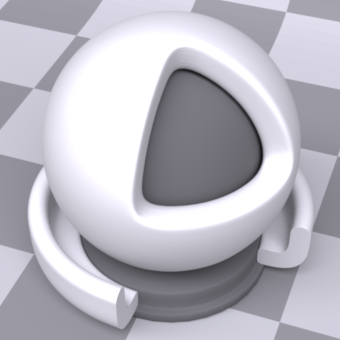}
             \put(-68.6,4.8) {\tikz[baseline] \node[fill=black, fill opacity=0.65, text opacity=1, text=white,inner sep=2pt] {\tiny 0.979/0.84};}            
             \put(-20,0){\includegraphics[width=20pt]{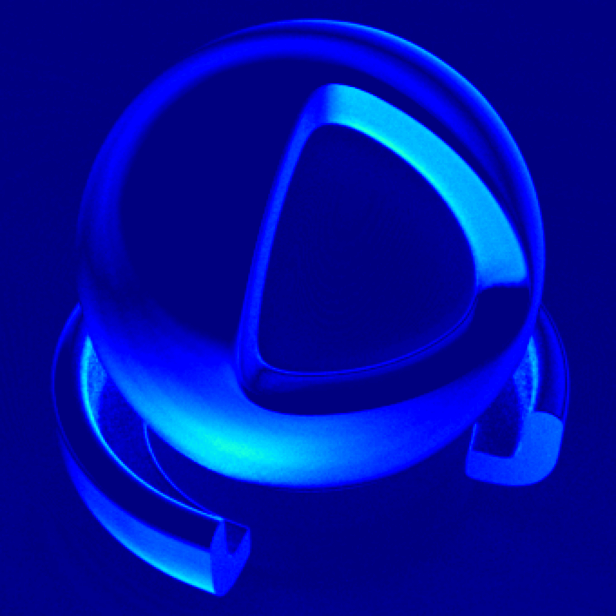}}
            \end{minipage}
            \begin{minipage}{0.95in}
            \includegraphics[width=0.95in]{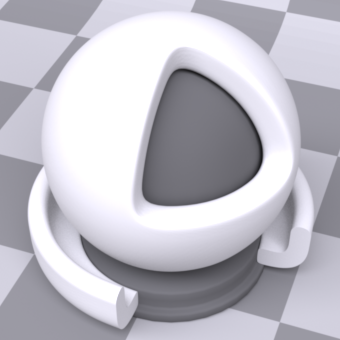}
             \put(-68.6,4.8) {\tikz[baseline] \node[fill=black, fill opacity=0.65, text opacity=1, text=white,inner sep=2pt] {\tiny 0.947/1.48};}            
             \put(-20,0){\includegraphics[width=20pt]{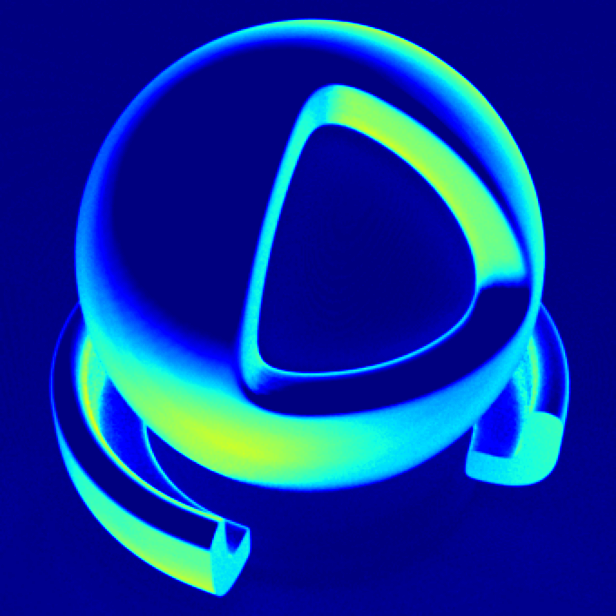}}
            \end{minipage}
            \begin{minipage}{0.95in}
            \includegraphics[width=0.95in]{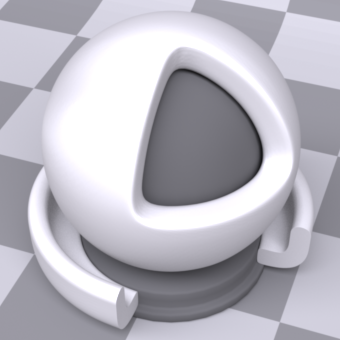}
             \put(-68.6,4.8) {\tikz[baseline] \node[fill=black, fill opacity=0.65, text opacity=1, text=white,inner sep=2pt] {\tiny 0.985/0.73};}            
             \put(-20,0){\includegraphics[width=20pt]{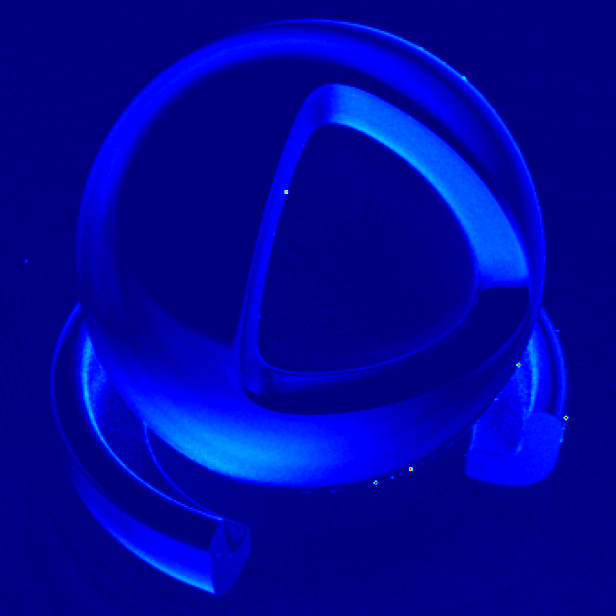}}
            \end{minipage}
            \begin{minipage}{0.95in}
            \includegraphics[width=0.95in]{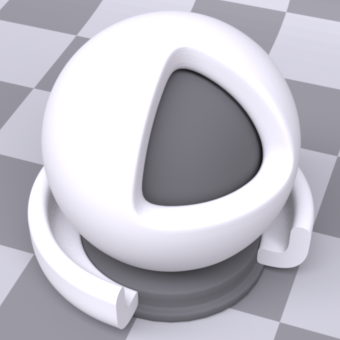}
             \put(-68.6,4.8) {\tikz[baseline] \node[fill=black, fill opacity=0.65, text opacity=1, text=white,inner sep=2pt] {\tiny 0.932/1.73};}            
             \put(-20,0){\includegraphics[width=20pt]{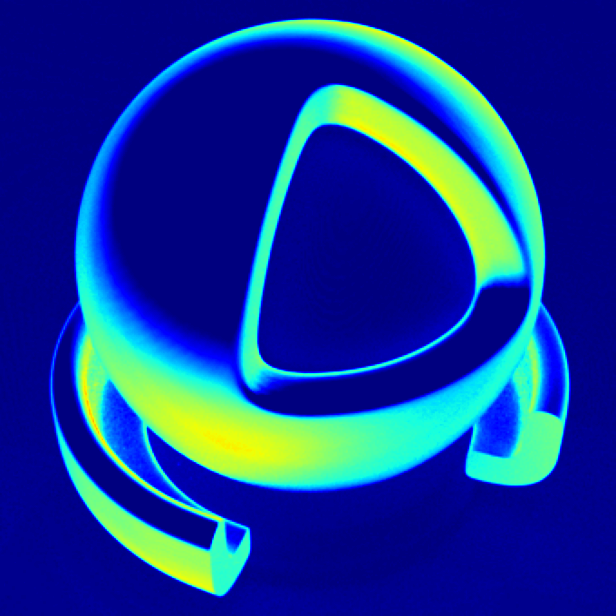}}
            \end{minipage}
        \end{minipage}
    \end{minipage}

    \begin{minipage}{7.1in}
        \begin{minipage}{0.03in}	
            \centering
            \rotatebox{90}{\tiny \uppercase{Vch\_ultra\_pink}}
        \end{minipage}	
        \hspace{-0.1in}
        \begin{minipage}{7.1in}
            \centering
            \begin{minipage}{0.95in}
            \includegraphics[width=0.95in]{fig_new/gt_EPFL/vch_ultra_pink_rgb.png}  
            \put(-68.6,4.8) {\tikz[baseline] \node[fill=black, fill opacity=0.65, text opacity=1, text=white,inner sep=2pt] {\tiny SSIM/$\Delta E_{ITP}$};}
             \end{minipage}
             \begin{minipage}{0.95in}               
            \includegraphics[width=0.95in]{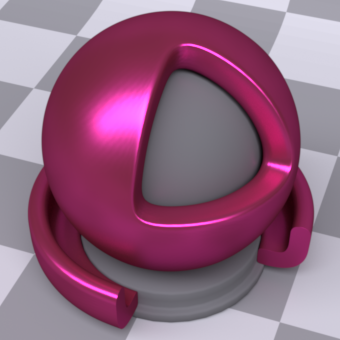}
             \put(-68.6,4.8) {\tikz[baseline] \node[fill=black, fill opacity=0.65, text opacity=1, text=white,inner sep=2pt] {\tiny 0.981/4.71};} 
             \put(-20,0){\includegraphics[width=20pt]{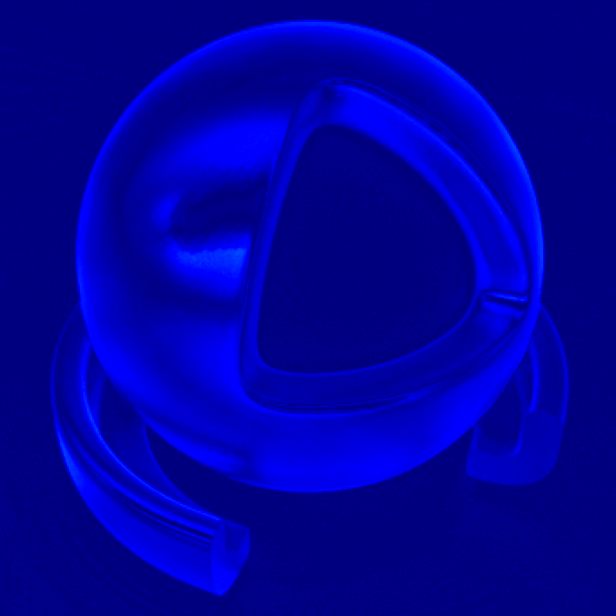}}
             \end{minipage}
             \begin{minipage}{0.95in}      
            \includegraphics[width=0.95in]{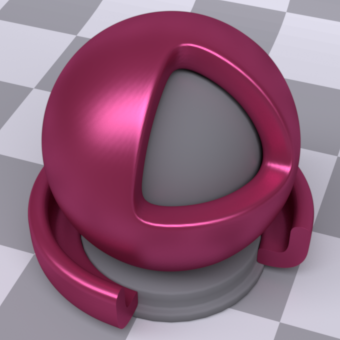}
             \put(-68.6,4.8) {\tikz[baseline] \node[fill=black, fill opacity=0.65, text opacity=1, text=white,inner sep=2pt] {\tiny 0.967/5.56};} 
             \put(-20,0){\includegraphics[width=20pt]{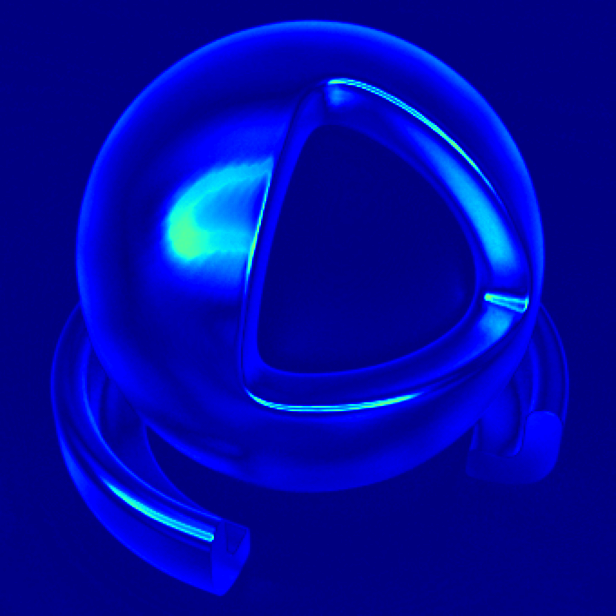}}
             \end{minipage}
             \begin{minipage}{0.95in}      
            \includegraphics[width=0.95in]{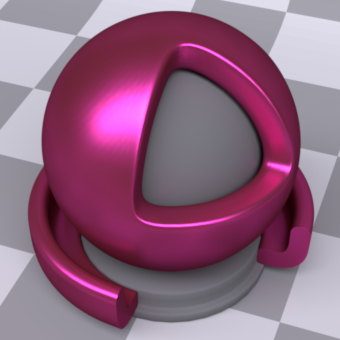}
             \put(-68.6,4.8) {\tikz[baseline] \node[fill=black, fill opacity=0.65, text opacity=1, text=white,inner sep=2pt] {\tiny 0.980/4.50};} 
             \put(-20,0){\includegraphics[width=20pt]{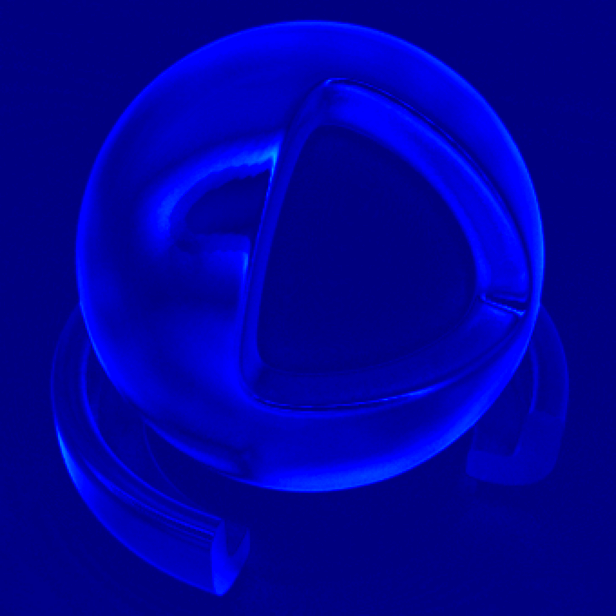}}
             \end{minipage}
             \begin{minipage}{0.95in}      
            \includegraphics[width=0.95in]{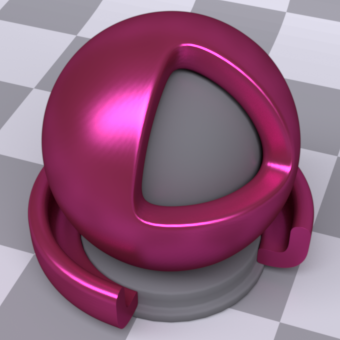}
             \put(-68.6,4.8) {\tikz[baseline] \node[fill=black, fill opacity=0.65, text opacity=1, text=white,inner sep=2pt] {\tiny 0.979/4.85};} 
             \put(-20,0){\includegraphics[width=20pt]{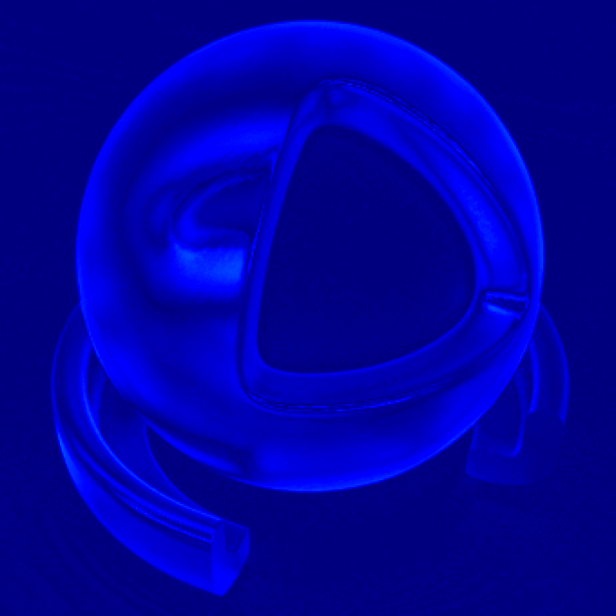}}
             \end{minipage}
             \begin{minipage}{0.95in}      
            \includegraphics[width=0.95in]{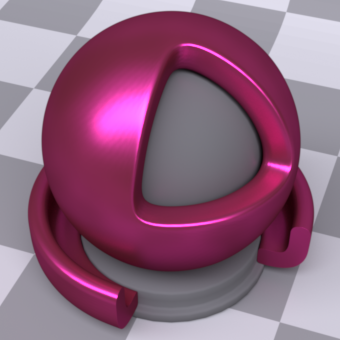}
             \put(-68.6,4.8) {\tikz[baseline] \node[fill=black, fill opacity=0.65, text opacity=1, text=white,inner sep=2pt] {\tiny 0.980/4.83};}
             \put(-20,0){\includegraphics[width=20pt]{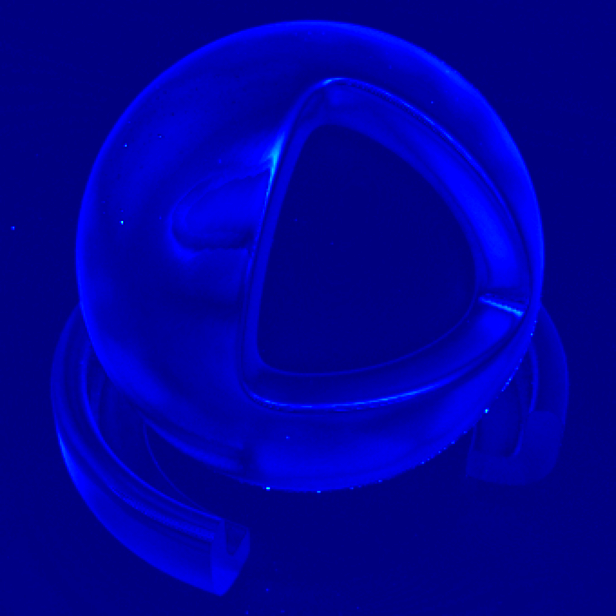}}
             \end{minipage}
             \begin{minipage}{0.95in}      
            \includegraphics[width=0.95in]{fig_new/analytical_gen/vch_ultra_pink_rgb.png}
             \put(-68.6,4.8) {\tikz[baseline] \node[fill=black, fill opacity=0.65, text opacity=1, text=white,inner sep=2pt] {\tiny 0.886/11.9};} 
             \put(-20,0){\includegraphics[width=20pt]{fig_new/analytical_gen_err/vch_ultra_pink_rgb.png}}
             \end{minipage}
        \end{minipage}
    \end{minipage}

    \begin{minipage}{\textwidth}
        \centering
        \hspace{-0.1in}
        \begin{minipage}[c]{0.95in}
            \centering
            \text{{}}
        \end{minipage}
        \begin{minipage}[c]{0.95in}
            \centering
            \text{ $\scriptstyle \mathcal{M}{+}\mathcal{S} \cdot \mathcal{D \hat{\cdot} \hat{\mathcal{F}}} \cdot \hat{\mathcal{G}} \mathcal{\hat{\cdot}} \hat{(\frac{1}{\mathcal{E}})}$}
        \end{minipage}
        \begin{minipage}[c]{0.95in}
            \centering
            \text{}
        \end{minipage}
        \begin{minipage}[c]{0.95in}
            \centering
            \text{ $\scriptstyle \mathcal{M}{+}\mathcal{S} \cdot \mathcal{D \cdot \hat{\mathcal{F}}} \cdot \mathcal{G} \hat{\cdot} \hat{(\frac{1}{\mathcal{E}})}$}
        \end{minipage}
        \begin{minipage}[c]{0.95in}
            \centering
            \text{}
        \end{minipage}
        \begin{minipage}[c]{0.95in}
            \centering
            \begin{tikzpicture}
                \node (txt) at (0,0) {\text{ $\scriptstyle \mathcal{M}{+}\hat{\mathcal{S}} \cdot \mathcal{D \hat{\cdot} \mathcal{F}} \cdot \hat{\mathcal{G}} \cdot \frac{1}{\mathcal{E}}$}};
            \end{tikzpicture}
        \end{minipage}
        \begin{minipage}[c]{0.95in}
            \centering
            \text{}
        \end{minipage}
    \end{minipage}

  \caption{Neural enhancement of other analytical BRDF models on measured BRDFs from EPFL dataset~\cite{Dupuy2018:Adaptive}. From the first column to last, the ground-truths, the results of enhanced/original Cook-Torrance model, enhanced/original Ward model, and enhanced/original FULL model in GenBRDF~\cite{Brady2014:Genbrdf}, respectively. \note{Quantitative errors in SSIM and $\Delta E_{ITP}$ ($\times 10^3$) are reported at the bottom-left of each related image}, and the error map is shown at the bottom-right. The enhanced model for each analytical one is provided at the bottom of the corresponding column.} 
    \label{fig:different_model}
\end{figure*}
\begin{figure*}[htbp]
    \centering
    \begin{minipage}{7.1in}
        \centering
         \begin{minipage}{0.95in}              
        \centering
        {\scriptsize Enhanced \\ \vspace{-0.05in} Cook-Torrance}
         \end{minipage}
         \begin{minipage}{0.95in}
        \centering
       {\scriptsize Cook-Torrance}
         \end{minipage}
         \begin{minipage}{0.95in}
        \centering
        {\scriptsize Enhanced Ward}
         \end{minipage}
         \begin{minipage}{0.95in}            
        \centering
        {\scriptsize Ward~\cite{Ward1992:Ward}}
         \end{minipage}
         \begin{minipage}{0.95in}  
        \centering
        {\scriptsize Enhanced GenBRDF}
         \end{minipage}
         \begin{minipage}{0.95in}     
        \centering
        {\scriptsize GenBRDF}
         \end{minipage}
         \begin{minipage}{0.25in}
         \phantom{111}
         \end{minipage}
    \end{minipage}
    
    \vspace{0.02in}
    
    \begin{minipage}{7.1in}
        \centering
         \begin{minipage}{0.95in}               
            \includegraphics[width=0.95in]{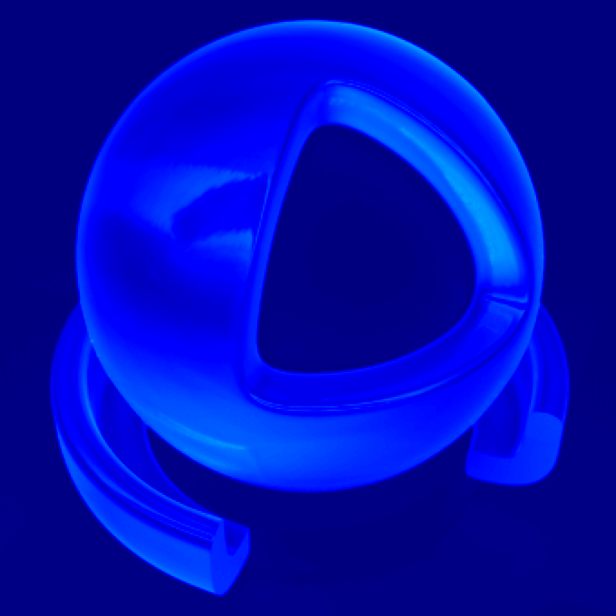}
         \end{minipage}
         \begin{minipage}{0.95in}      
            \includegraphics[width=0.95in]{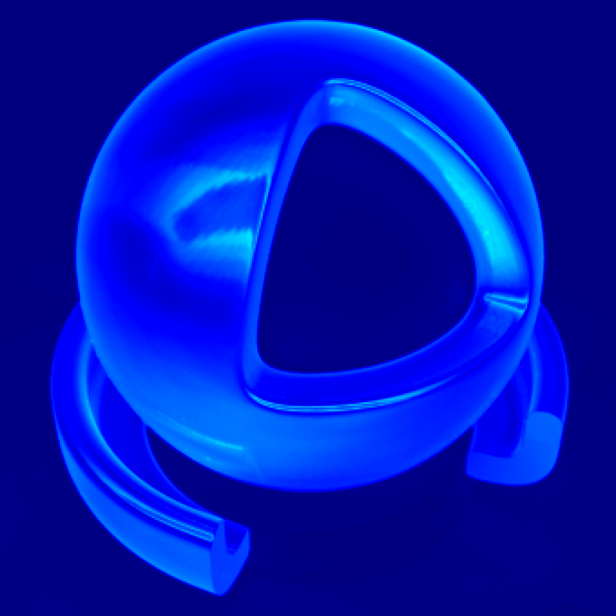}
         \end{minipage}
         \begin{minipage}{0.95in}      
            \includegraphics[width=0.95in]{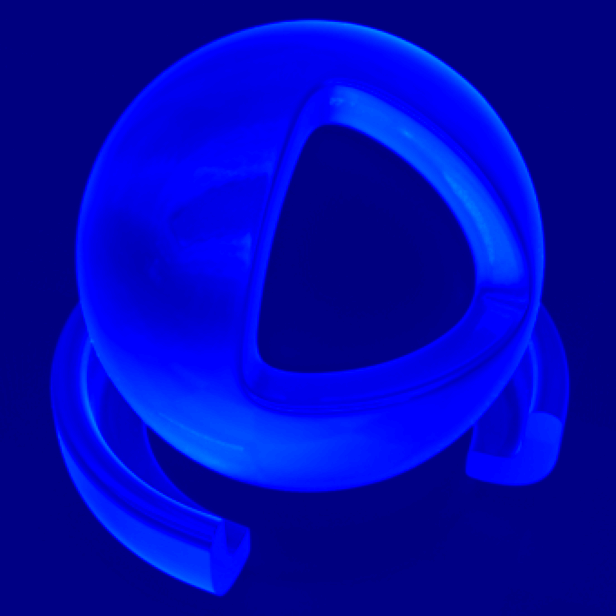}
         \end{minipage}
         \begin{minipage}{0.95in}      
            \includegraphics[width=0.95in]{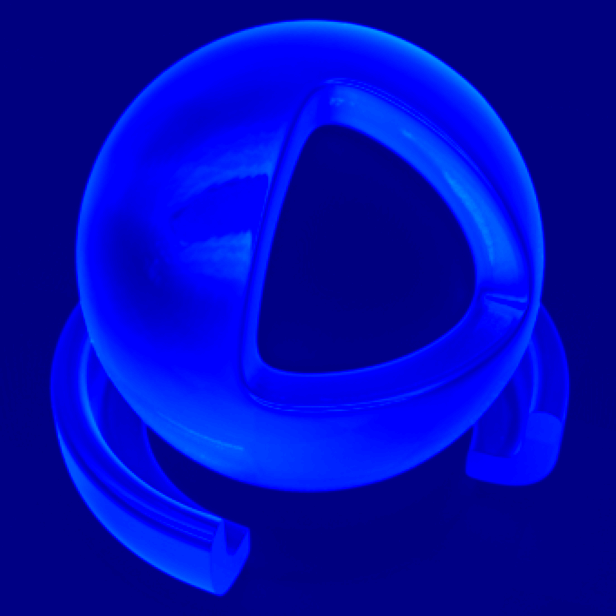}
         \end{minipage}
         \begin{minipage}{0.95in}      
            \includegraphics[width=0.95in]{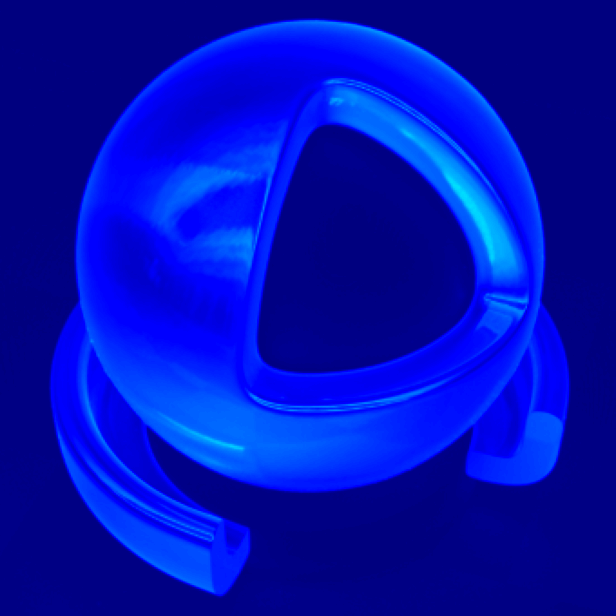}
         \end{minipage}
         \begin{minipage}{0.95in}      
            \includegraphics[width=0.95in]{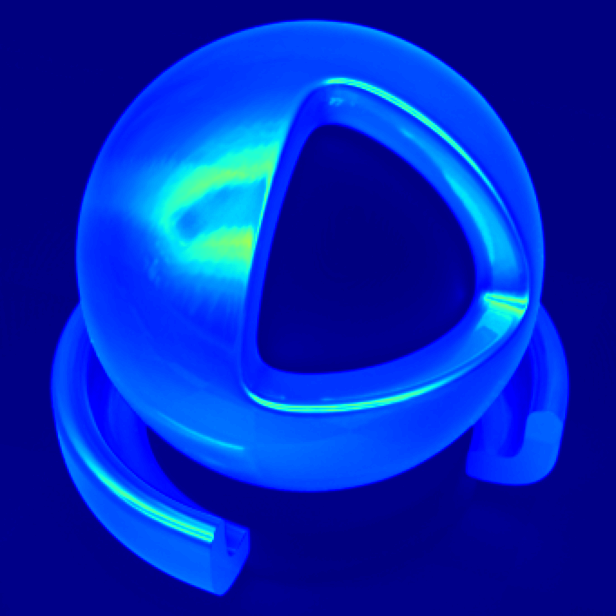}
         \end{minipage}
         \begin{minipage}{0.25in}      
            \includegraphics[width=0.95in,height=0.95in,keepaspectratio]{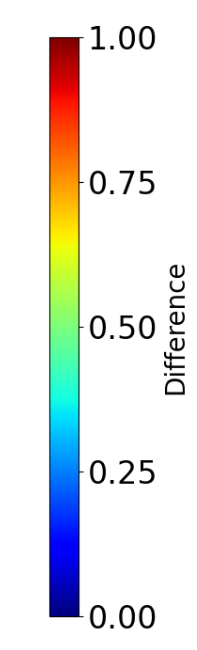}
         \end{minipage}
    \end{minipage}

        \begin{minipage}{\textwidth}
        \centering
        \begin{minipage}[c]{0.95in}
            \centering
            \text{ $\scriptstyle \mathcal{M}{+}\mathcal{S} \cdot \mathcal{D \hat{\cdot} \hat{\mathcal{F}}} \cdot \hat{\mathcal{G}} \mathcal{\hat{\cdot}} \hat{(\frac{1}{\mathcal{E}})}$}
        \end{minipage}
        \begin{minipage}[c]{0.95in}
            \centering
            \text{ $\scriptstyle \mathcal{M}{+}\mathcal{S} \cdot \mathcal{D \cdot \mathcal{F}} \cdot \mathcal{G} \cdot \frac{1}{\mathcal{E}}$}
        \end{minipage}
        \begin{minipage}[c]{0.95in}
            \centering
            \text{ $\scriptstyle \mathcal{M}{+}\mathcal{S} \cdot \mathcal{D \cdot \hat{\mathcal{F}}} \cdot \mathcal{G} \hat{\cdot} \hat{(\frac{1}{\mathcal{E}})}$}
        \end{minipage}
        \begin{minipage}[c]{0.95in}
            \centering
            \text{ $\scriptstyle \mathcal{M}{+}\mathcal{S} \cdot \mathcal{D \cdot \mathcal{F}} \cdot \mathcal{G} \cdot \frac{1}{\mathcal{E}}$}
        \end{minipage}
        \begin{minipage}[c]{0.95in}
            \centering
            \begin{tikzpicture}
                \node (txt) at (0,0) {\text{ $\scriptstyle \mathcal{M}{+}\hat{\mathcal{S}} \cdot \mathcal{D \hat{\cdot} \mathcal{F}} \cdot \hat{\mathcal{G}} \cdot \frac{1}{\mathcal{E}}$}};
            \end{tikzpicture}
        \end{minipage}
        \begin{minipage}[c]{0.95in}
            \centering
                \begin{tikzpicture}
                \node (txt) at (0,0) {\text{$\scriptstyle \mathcal{M}{+}\mathcal{S} \cdot \mathcal{D \cdot \mathcal{F}} \cdot \mathcal{G} \cdot \frac{1}{\mathcal{E}}$}};
            \end{tikzpicture}
        \end{minipage}
        \begin{minipage}[c]{0.25in}
            \centering
            \text{{}}
        \end{minipage}
    \end{minipage}

    \caption{Average error map for all BRDFs from the EPFL dataset~\cite{Dupuy2018:Adaptive} across three analytical BRDF models and their enhanced versions. From the first column to the last, the results correspond to the boosted/original Cook-Torrance model, the enhanced/original Ward model, and the enhanced/original FULL model in GenBRDF~\cite{Brady2014:Genbrdf}, respectively.}
    
\label{fig:ssim_map}
\end{figure*}

\textbf{Enhancing Other Models.} In theory, our framework can improve any analytical BRDF model. We show the increased quality of the enhanced versions of Cook-Torrance/Ward BRDF and one model proposed in GenBRDF~\cite{Brady2014:Genbrdf} in~\figref{fig:different_model}. The average SSIM for all BRDFs from the EPFL dataset~\cite{Dupuy2018:Adaptive} across three BRDF models and their enhanced versions is also shown in ~\figref{fig:ssim_map}. 

\begin{figure}[htbp!]
    \centering	
    \begin{minipage}{3.4in}
    \begin{minipage}{0.02in}	
            \centering
                \vspace{0.2in}
                \rotatebox{90}{\scriptsize \textsc{ }}
            \end{minipage}
        \begin{minipage}{3.4in}
            \centering
            
            \begin{minipage}{0.18\linewidth}
                \centering
                 \subcaption{\scriptsize Ground-Truth}
            \end{minipage}			
            \begin{minipage}{0.18\linewidth}
                \centering
                 \subcaption{\scriptsize Ours}
            \end{minipage}		
            \begin{minipage}{0.18\linewidth}
                \centering
                 \subcaption{\scriptsize Residual (1x)}
            \end{minipage}		
            \begin{minipage}{0.18\linewidth}
                \centering
                 \subcaption{\scriptsize Residual (5x)}
            \end{minipage}	
            \begin{minipage}{0.18\linewidth}
                \centering
                 \subcaption{\tiny All Neural Modules}
            \end{minipage}	
        \end{minipage}
    \end{minipage}
    
    \begin{minipage}{3.4in}
        \begin{minipage}{0.02in}	
            \centering
                \rotatebox{90}{{\fontsize{4pt}{5pt}\selectfont \uppercase{Millennium\_falcon}}}
            \end{minipage}	
             \begin{minipage}{3.4in}	
                \centering
            \begin{minipage}{0.18\linewidth}
                \centering
                \includegraphics[width=\linewidth]{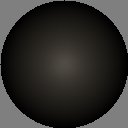}
            \end{minipage}
            \begin{minipage}{0.18\linewidth}
                \centering
                \includegraphics[width=\linewidth]{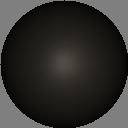}
                \put(-34.5,2.5) {\small \color{white} 0.0014}
            \end{minipage}
            \begin{minipage}{0.18\linewidth}
                \includegraphics[width=\linewidth]{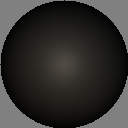}
                \put(-34.5,2.5) {\small \color{white} 0.0044}
            \end{minipage}
            \begin{minipage}{0.18\linewidth}
                \includegraphics[width=\linewidth]{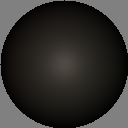}
                \put(-34.5,2.5) {\small \color{white} 0.0038}
            \end{minipage}
            \begin{minipage}{0.18\linewidth}
                \includegraphics[width=\linewidth]{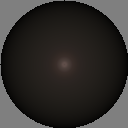}
                \put(-34.5,2.5) {\small \color{white} 0.0028}
            \end{minipage}
        \end{minipage}	
    \end{minipage}	

    \begin{minipage}{3.4in}
        \begin{minipage}{0.02in}	
            \centering
                \rotatebox{90}{\tiny \uppercase{Silk\_blue}}
        \end{minipage}	
         \begin{minipage}{3.4in}	
            \centering
            \begin{minipage}{0.18\linewidth}
            \includegraphics[width=\linewidth]{fig_new/gt_EPFL_top/vch_silk_blue_rgb.png}
            \end{minipage}	
            \begin{minipage}{0.18\linewidth}
            \includegraphics[width=\linewidth]{fig_new/boosted_ggx_top/vch_silk_blue_rgb.png}
            \put(-34.5,2.5) {\small \color{white} 0.0056}
            \end{minipage}	
            \begin{minipage}{0.18\linewidth}
            \includegraphics[width=\linewidth]{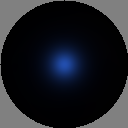}
            \put(-34.5,2.5) {\small \color{white} 0.0115}
            \end{minipage}	
            \begin{minipage}{0.18\linewidth}
            \includegraphics[width=\linewidth]{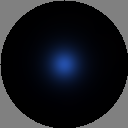}
            \put(-34.5,2.5) {\small \color{white} 0.0109}
            \end{minipage}
            \begin{minipage}{0.18\linewidth}
            \includegraphics[width=\linewidth]{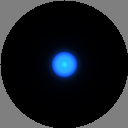}
            \put(-34.5,2.5) {\small \color{white} 0.0091}
            \end{minipage}
        \end{minipage}	
    \end{minipage}	

    \begin{minipage}{3.4in}
        \begin{minipage}{0.02in}	
            \centering
                \rotatebox{90}{\tiny \uppercase{Golden\_yellow}}
        \end{minipage}	
         \begin{minipage}{3.4in}	
            \centering
            \begin{minipage}{0.18\linewidth}
            \includegraphics[width=\linewidth]{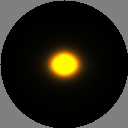}
            \end{minipage}
            \begin{minipage}{0.18\linewidth}
            \includegraphics[width=\linewidth]{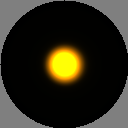}
            \put(-34.5,2.5) {\small \color{white} 0.0097}
            \end{minipage}
            \begin{minipage}{0.18\linewidth}
            \includegraphics[width=\linewidth]{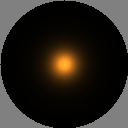}
            \put(-34.5,2.5) {\small \color{white} 0.0113}
            \end{minipage}
            \begin{minipage}{0.18\linewidth}
            \includegraphics[width=\linewidth]{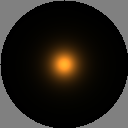}
            \put(-34.5,2.5) {\small \color{white} 0.0110}
            \end{minipage}
            \begin{minipage}{0.18\linewidth}
            \includegraphics[width=\linewidth]{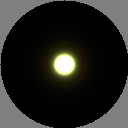}
            \put(-34.5,2.5) {\small \color{white} 0.0110}
            \end{minipage}
        \end{minipage}	
    \end{minipage}	
    
   \caption{\note{Comparisons with residual networks and all neural modules. From the left column to right, the ground-truths, results of our enhanced GGX, GGX model appended with a residual network roughly the size of our model (\{32,64,64,16\}), another residual network with about 5 times the size (\{32,64,128,128,64\}) and a model with all nodes/operators replaced by neural modules. Quantitative error is reported at the bottom of each related image, computed as the mean absolute value of the logarithmic loss.}} 
    \label{fig:compare_resnet}
\end{figure}

\textbf{Residual Networks.} A straightforward alternative to enhancing an analytical model is to append a residual neural network at its output. As shown in~\figref{fig:compare_resnet}, we compare our method against two such residual MLPs: one with a similar size as our enhanced model (architecture: \{32, 64, 64, 16\}) and another five times larger (\{32, 64, 128, 128, 64\}). Both residual networks take as input the concatenation of the lighting-view direction pair, analytical BRDF parameters, and neural BRDF parameters, similar to the first input configuration shown in~\figref{fig:module}. The analytical and neural BRDF parameters, along with the residual network, are jointly optimized to minimize the loss between the output and the ground truth. Despite the larger capacity, neither residual network matches the accuracy of our method. This demonstrates the value of our framework: the rear-end of an analytical model might not be the best place to plug in neural modules; our framework automatically explores and finds suitable locations for neural replacements.

\note{\textbf{All Neural Modules.}
We compare with a model where all nodes and operators are replaced with neural modules. In~\figref{fig:compare_resnet}, despite its larger network size, our enhanced model outperforms this all-neural-node model in all materials. Note that such a model also loses the semantics of parameters of the original analytic model.}



\begin{figure}[htbp]
    \centering
    \includegraphics[width = 0.9\linewidth]{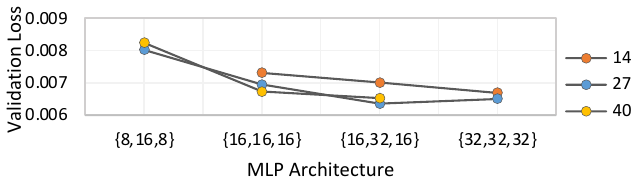}
    \caption{Validation losses of different variants of our model. The horizontal axis displays different neural module architectures, ordered in increasing size. We also validate different parameter numbers.} 
\label{fig:loss}
\end{figure}

\begin{figure}[hbt]
    \centering		
    \begin{minipage}{3.4in}
        \begin{minipage}{0.02in}
        \end{minipage}	
        \begin{minipage}{3.4in}
            \centering
            \begin{minipage}{0.23\linewidth} 
            \centering
             \subcaption{\tiny \shortstack {\uppercase{Chm} \\ \uppercase{light\_blue}}}
             \end{minipage}
             \begin{minipage}{0.23\linewidth}              
            \centering
            \subcaption{\tiny \shortstack {\uppercase{Ilm\_aniso} \\ \uppercase{tarkin\_tunic}}}
             \end{minipage}
             \begin{minipage}{0.23\linewidth}
            \centering
            \subcaption{\tiny \shortstack {\uppercase{Ilm\_darth} \\ \uppercase{vader\_pants}}}
             \end{minipage}
             \begin{minipage}{0.23\linewidth}
            \centering
            \subcaption{\tiny \shortstack {\uppercase{Vch} \\ \uppercase{ultra\_pink}}}
             \end{minipage}
        \end{minipage}
    \end{minipage}
    
    \vspace{0.01in}
    
    \begin{minipage}{3.4in}
        \begin{minipage}{0.02in}	
            \centering
            \rotatebox{90}{\scriptsize Ground-Truth}
        \end{minipage}	
        \begin{minipage}{3.4in}	
            \centering
            \begin{minipage}{0.23\linewidth}
                \centering
                \includegraphics[width=\linewidth]{fig_new/gt_EPFL/chm_light_blue_rgb.png}
                \put(-56,4.8) {\tikz[baseline] \node[fill=black, fill opacity=0.65, text opacity=1, text=white,inner sep=2pt] {\tiny SSIM/$\Delta E_{ITP}$};}
            \end{minipage}
            \begin{minipage}{0.23\linewidth}
                \centering
                \includegraphics[width=\linewidth]{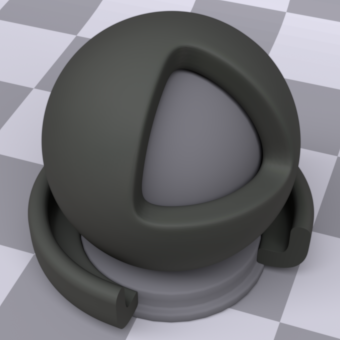}
                \put(-56,4.8) {\tikz[baseline] \node[fill=black, fill opacity=0.65, text opacity=1, text=white,inner sep=2pt] {\tiny SSIM/$\Delta E_{ITP}$};}
            \end{minipage}
            \begin{minipage}{0.23\linewidth}
                \centering
                \includegraphics[width=\linewidth]{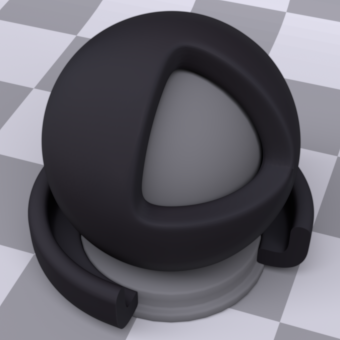}
                \put(-56,4.8) {\tikz[baseline] \node[fill=black, fill opacity=0.65, text opacity=1, text=white,inner sep=2pt] {\tiny SSIM/$\Delta E_{ITP}$};}
            \end{minipage}
            \begin{minipage}{0.23\linewidth}
                \centering
                \includegraphics[width=\linewidth]{fig_new/gt_EPFL/vch_ultra_pink_rgb.png}
                \put(-56,4.8) {\tikz[baseline] \node[fill=black, fill opacity=0.65, text opacity=1, text=white,inner sep=2pt] {\tiny SSIM/$\Delta E_{ITP}$};}
            \end{minipage}
        \end{minipage}
    \end{minipage}

    \begin{minipage}{3.4in}
        \begin{minipage}{0.02in}	
            \centering
            \rotatebox{90}{\scriptsize \{45,32,32,32\}}
        \end{minipage}	
        \begin{minipage}{3.4in}	
            \centering
            \begin{minipage}{0.23\linewidth}
                \includegraphics[width=\linewidth]{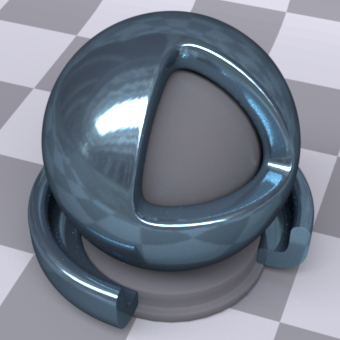}
                \put(-56,4.8) {\tikz[baseline] \node[fill=black, fill opacity=0.65, text opacity=1, text=white,inner sep=2pt] {\tiny 0.970/2.80};} 
                \put(-18,0){\includegraphics[width=18pt]{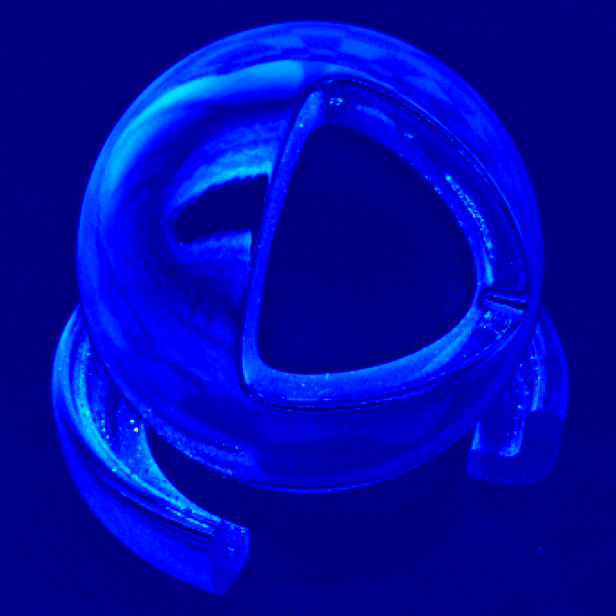}}
            \end{minipage}
            \begin{minipage}{0.23\linewidth}
                \includegraphics[width=\linewidth]{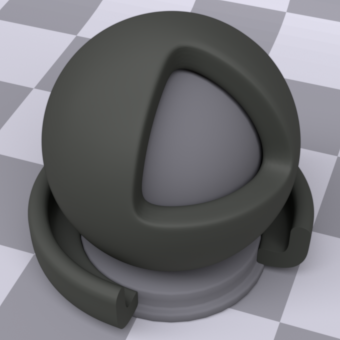}
                \put(-56,4.8) {\tikz[baseline] \node[fill=black, fill opacity=0.65, text opacity=1, text=white,inner sep=2pt] {\tiny 0.996/0.65};} 
                \put(-18,0){\includegraphics[width=18pt]{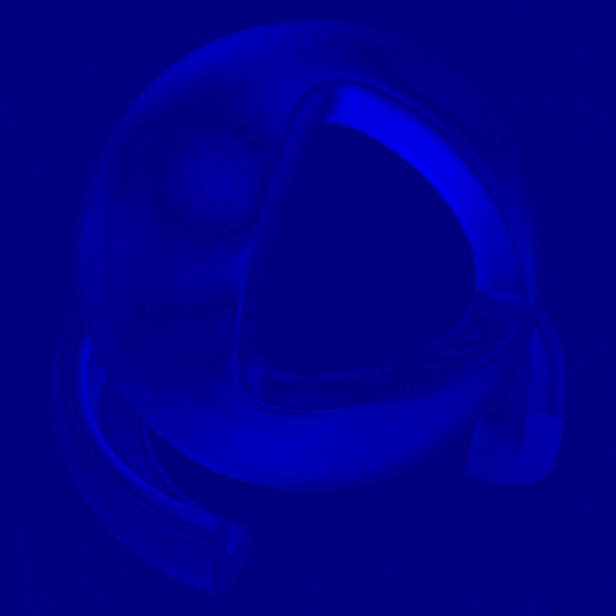}}
            \end{minipage}
            \begin{minipage}{0.23\linewidth}
                \includegraphics[width=\linewidth]{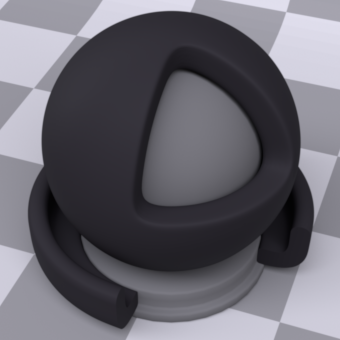}
                \put(-56,4.8) {\tikz[baseline] \node[fill=black, fill opacity=0.65, text opacity=1, text=white,inner sep=2pt] {\tiny 0.990/0.99};} 
                \put(-18,0){\includegraphics[width=18pt]{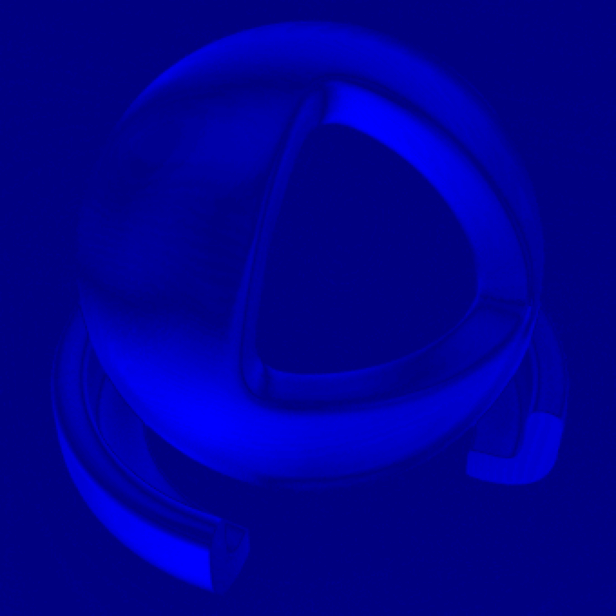}}
            \end{minipage}
            \begin{minipage}{0.23\linewidth}
                \includegraphics[width=\linewidth]{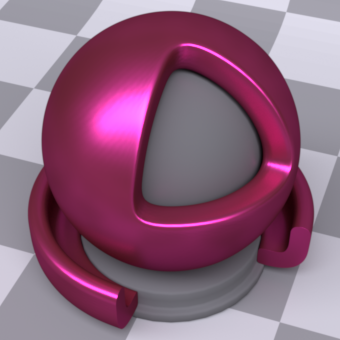}
                \put(-56,4.8) {\tikz[baseline] \node[fill=black, fill opacity=0.65, text opacity=1, text=white,inner sep=2pt] {\tiny 0.979/4.37};} 
                \put(-18,0){\includegraphics[width=18pt]{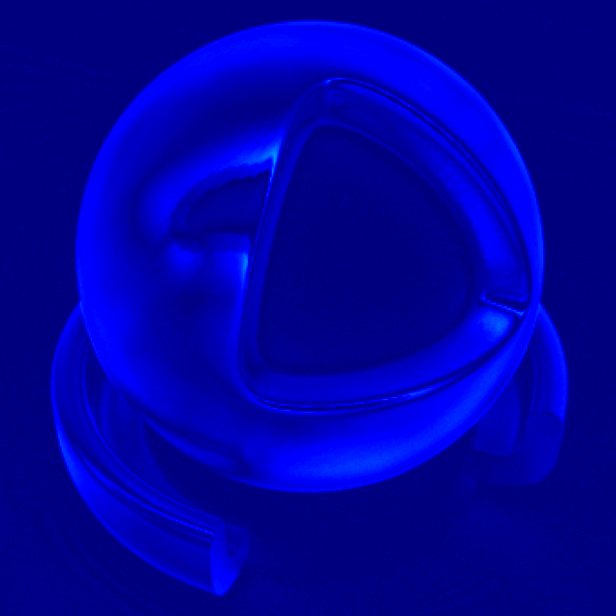}}
            \end{minipage}
        \end{minipage}
    \end{minipage}

    \begin{minipage}{3.4in}
        \begin{minipage}{0.02in}	
            \centering
            \rotatebox{90}{\scriptsize \{45,16,32,16\}}
        \end{minipage}	
        \begin{minipage}{3.4in}	
            \centering
            \begin{minipage}{0.23\linewidth}
                \includegraphics[width=\linewidth]{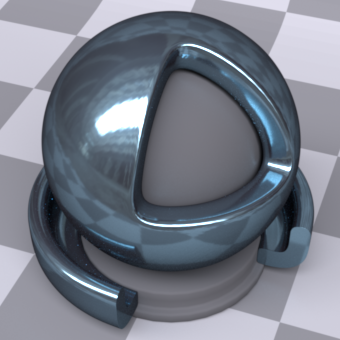}
                \put(-56,4.8) {\tikz[baseline] \node[fill=black, fill opacity=0.65, text opacity=1, text=white,inner sep=2pt] {\tiny 0.984/2.37};} 
                \put(-18,0){\includegraphics[width=18pt]{fig_new/boosted_ggx_err/chm_light_blue_rgb.png}}
            \end{minipage}
            \begin{minipage}{0.23\linewidth}
                \includegraphics[width=\linewidth]{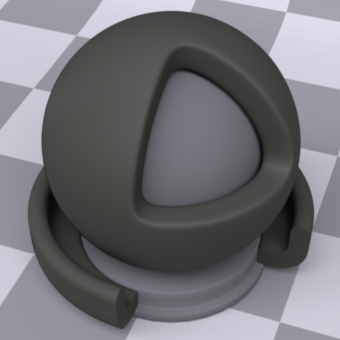}
                \put(-56,4.8) {\tikz[baseline] \node[fill=black, fill opacity=0.65, text opacity=1, text=white,inner sep=2pt] {\tiny 0.994/0.74};} 
                \put(-18,0){\includegraphics[width=18pt]{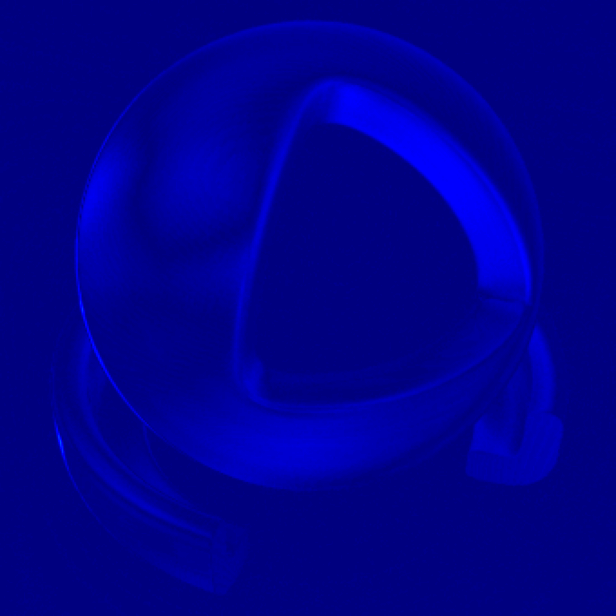}}
            \end{minipage}
            \begin{minipage}{0.23\linewidth}
                \includegraphics[width=\linewidth]{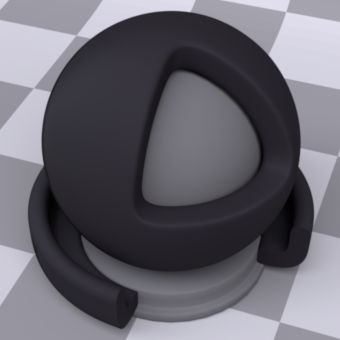}
                \put(-56,4.8) {\tikz[baseline] \node[fill=black, fill opacity=0.65, text opacity=1, text=white,inner sep=2pt] {\tiny 0.988/0.99};} 
                \put(-18,0){\includegraphics[width=18pt]{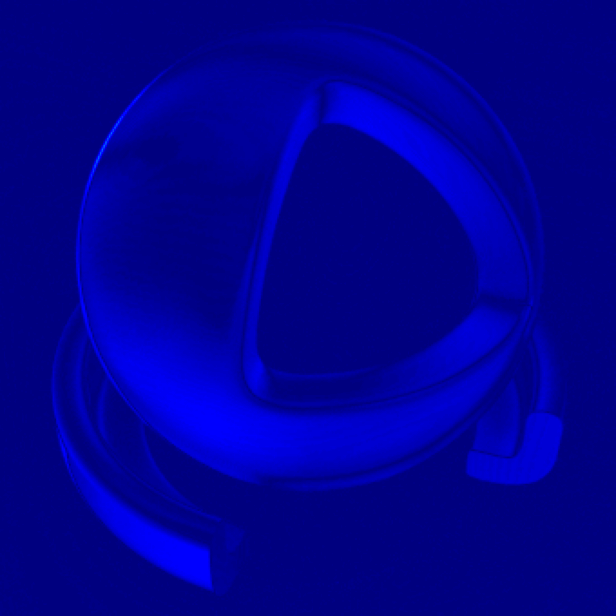}}
            \end{minipage}
            \begin{minipage}{0.23\linewidth}
                \includegraphics[width=\linewidth]{fig_new/boosted_ggx/vch_ultra_pink_rgb.png}
                \put(-56,4.8) {\tikz[baseline] \node[fill=black, fill opacity=0.65, text opacity=1, text=white,inner sep=2pt] {\tiny 0.980/4.96};} 
                \put(-18,0){\includegraphics[width=18pt]{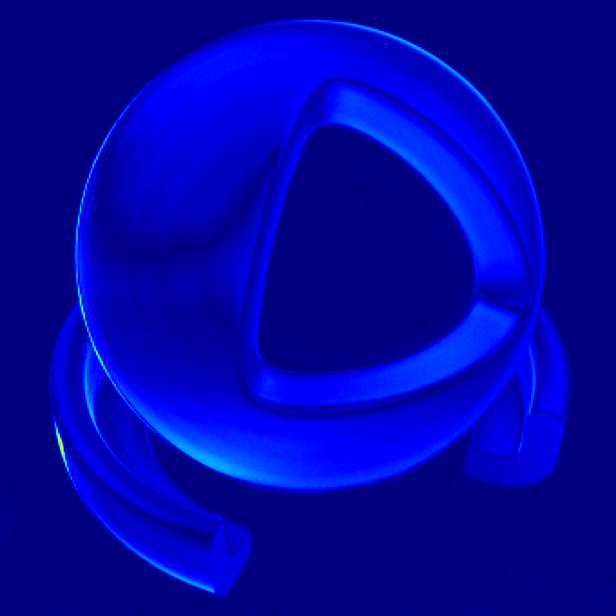}}
            \end{minipage}
        \end{minipage}
    \end{minipage}
    
    \begin{minipage}{3.4in}
        \begin{minipage}{0.02in}	
            \centering
            \rotatebox{90}{\scriptsize \{45,16,16,16\}}
        \end{minipage}	
        \begin{minipage}{3.4in}	
            \centering
            \begin{minipage}{0.23\linewidth}
                \includegraphics[width=\linewidth]{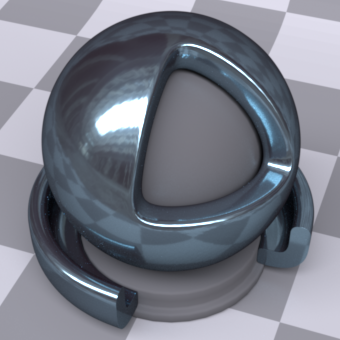}
                \put(-56,4.8) {\tikz[baseline] \node[fill=black, fill opacity=0.65, text opacity=1, text=white,inner sep=2pt] {\tiny 0.981/2.71};} 
                \put(-18,0){\includegraphics[width=18pt]{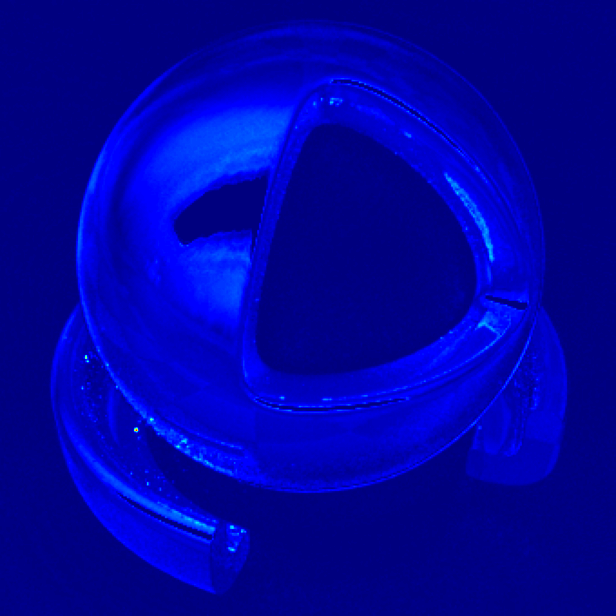}}
            \end{minipage}
            \begin{minipage}{0.23\linewidth}
                \includegraphics[width=\linewidth]{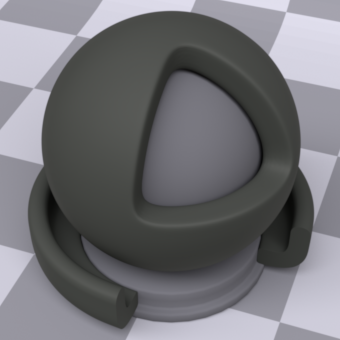}
                \put(-56,4.8) {\tikz[baseline] \node[fill=black, fill opacity=0.65, text opacity=1, text=white,inner sep=2pt] {\tiny 0.994/0.76};} 
                \put(-18,0){\includegraphics[width=18pt]{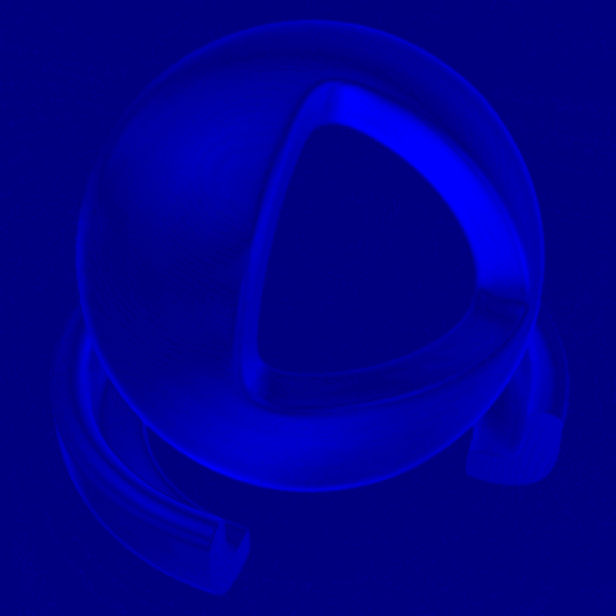}}
            \end{minipage}
            \begin{minipage}{0.23\linewidth}
                \includegraphics[width=\linewidth]{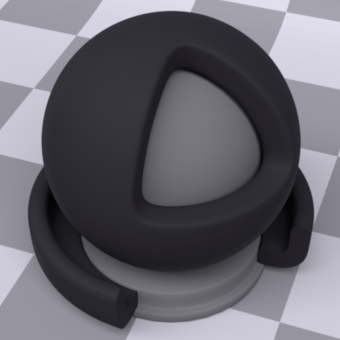}
                \put(-56,4.8) {\tikz[baseline] \node[fill=black, fill opacity=0.65, text opacity=1, text=white,inner sep=2pt] {\tiny 0.987/1.08};} 
                \put(-18,0){\includegraphics[width=18pt]{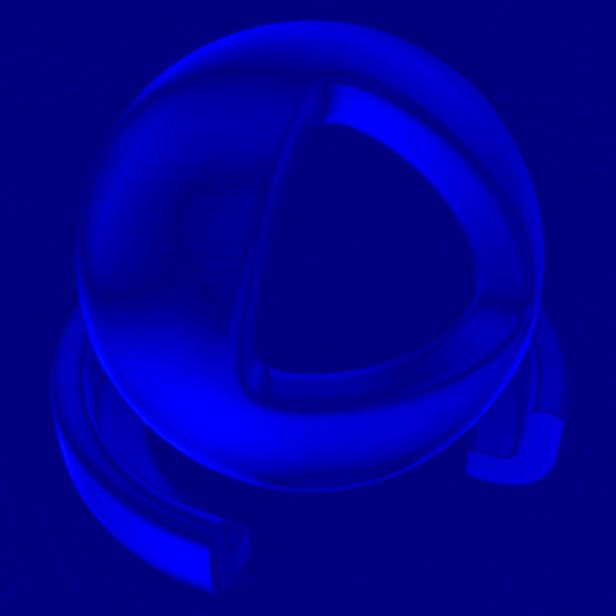}}
            \end{minipage}
            \begin{minipage}{0.23\linewidth}
                \includegraphics[width=\linewidth]{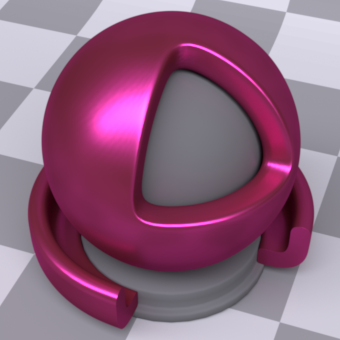}
                \put(-56,4.8) {\tikz[baseline] \node[fill=black, fill opacity=0.65, text opacity=1, text=white,inner sep=2pt] {\tiny 0.975/4.97};} 
                \put(-18,0){\includegraphics[width=18pt]{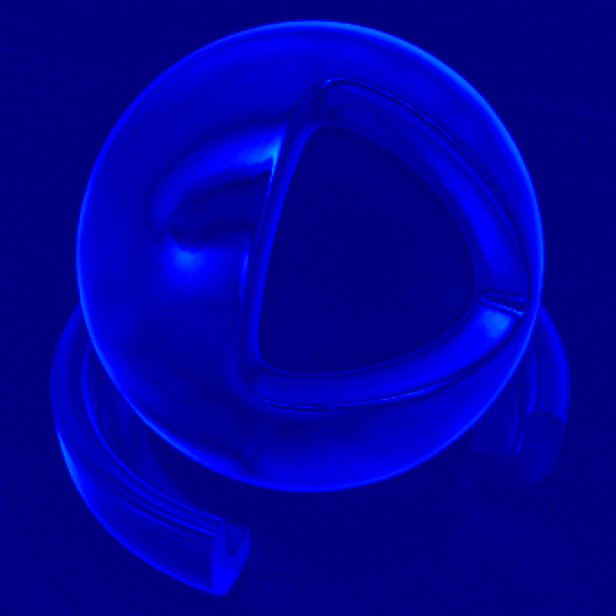}}
            \end{minipage}
        \end{minipage}
    \end{minipage}
    
    \begin{minipage}{3.4in}
        \begin{minipage}{0.02in}	
            \centering
            \rotatebox{90}{\scriptsize \{45,8,16,8\}}
        \end{minipage}	
        \begin{minipage}{3.4in}	
            \centering
            \begin{minipage}{0.23\linewidth}
                \includegraphics[width=\linewidth]{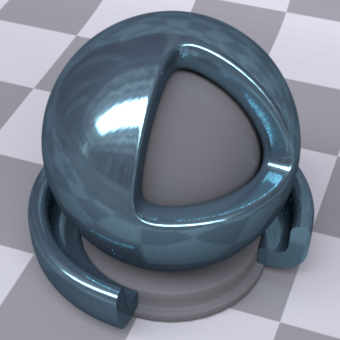}
                \put(-56,4.8) {\tikz[baseline] \node[fill=black, fill opacity=0.65, text opacity=1, text=white,inner sep=2pt] {\tiny 0.954/3.24};} 
                \put(-18,0){\includegraphics[width=18pt]{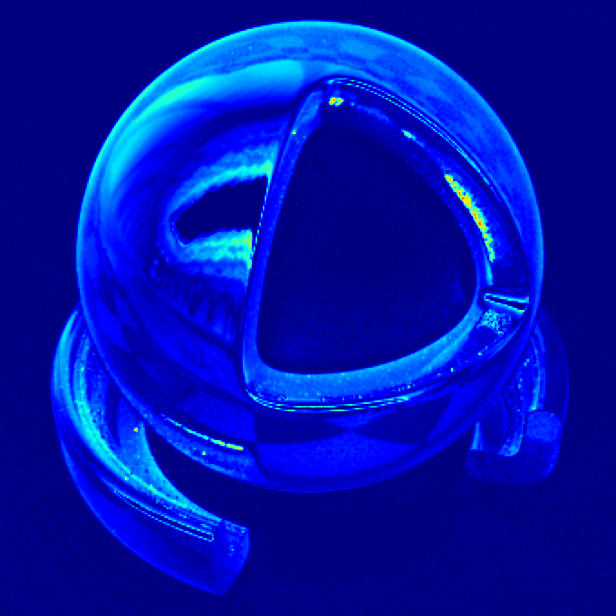}}
            \end{minipage}
            \begin{minipage}{0.23\linewidth}
                \includegraphics[width=\linewidth]{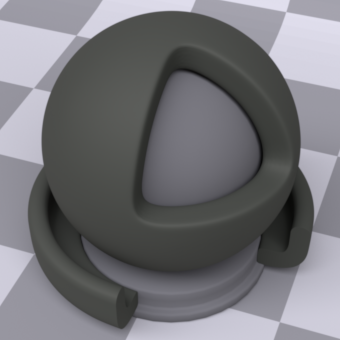}
                \put(-56,4.8) {\tikz[baseline] \node[fill=black, fill opacity=0.65, text opacity=1, text=white,inner sep=2pt] {\tiny 0.990/0.99};} 
                \put(-18,0){\includegraphics[width=18pt]{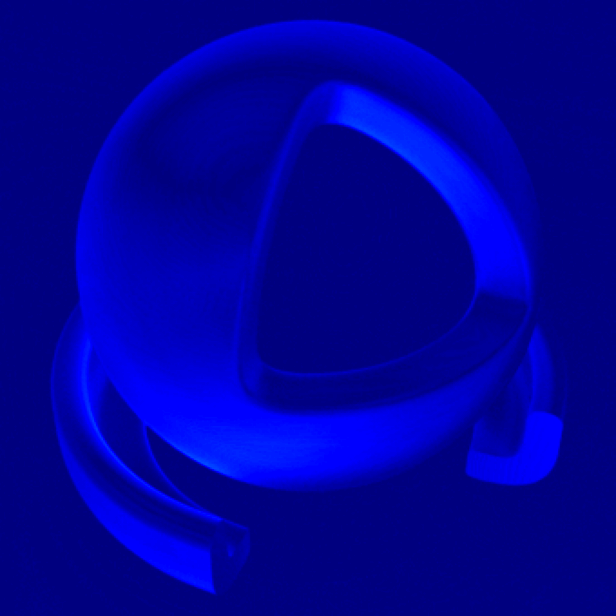}}
            \end{minipage}
            \begin{minipage}{0.23\linewidth}
                \includegraphics[width=\linewidth]{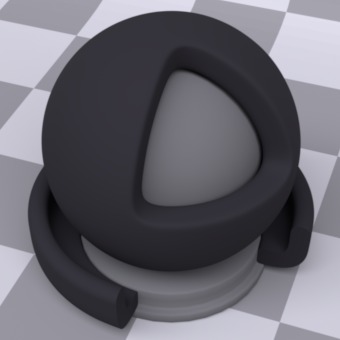}
                \put(-56,4.8) {\tikz[baseline] \node[fill=black, fill opacity=0.65, text opacity=1, text=white,inner sep=2pt] {\tiny 0.982/1.39};} 
                \put(-18,0){\includegraphics[width=18pt]{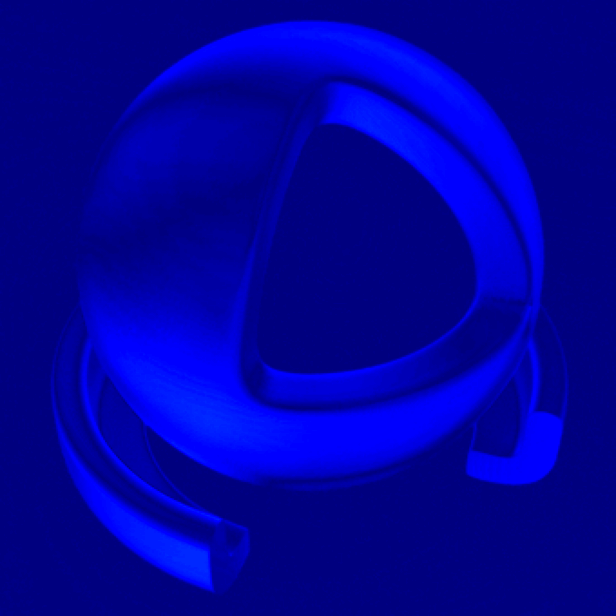}}
            \end{minipage}
            \begin{minipage}{0.23\linewidth}
                \includegraphics[width=\linewidth]{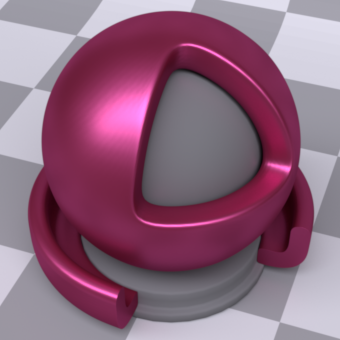}
                \put(-56,4.8) {\tikz[baseline] \node[fill=black, fill opacity=0.65, text opacity=1, text=white,inner sep=2pt] {\tiny 0.973/5.14};} 
                \put(-18,0){\includegraphics[width=18pt]{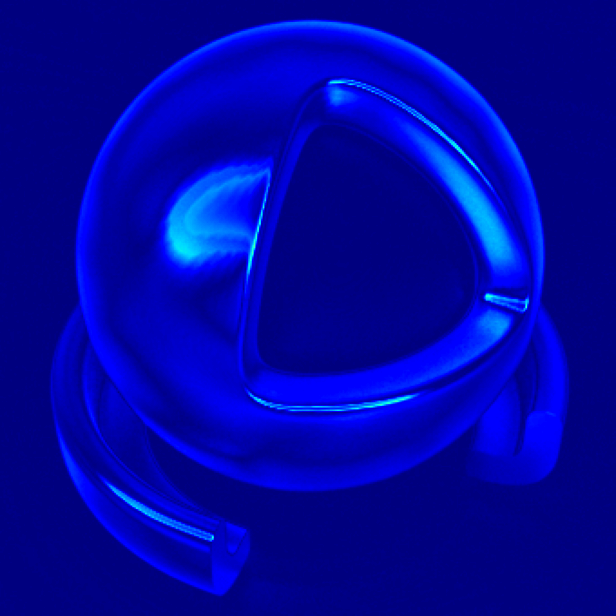}}
            \end{minipage}
        \end{minipage}
    \end{minipage}
    
    \caption{Impact of neural module size over reconstruction quality. From the top row to the bottom, the ground-truths, and reconstruction results from our enhanced GGX with decreasing neural module size. The architecture of each neural module is indicated on the left of each row. \note{Quantitative errors in SSIM and $\Delta E_{ITP}$ ($\times10^3$) are reported at the bottom-left of each related image}, and the error map at the bottom-right.}
\label{fig:network_size}
\end{figure}

\textbf{Neural Module Size.}
We evaluate the impact of neural module size over reconstruction quality both quantitatively (\figref{fig:loss}) and qualitatively (\figref{fig:network_size}). We test four 4-layer MLPs with different sizes/architectures: \{45, 8, 16, 8\}, \{45, 16, 16, 16\}, \{45, 16, 32, 16\}, and \{45, 32, 32, 32\}. Our current choice strikes a good balance between fitting accuracy and network size.

\begin{figure}[htbp]
    \centering	
    \begin{minipage}{3.4in}
    \begin{minipage}{0.02in}	
            \centering
                \vspace{0.2in}
                \rotatebox{90}{\scriptsize \textsc{ }}
            \end{minipage}
        \begin{minipage}{3.4in}
            \centering
            \begin{minipage}{0.23\linewidth}
                \centering
                \subcaption{\scriptsize Ground-Truth}
            \end{minipage}			
            \begin{minipage}{0.23\linewidth}
                \centering
               \subcaption{\scriptsize 40}
            \end{minipage}		
            \begin{minipage}{0.23\linewidth}
                \centering
                \subcaption{\scriptsize 27}
            \end{minipage}		
            \begin{minipage}{0.23\linewidth}
                \centering
               \subcaption{\scriptsize 14}
            \end{minipage}	
        \end{minipage}
    \end{minipage}

    \begin{minipage}{3.4in}
        \begin{minipage}{0.02in}	
            \centering
            \rotatebox{90}{\tiny \uppercase{Brushed\_al}}
        \end{minipage}	
        \begin{minipage}{3.4in}	
            \centering
            \begin{minipage}{0.23\linewidth}
                \includegraphics[width=\linewidth]{fig_new/gt_EPFL/aniso_brushed_aluminium_1_rgb.png}
                \put(-56,4.8) {\tikz[baseline] \node[fill=black, fill opacity=0.65, text opacity=1, text=white,inner sep=2pt] {\tiny SSIM/$\Delta E_{ITP}$};}
            \end{minipage}
            \begin{minipage}{0.23\linewidth}
                \centering
                \includegraphics[width=\linewidth]{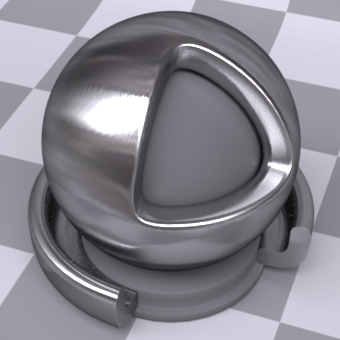}
                \put(-56,4.8) {\tikz[baseline] \node[fill=black, fill opacity=0.65, text opacity=1, text=white,inner sep=2pt] {\tiny 0.981/0.85};} 
                \put(-18,0){\includegraphics[width=18pt]{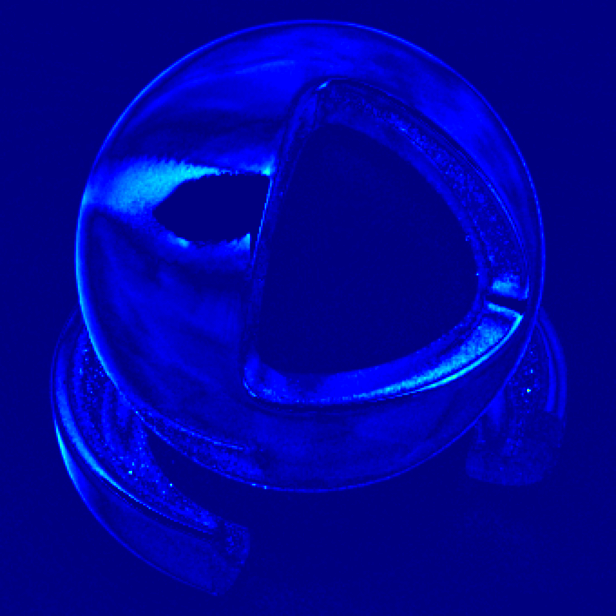}}
            \end{minipage}
            \begin{minipage}{0.23\linewidth}
                \centering
                \includegraphics[width=\linewidth]{fig_new/boosted_ggx/aniso_brushed_aluminium_1_rgb.png}
                \put(-56,4.8) {\tikz[baseline] \node[fill=black, fill opacity=0.65, text opacity=1, text=white,inner sep=2pt] {\tiny 0.977/0.96};} 
                \put(-18,0){\includegraphics[width=18pt]{fig_new/boosted_ggx_err/aniso_brushed_aluminium_1_rgb.png}}
            \end{minipage}
            \begin{minipage}{0.23\linewidth}
                \centering
                \includegraphics[width=\linewidth]{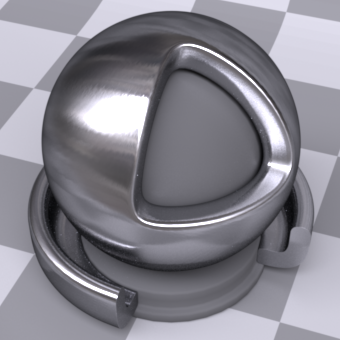}
                \put(-56,4.8) {\tikz[baseline] \node[fill=black, fill opacity=0.65, text opacity=1, text=white,inner sep=2pt] {\tiny 0.974/1.37};} 
                \put(-18,0){\includegraphics[width=18pt]{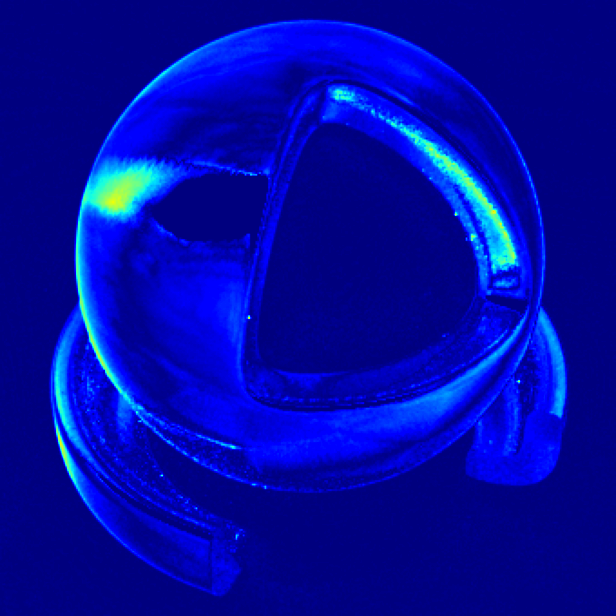}}
            \end{minipage}
        \end{minipage}
    \end{minipage}
    
    \begin{minipage}{3.4in}
        \begin{minipage}{0.02in}	
            \centering
            \rotatebox{90}{\tiny \uppercase{Felt\_white}}
        \end{minipage}	
        \begin{minipage}{3.4in}	
            \centering
            \begin{minipage}{0.23\linewidth}
                \includegraphics[width=\linewidth]{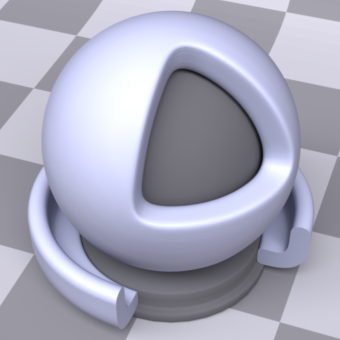}
                \put(-56,4.8) {\tikz[baseline] \node[fill=black, fill opacity=0.65, text opacity=1, text=white,inner sep=2pt] {\tiny SSIM/$\Delta E_{ITP}$};}
            \end{minipage}
            \begin{minipage}{0.23\linewidth}
                \centering
                \includegraphics[width=\linewidth]{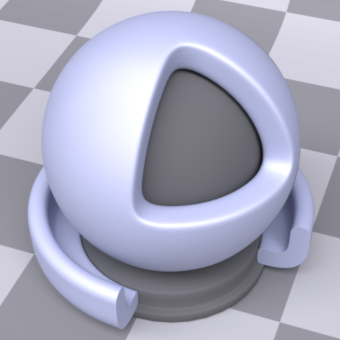}
                \put(-56,4.8) {\tikz[baseline] \node[fill=black, fill opacity=0.65, text opacity=1, text=white,inner sep=2pt] {\tiny 0.974/1.56};} 
                \put(-18,0){\includegraphics[width=18pt]{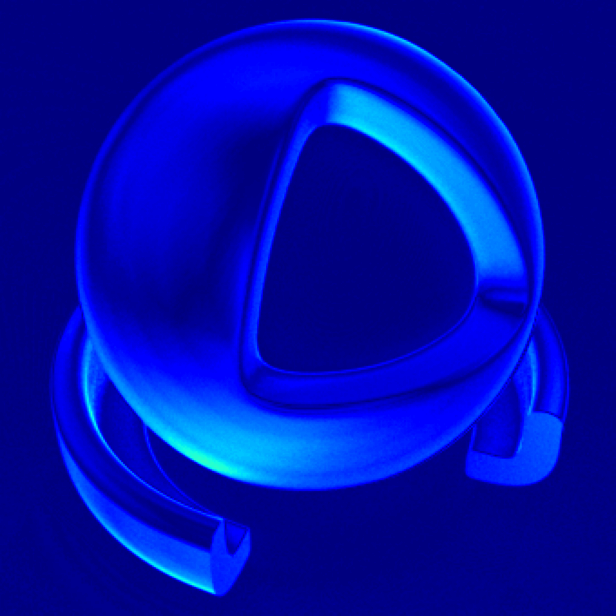}}
            \end{minipage}
            \begin{minipage}{0.23\linewidth}
                \centering
                \includegraphics[width=\linewidth]{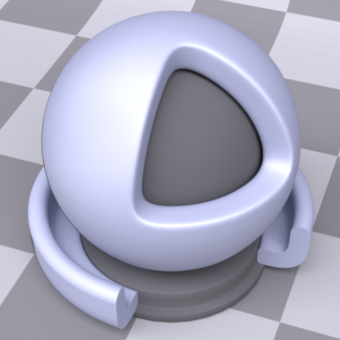}
                \put(-56,4.8) {\tikz[baseline] \node[fill=black, fill opacity=0.65, text opacity=1, text=white,inner sep=2pt] {\tiny 0.976/1.33};} 
                \put(-18,0){\includegraphics[width=18pt]{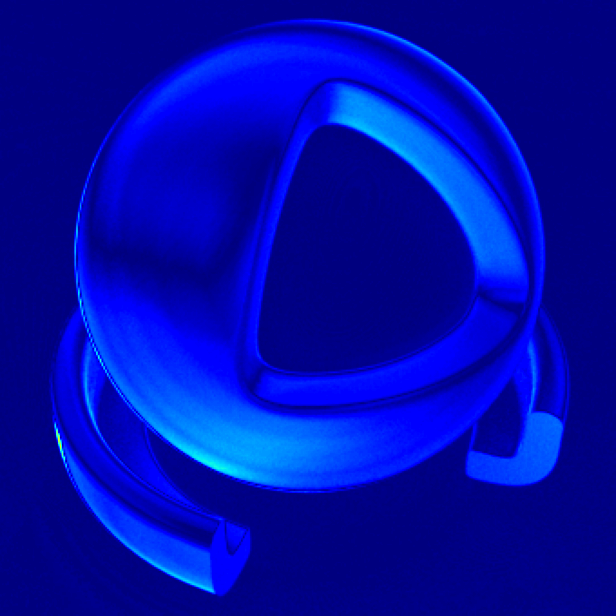}}
            \end{minipage}
            \begin{minipage}{0.23\linewidth}
                \centering
                \includegraphics[width=\linewidth]{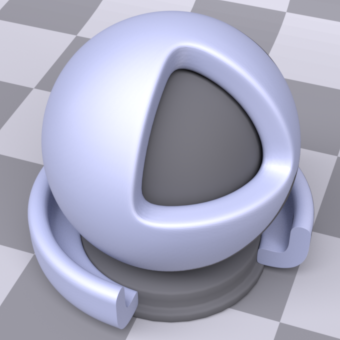}
                \put(-56,4.8) {\tikz[baseline] \node[fill=black, fill opacity=0.65, text opacity=1, text=white,inner sep=2pt] {\tiny 0.968/1.55};} 
                \put(-18,0){\includegraphics[width=18pt]{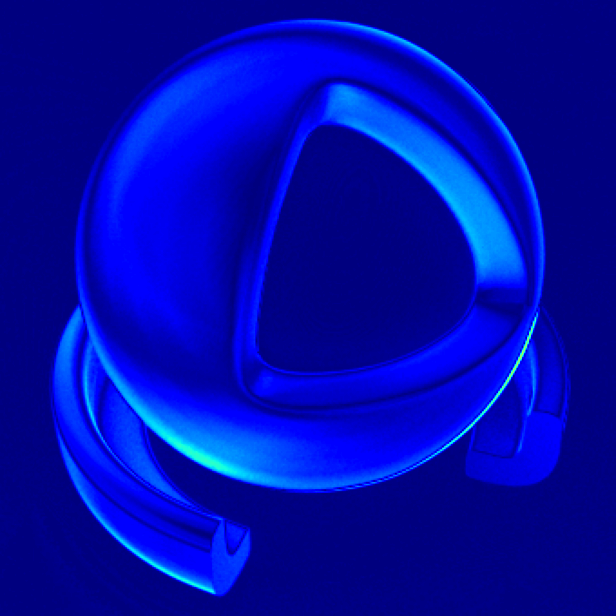}}
            \end{minipage}
        \end{minipage}
    \end{minipage}

    \begin{minipage}{3.4in}
        \begin{minipage}{0.02in}	
            \centering
            \rotatebox{90}{\tiny \uppercase{Silk\_blue}}
        \end{minipage}	
        \begin{minipage}{3.4in}	
            \centering
            \begin{minipage}{0.23\linewidth}
                \includegraphics[width=\linewidth]{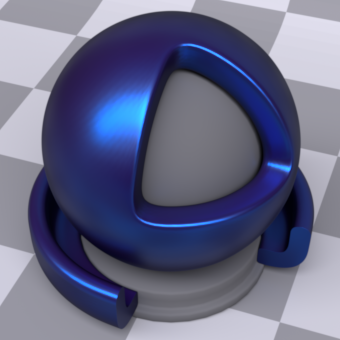}
                \put(-56,4.8) {\tikz[baseline] \node[fill=black, fill opacity=0.65, text opacity=1, text=white,inner sep=2pt] {\tiny SSIM/$\Delta E_{ITP}$};}
            \end{minipage}
            \begin{minipage}{0.23\linewidth}
                \includegraphics[width=\linewidth]{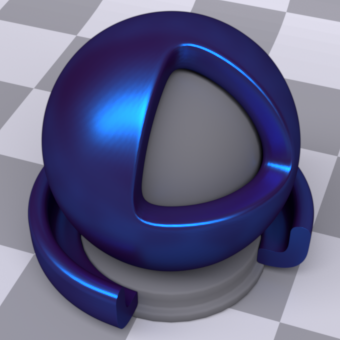}
                \put(-56,4.8) {\tikz[baseline] \node[fill=black, fill opacity=0.65, text opacity=1, text=white,inner sep=2pt] {\tiny 0.969/6.24};} 
                \put(-18,0){\includegraphics[width=18pt]{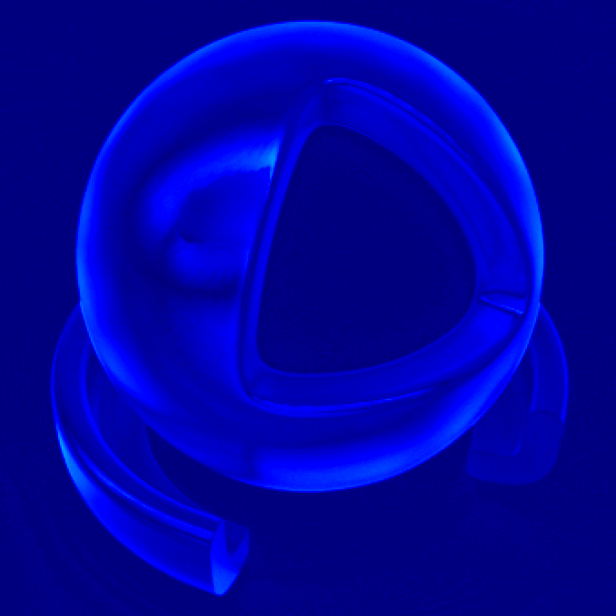}}
            \end{minipage}
            \begin{minipage}{0.23\linewidth}
                \includegraphics[width=\linewidth]{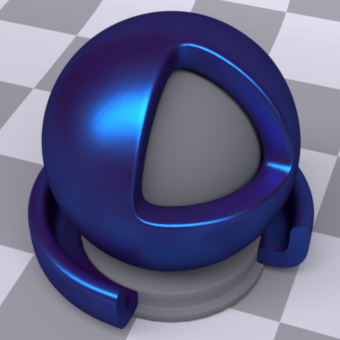}
                \put(-56,4.8) {\tikz[baseline] \node[fill=black, fill opacity=0.65, text opacity=1, text=white,inner sep=2pt] {\tiny 0.970/5.85};} 
                \put(-18,0){\includegraphics[width=18pt]{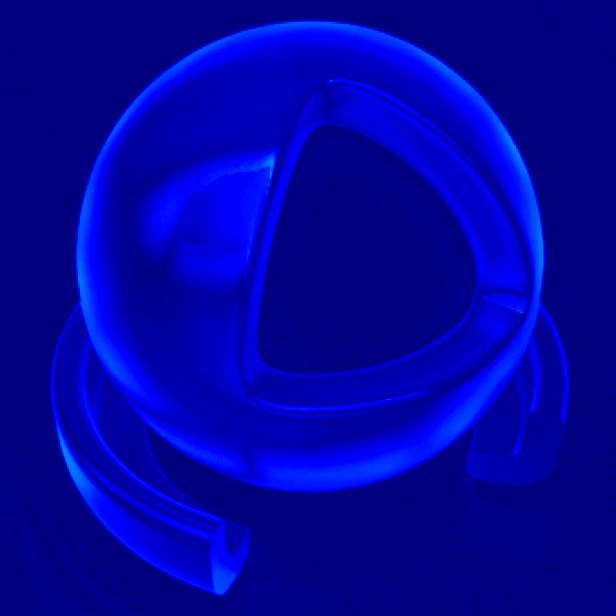}}
            \end{minipage}
            \begin{minipage}{0.23\linewidth}
                \includegraphics[width=\linewidth]{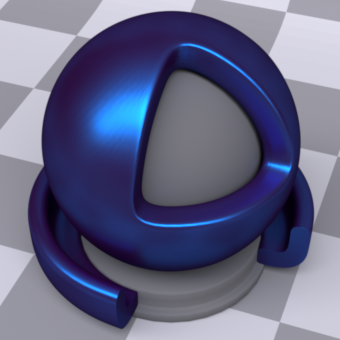}
                \put(-56,4.8) {\tikz[baseline] \node[fill=black, fill opacity=0.65, text opacity=1, text=white,inner sep=2pt] {\tiny 0.965/6.32};} 
                \put(-18,0){\includegraphics[width=18pt]{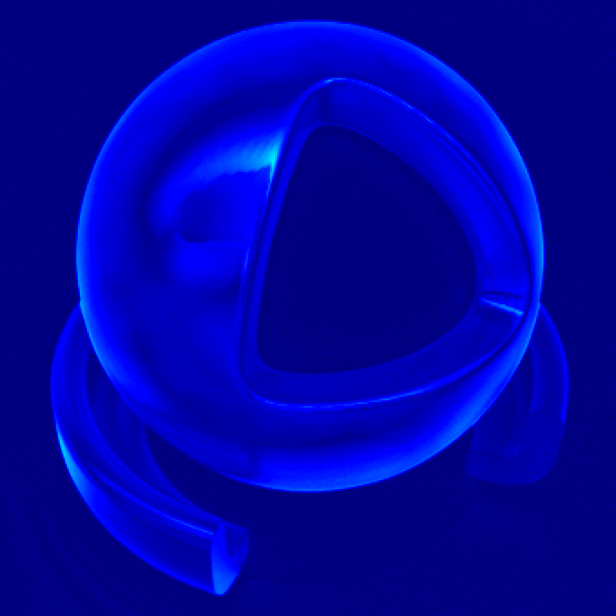}}
            \end{minipage}
        \end{minipage}
    \end{minipage}
    
    \caption{Impact of neural parameter number over reconstruction quality. The first column are ground-truths, and the results of our enhanced GGX with different neural parameter numbers. \note{Quantitative errors in SSIM and $\Delta E_{ITP}$ ($\times10^3$) are reported at the bottom-left of each related image}, and the error map is shown at the bottom-right.}
\label{fig:parameters_dim}
\end{figure}

\textbf{Neural Parameter Number.}
We further investigate the impact of the number of neural parameters over reconstruction quality, qualitatively (\figref{fig:parameters_dim}) and quantitatively (\figref{fig:loss}). We test 3 different neural BRDF parameters. Our current choice is made after balancing reconstruction quality and the footprint of neural parameters.

\begin{figure}[htbp!]

    \centering
    \begin{minipage}{2.8in}
        \begin{minipage}{0.02in}	
            \centering
                \vspace{0.2in}
                \rotatebox{90}{\scriptsize \textsc{ }}
        \end{minipage}
        \begin{minipage}{2.8in}
            \centering
            \begin{minipage}{0.31\linewidth}
                \centering
                \subcaption{\scriptsize Ground-Truth}
            \end{minipage}			
            \begin{minipage}{0.31\linewidth}
                \centering
               \subcaption{\tiny operator+nodes enhanced}
            \end{minipage}		
            \begin{minipage}{0.31\linewidth}
                \centering
                \subcaption{\tiny nodes enhanced only}
            \end{minipage}		
        \end{minipage}
    \end{minipage}

    \begin{minipage}{2.6in}
        \begin{minipage}{0.02in}	
            \centering
            \rotatebox{90}{\tiny \uppercase{Satin\_white}}
        \end{minipage}	
        \begin{minipage}{2.6in}	
            \centering
            \begin{minipage}{0.31\linewidth}
                \includegraphics[width=\linewidth]{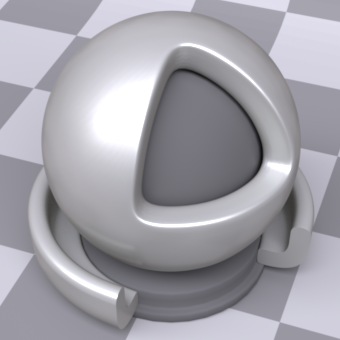}
                \put(-58,4.8) {\tikz[baseline] \node[fill=black, fill opacity=0.65, text opacity=1, text=white,inner sep=2pt] {\tiny SSIM/$\Delta E_{ITP}$};}
            \end{minipage}
            \begin{minipage}{0.31\linewidth}
                \centering
                \includegraphics[width=\linewidth]{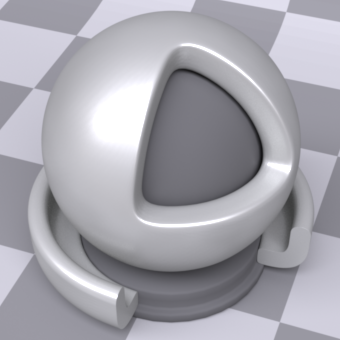}
                \put(-58,4.8) {\tikz[baseline] \node[fill=black, fill opacity=0.65, text opacity=1, text=white,inner sep=2pt] {\tiny 0.983/0.91};} 
                \put(-18,0){\includegraphics[width=18pt]{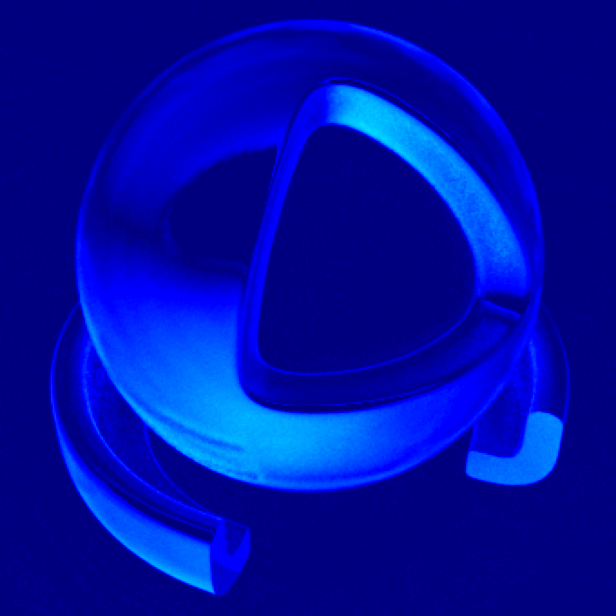}}
            \end{minipage}
            \begin{minipage}{0.31\linewidth}
                \centering
                \includegraphics[width=\linewidth]{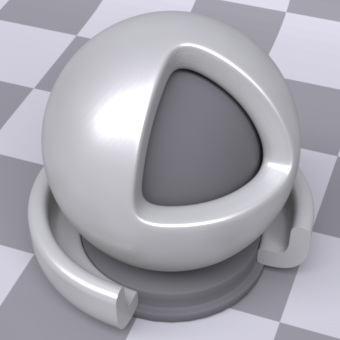}
                \put(-58,4.8) {\tikz[baseline] \node[fill=black, fill opacity=0.65, text opacity=1, text=white,inner sep=2pt] {\tiny 0.969/1.06};} 
                \put(-18,0){\includegraphics[width=18pt]{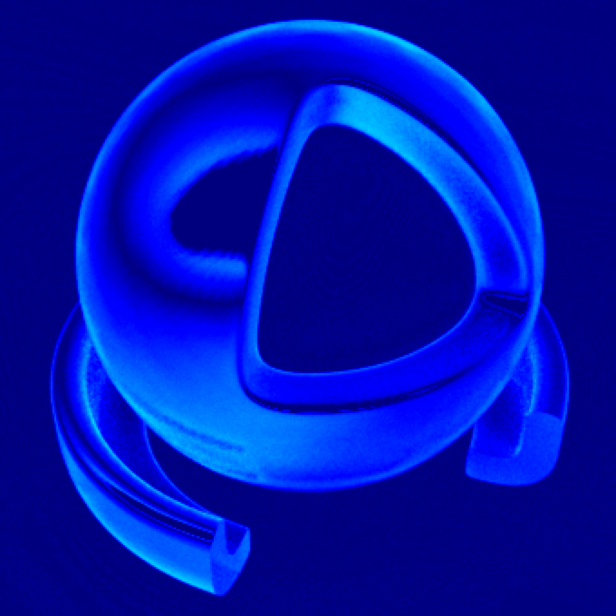}}
            \end{minipage}
        \end{minipage}
    \end{minipage}

    \begin{minipage}{2.6in}
        \begin{minipage}{0.02in}	
            \centering
            \rotatebox{90}{\tiny \uppercase{Golden\_yellow}}
        \end{minipage}	
        \begin{minipage}{2.6in}	
            \centering
            \begin{minipage}{0.31\linewidth}
                \includegraphics[width=\linewidth]{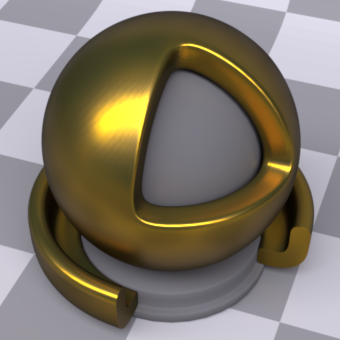}
                \put(-58,4.8) {\tikz[baseline] \node[fill=black, fill opacity=0.65, text opacity=1, text=white,inner sep=2pt] {\tiny SSIM/$\Delta E_{ITP}$};}
            \end{minipage}
            \begin{minipage}{0.31\linewidth}
                \includegraphics[width=\linewidth]{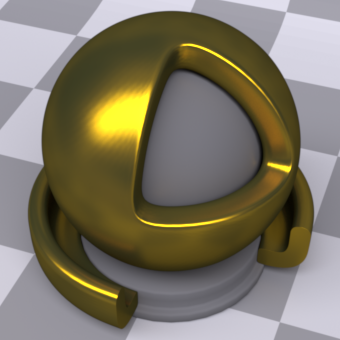}
                \put(-58,4.8) {\tikz[baseline] \node[fill=black, fill opacity=0.65, text opacity=1, text=white,inner sep=2pt] {\tiny 0.973/6.36};} 
                \put(-18,0){\includegraphics[width=18pt]{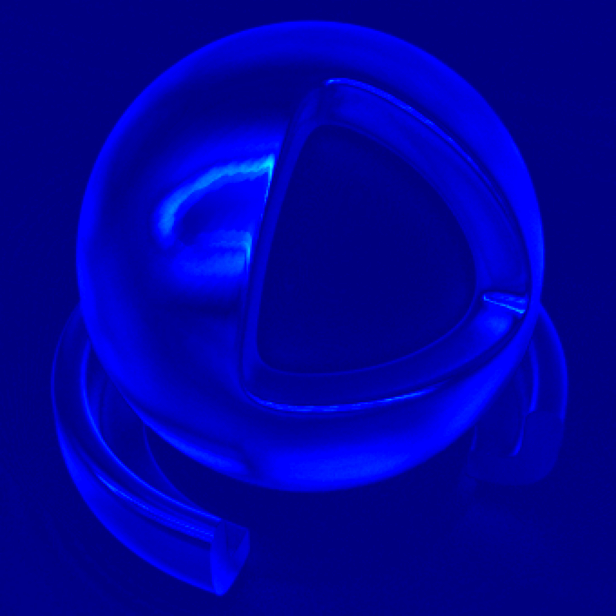}}
            \end{minipage}
            \begin{minipage}{0.31\linewidth}
                \includegraphics[width=\linewidth]{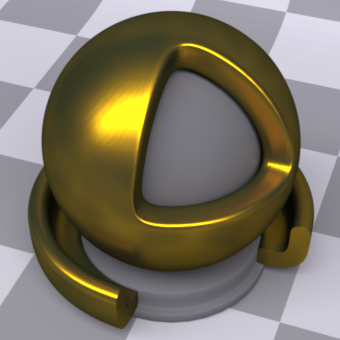}
                \put(-58,4.8) {\tikz[baseline] \node[fill=black, fill opacity=0.65, text opacity=1, text=white,inner sep=2pt] {\tiny 0.968/6.48};} 
                \put(-18,0){\includegraphics[width=18pt]{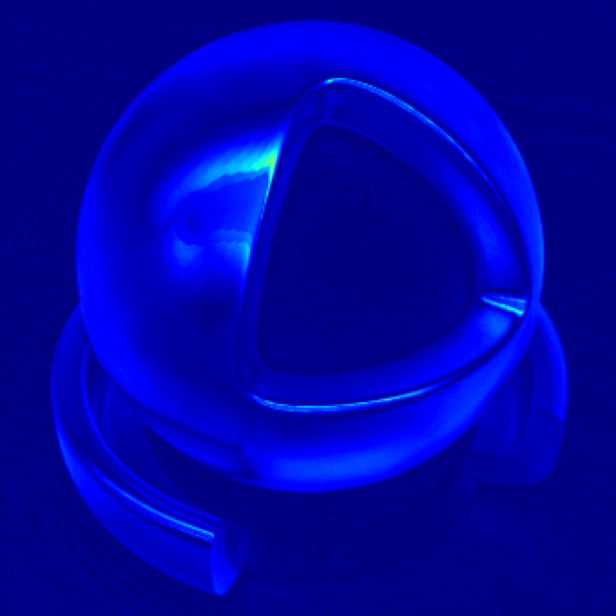}}
            \end{minipage}
        \end{minipage}
    \end{minipage}

    \begin{minipage}{2.6in}
        \centering
        \hspace{-0.1in}
        \begin{minipage}{0.02in}	
            \centering
            \rotatebox{90}{}
        \end{minipage}	
        \begin{minipage}{2.6in}	
            \centering
            \begin{minipage}{0.31\linewidth}
                \centering
                \text{}
            \end{minipage}
            \begin{minipage}{0.31\linewidth}
                \centering
                \text{ $\scriptstyle \mathcal{M}{+}\mathcal{S} \cdot \mathcal{D {\cdot} \hat{\mathcal{F}}} \hat{\cdot} \hat{\mathcal{G}} \mathcal{{\cdot}} \hat{(\frac{1}{\mathcal{E}})}$}
            \end{minipage}
            \begin{minipage}{0.31\linewidth}
                \centering
                \text{ $\scriptstyle \mathcal{M}{+}\mathcal{S} \cdot \mathcal{D {\cdot} {\mathcal{F}}} \cdot \hat{\mathcal{G}} \mathcal{{\cdot}} \hat{(\frac{1}{\mathcal{E}})}$}
            \end{minipage}
        \end{minipage}
    \end{minipage}
    
    \caption{\note{Impact of module granularity over reconstruction. From the first column to last, the ground-truths, GGX model with nodes and operators enhanced, GGX model with nodes enhanced only. Quantitative errors in SSIM and $\Delta E_{ITP}$ ($\times10^3$) are reported at the bottom-left of each related image, and the error map is shown at the bottom-right.}}
\label{fig:granularity}
\end{figure}
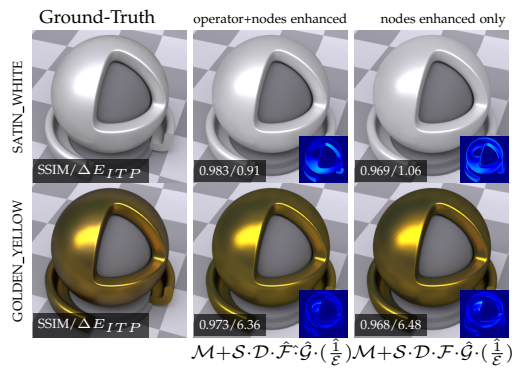

\textbf{Neural Module Granularity.}
\note{In~\figref{fig:granularity}, we show two enhanced models with different module granularity: one model allows both nodes and operators to be enhanced, while the other only nodes replaceable with neural modules. The former shows a higher reconstruction quality due to its smaller granularity.}

\begin{figure}[htbp]
    \centering
    \begin{minipage}{3.4in}
    \begin{minipage}{0.02in}	
            \centering
                \vspace{0.2in}
                \rotatebox{90}{\scriptsize \textsc{ }}
            \end{minipage}
        \begin{minipage}{3.4in}
            \centering
            
            \begin{minipage}{0.23\linewidth}
                \centering
                \subcaption{\scriptsize \shortstack{Original}}
            \end{minipage}			
            \begin{minipage}{0.23\linewidth}
                \centering
                \subcaption{\scriptsize \shortstack{Edited $\rho_d$}}
            \end{minipage}		
            \begin{minipage}{0.23\linewidth}
                \centering
                \subcaption{\scriptsize \shortstack{$0\times\rho_d$}}
            \end{minipage}		
            \begin{minipage}{0.23\linewidth}
                \centering
                \subcaption{\scriptsize \shortstack{$4\times\rho_s$}}
            \end{minipage}	
        \end{minipage}
    \end{minipage}

    \centering		
    \begin{minipage}{3.4in}
        \begin{minipage}{0.02in}	
            \centering
                \rotatebox{90}{\tiny \uppercase{Leaf\_maple}}
            \end{minipage}	
             \begin{minipage}{3.4in}	
                \centering
            \begin{minipage}{0.23\linewidth}
                \includegraphics[width=\linewidth]{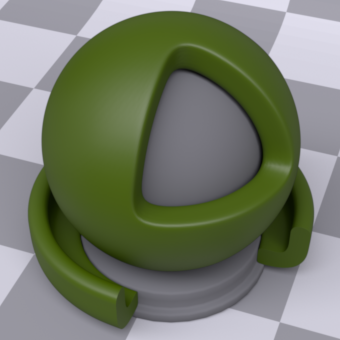}
            \end{minipage}
            \begin{minipage}{0.23\linewidth}
                \centering
                \includegraphics[width=\linewidth]{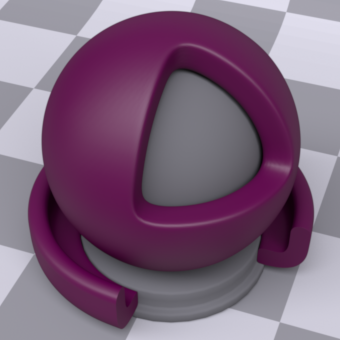}
            \end{minipage}
            \begin{minipage}{0.23\linewidth}
                \centering
                \includegraphics[width=\linewidth]{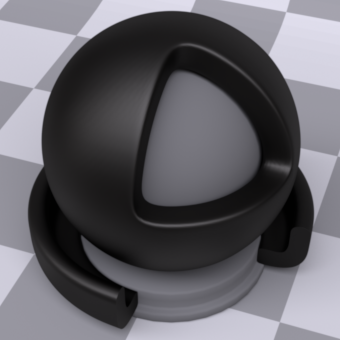}
            \end{minipage}
            \begin{minipage}{0.23\linewidth}
                \centering
                \includegraphics[width=\linewidth]{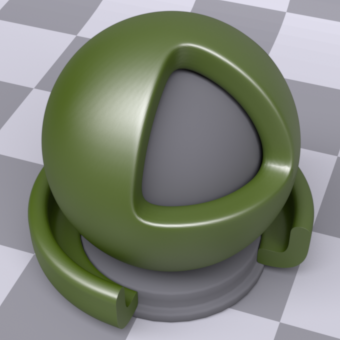}
            \end{minipage}
        \end{minipage}	
    \end{minipage}

   \caption{\note{Editing of our enhanced GGX model. From the first image to last, original, changing $\rho_d$ to a different color, setting $\rho_d$ to 0, and increasing $\rho_s$ by 4$\times$, respectively.}}
    \label{fig:pd}
\end{figure}

\textbf{Number of Epochs between Enhancement State Changes.}
We evaluate the impact of the number of epochs between consecutive state changes (\sec{sec:HOS}). When we set the number to 10, the order of boosted terms is $\mathcal{E}$, $\mathcal{G}$, $\mathcal{F}$, $\mathcal{\times}($of $\mathcal{D \times F})$; when 30 is set, the order is 
$\mathcal{E}$, $\mathcal{G}$, $\mathcal{\times}($of $ \mathcal{F \times G})$, $\mathcal{F}$; when 50 is set, the order is $\mathcal{E}$, $\mathcal{G}$, $\mathcal{F}$, $\mathcal{\times}($of $\mathcal{D \times F})$. Note that the enhanced models of the three cases are almost the same, despite the differences in the number of epochs between state changes.

\textbf{Material Editing.}
In~\figref{fig:pd}, we show material editing results of our enhanced GGX model. Since both diffuse and specular albedos remain in the model after our enhancement, we can directly change them for editing purposes. \xzs{Note that no analytical refitting is needed for importance sampling the edited materials in the figure.}


\begin{figure}[htbp!]
    \centering	
    \begin{minipage}{3.4in}
    \begin{minipage}{0.02in}	
            \centering
                \vspace{0.2in}
                \rotatebox{90}{\scriptsize \textsc{ }}
            \end{minipage}
        \begin{minipage}{3.4in}
            \centering
            \begin{minipage}{0.23\linewidth}
                \centering
               \subcaption{\scriptsize Ground-Truth}
            \end{minipage}			
            \begin{minipage}{0.23\linewidth}
                \centering
                \subcaption{\scriptsize Ours}
            \end{minipage}		
            \begin{minipage}{0.23\linewidth}
                \centering
                \subcaption{\scriptsize NBRDF~\cite{Sztrajman2021:NBRDF}}
            \end{minipage}		
            \begin{minipage}{0.23\linewidth}
                \centering
                \subcaption{\scriptsize NLBRDF~\cite{FAN2022:NLBRDF}}
            \end{minipage}	
        \end{minipage}
    \end{minipage}
    
    \begin{minipage}{3.4in}
        \begin{minipage}{0.02in}	
            \centering
            \rotatebox{90}{\text {\tiny \uppercase{green\_malachite}}}
        \end{minipage}	
        \begin{minipage}{3.4in}	
            \centering
            \begin{minipage}{0.23\linewidth}
                \centering
                \includegraphics[width=\linewidth]{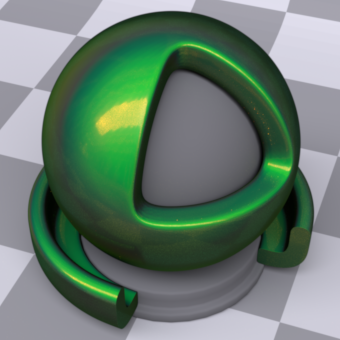}
                \put(-56,4.8) {\tikz[baseline] \node[fill=black, fill opacity=0.65, text opacity=1, text=white,inner sep=2pt] {\tiny SSIM/$\Delta E_{ITP}$};}
            \end{minipage}
            \begin{minipage}{0.23\linewidth}
                \centering
                \includegraphics[width=\linewidth]{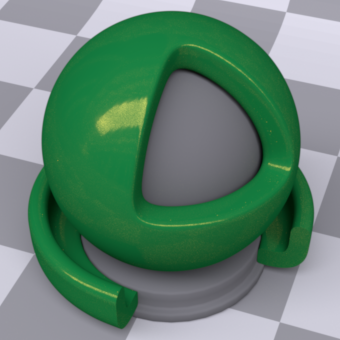}
                \put(-56,4.8) {\tikz[baseline] \node[fill=black, fill opacity=0.65, text opacity=1, text=white,inner sep=2pt] {\tiny 0.898/13.5};} 
                \put(-18,0){\includegraphics[width=18pt]{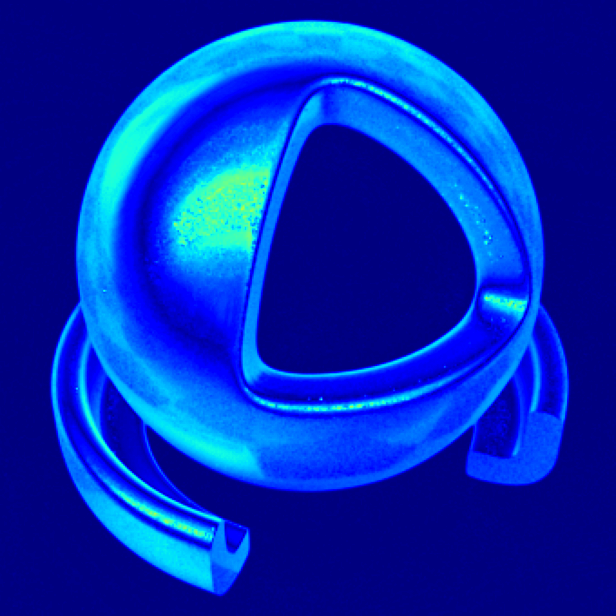}}
            \end{minipage}
            \begin{minipage}{0.23\linewidth}
                \centering
                \includegraphics[width=\linewidth]{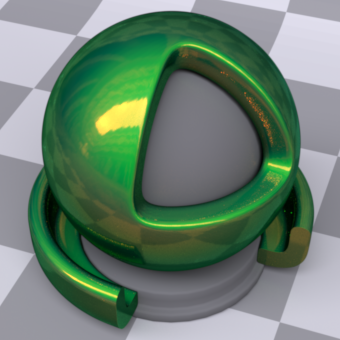}
                \put(-56,4.8) {\tikz[baseline] \node[fill=black, fill opacity=0.65, text opacity=1, text=white,inner sep=2pt] {\tiny 0.927/7.20};} 
                \put(-18,0){\includegraphics[width=18pt]{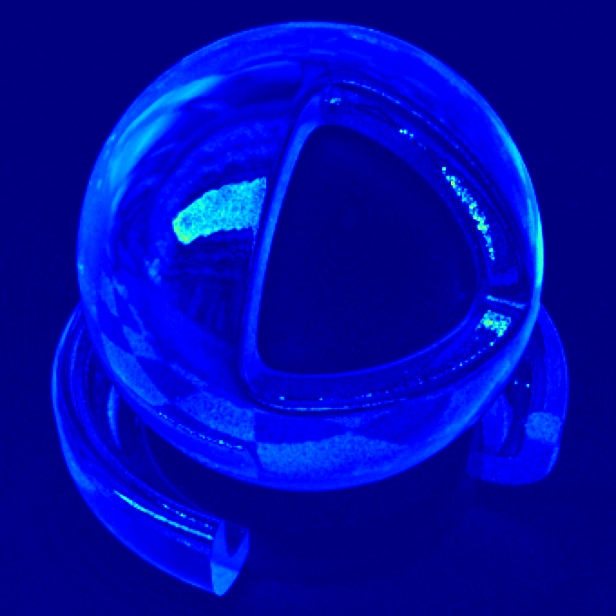}}
            \end{minipage}
            \begin{minipage}{0.23\linewidth}
                \centering
                \includegraphics[width=\linewidth]{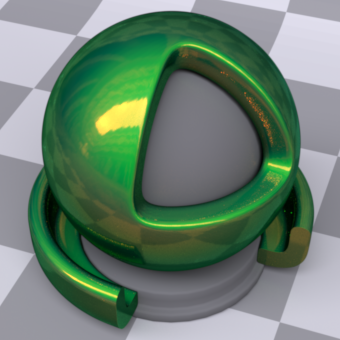}
                \put(-56,4.8) {\tikz[baseline] \node[fill=black, fill opacity=0.65, text opacity=1, text=white,inner sep=2pt] {\tiny 0.904/10.3};} 
                \put(-18,0){\includegraphics[width=18pt]{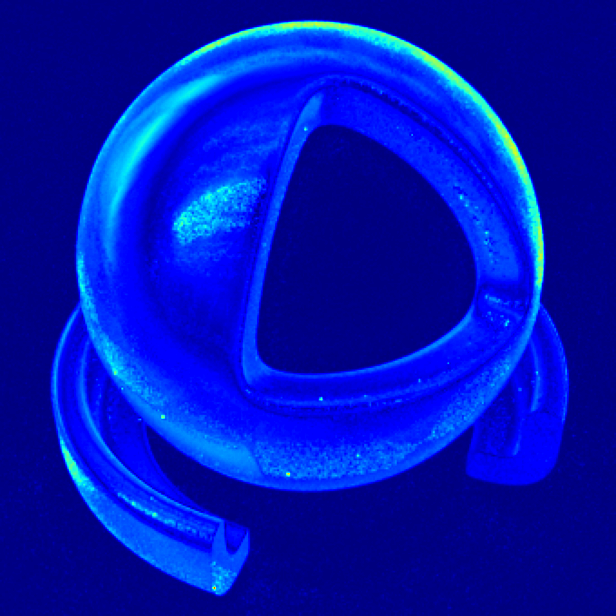}}
            \end{minipage}
        \end{minipage}
    \end{minipage}
   
  \caption{Limitation. We cannot accurately fit a sample, which cannot be well represented by the input analytical model, and substantially deviates from the training data, such as the color-changing BRDF here. While NBRDF~\cite{Sztrajman2021:NBRDF} can produce a good fitting, its network is dedicated to a particular BRDF only.}
\label{fig:color_change}
\end{figure}
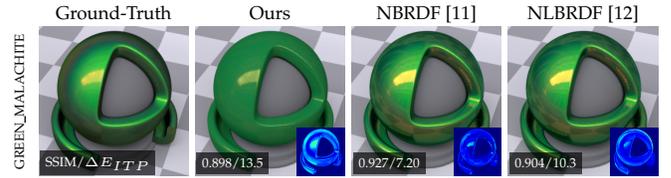

\section{Limitations \& Future Work}
\label{sec:future_work}
Our work is subject to a number of limitations. First, a neural enhanced model cannot generalize to cases that (1) cannot be represented by the input analytical model, and (2) are substantially different from the training data, \note{as shown in examples in~\figref{fig:color_change} and the supplemental material}. Next, our framework takes as input an analytical model, and does not propose new computation graphs. It will be interesting to combine existing generative approaches (e.g.,~\cite{Brady2014:Genbrdf}) with ours, to further pursue more efficient models. In addition, unlike the analytical parameters, the neural ones do not have immediate semantics. We are interested in applying related techniques~\cite{Zheng2023:CompactBRDF,guo2020materialgan} to support material editing with neural parameters. Finally, while we obtain high-quality visual results with no explicit constraint on energy conservation or reciprocity, it might be useful to strictly enforce the physical correctness.



In the future, we hope that our key idea of neural enhancement can offer a novel, alternative approach other than pure analytical or neural methods, which may bring unique benefits to other fields (e.g., procedural modeling), by making more use of domain knowledge embedded in existing models. Moreover, we are intrigued in automatically determining the granularity of computation nodes from an input analytical model, as well as applying new tools (e.g.,~\cite{liu2024kan}) to further improve the efficiency of neural modules. Finally, it is a promising direction to combine our idea with neural architecture search.




\ifCLASSOPTIONcompsoc
\section*{Acknowledgments}
\else
\section*{Acknowledgment}
\fi
This work is partially supported by NSF China (62332015, 62227806 \& 62421003), the Fundamental Research Funds for the Central Universities (226-2023-00145), a gift from Adobe, the XPLORER PRIZE, Information Technology Center, State Key Lab of CAD\&CG, Zhejiang University, the Central Government-Guided Fund for Local Science and Technology Development (2025ZY01034) and Provincial Key Laboratory Program (2025E10048).

\bibliographystyle{ieeetr}
\bibliography{reference}


\begin{IEEEbiography}[{\includegraphics[width=1in,height=1.25in,clip,keepaspectratio]{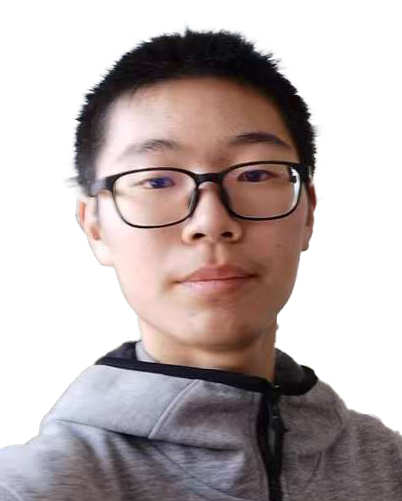}}]{Xuanzhe Shen} is a master student in the State Key Lab of CAD~\&~CG, Zhejiang University. He received his B.Eng. degree from the same university in 2024. His research interests include appearance reconstruction and rendering.
\end{IEEEbiography}

\begin{IEEEbiography}[{\includegraphics[width=1in,height=1.25in,clip,keepaspectratio]{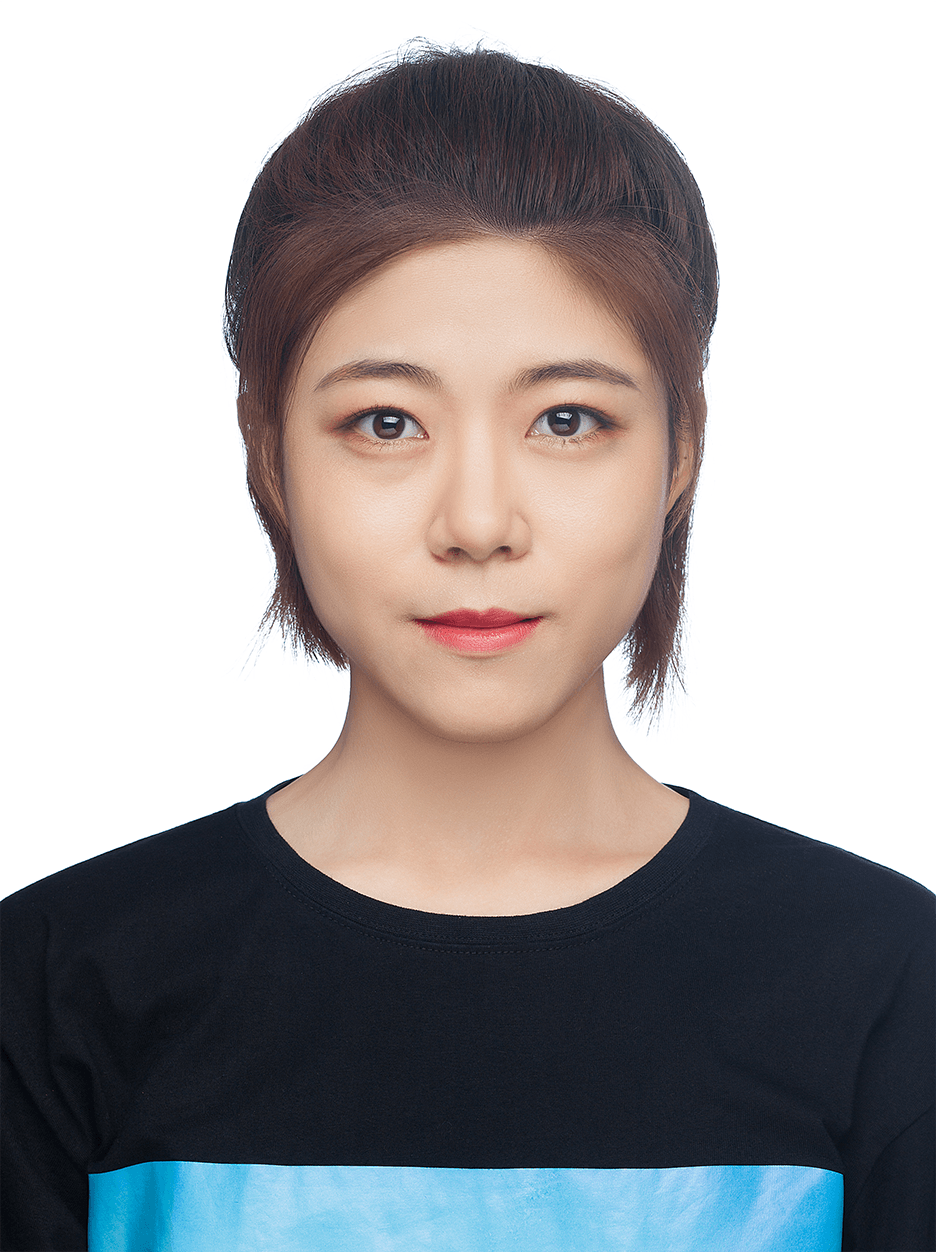}}]{Xiaohe Ma} is currently a Senior Research Scientist at Meshy AI. She received her Ph.D. degree in Computer Science from the State Key Lab of CAD \& CG, Zhejiang University, in 2025, and her B.Eng. degree from the School of Data and Computer Science, Sun Yat-sen University, in 2019. Her research interests include texture generation and 3D/4D generation with learning-based techniques.
\end{IEEEbiography}

\begin{IEEEbiography}[{\includegraphics[width=1in,height=1.25in,clip,keepaspectratio]{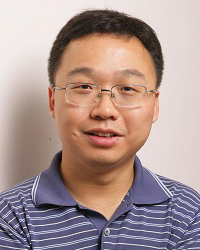}}]{Kun Zhou} is a Cheung Kong Professor of Computer Science at Zhejiang University, the Director of the State Key Lab of CAD\&CG and Hangzhou Research Institute of Holographic and AI Technology. He received his Ph.D. degree from Zhejiang University, and then spent six years with Microsoft Research Asia, serving as a lead researcher of the graphics group before returning to Zhejiang University. He was named one of the world's top 35 young innovators by MIT Technology Review (2011), and received an Asiagraphics Outstanding Technical Contributions Award (2022) and an ACM SIGGRAPH Test-of-Time Award (2024). He is a Fellow of IEEE and ACM.
\end{IEEEbiography}

\begin{IEEEbiography}[{\includegraphics[width=1in,height=1.25in,clip,keepaspectratio]{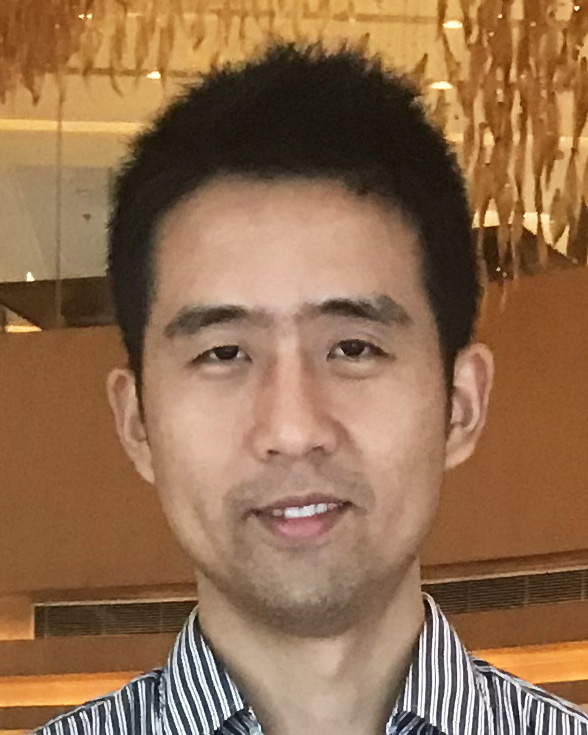}}]{Hongzhi Wu}
is a professor in the State Key
Lab of CAD \& CG, Zhejiang University. He received B.Sc. in computer science from Fudan
University, and Ph.D. in computer science from
Yale University. His current research interests
include high-density illumination multiplexing devices and differentiable acquisition. Hongzhi is
a recipient of Excellent Young Scholars, NSF
China. He is on the editorial board of IEEE
TVCG.
\end{IEEEbiography}

\end{document}